\documentclass[11pt]{article}
\pdfoutput=1
\usepackage{jcapmod}

\usepackage{tikz}
\usetikzlibrary{calc}
\usepackage{booktabs}
\usepackage{bookmark}
\usepackage{dsfont}
\usepackage{geometry}
\usepackage[english]{babel}
\usepackage{amsmath,amssymb,amsbsy,amstext, amsthm, simplewick}
\usepackage{hyperref}
\usepackage{graphicx}
\usepackage{amsfonts}
\usepackage[small]{caption}
\usepackage{upgreek}
\usepackage{hyperref}
\usepackage{enumerate}   
\usepackage{slashed}
\usepackage[makeroom]{cancel}
\usepackage{tabularx}
\usepackage[normalem]{ulem}

\usepackage[export]{adjustbox}
\usepackage{bm}

\usepackage{tikz}
\usepackage{tikz-feynman}
\tikzfeynmanset{
  graviton/.style={
    double,
    decoration={snake, amplitude=0.6mm, segment length=2mm},
    decorate
  }
}


\definecolor{Red}{RGB}{214, 39, 40}
\definecolor{Blue}{RGB} {31, 119, 180}
\definecolor{Orange}{RGB}{255, 153, 51}
\definecolor{Purple}{RGB}{178, 102, 255}
\definecolor{Green}{RGB}{44, 160, 44}
\definecolor{regal}{RGB}{90,0,120}

\definecolor{darkblue}{rgb}{0.15,0.35,0.55}
\definecolor{reddish}{rgb}{0.65, 0.2, 0.2}
\usepackage[linktocpage=true]{hyperref}
\hypersetup{
colorlinks=true,
citecolor=darkblue,
linkcolor=reddish,
urlcolor=darkblue,
pdfauthor={},
pdftitle={},
pdfsubject={}
}

\newcolumntype{L}[1]{>{\raggedright\arraybackslash}p{#1}}
\newcolumntype{C}[1]{>{\centering\arraybackslash}p{#1}}
\newcolumntype{R}[1]{>{\raggedleft\arraybackslash}p{#1}}
\newcolumntype{M}[1]{>{\centering\arraybackslash}m{#1}}
\newcolumntype{N}{@{}m{0pt}@{}}

\makeatletter
\g@addto@macro\bfseries{\boldmath}
\makeatother

\def\d{{\rm d}}

\def\pQ{\widehat{Q}}
\def\tX{\widetilde{X}}
\def\tO{\widetilde{\cal O}}

\usepackage{fontawesome5}

\makeatletter
\newcommand{\github}[1]{%
  \href{#1}{\faGithub}%
}
\makeatother

\NewDocumentCommand{\colornucleus}{omme{_^}}{%
 \begingroup\colorlet{currcolor}{.}%
  \IfValueTF{#1}
  {\textcolor[#1]{#2}}
  {\textcolor{#2}}
  {%
 #3
  \IfValueT{#4}{_{\textcolor{currcolor}{#4}}}
 \IfValueT{#5}{^{\textcolor{currcolor}{#5}}}
 }
}

\newcommand{\la}{\langle}
\newcommand{\ra}{\rangle}
\newcommand{\cO}{\mathcal{O}}

\def\wt{\widetilde}

\usepackage{colortbl}
\definecolor{lightgreen}{cmyk}{0.2, 0, 0.2, 0.2}
\definecolor{lightgray}{cmyk}{0.1,0.2,0,0.1}
\definecolor{lightgray2}{cmyk}{0.1,0.1,0,0.1}
\definecolor{repBlue}{RGB}{31, 119, 180}
\definecolor{repRed}{RGB}{	214, 39, 40}
\definecolor{repGreen}{RGB}{44, 160, 44}
\definecolor{repOrange}{RGB}{255, 127, 14}
\definecolor{repViolet}{RGB}{102,0,204}

\makeatletter
\newlength{\apb@width}
\newcommand{\autoparbox}[2][c]{\settowidth{\apb@width}{#2}\parbox[#1]{\apb@width}{#2}}

\makeatother

\usepackage[framemethod=default]{mdframed}
\newmdenv[skipabove=7pt,
skipbelow=7pt,
rightline=true,
leftline=true,
topline=true,
bottomline=true,
backgroundcolor=gray!10,
linecolor=black,
innerleftmargin=5pt,
innerrightmargin=5pt,
innertopmargin=5pt,
innerbottommargin=5pt,
leftmargin=0cm,
rightmargin=0cm,
linewidth=1pt]{eBox}

\usetikzlibrary{shapes.misc}

\numberwithin{equation}{section}

\usepackage{slashed}

\def\beq{\begin{equation}}
	\def\eeq{\end{equation}}
\def\bea{\begin{eqnarray}}
	\def\eea{\end{eqnarray}}
\def\be{\begin{equation}}
	\def\ee{\end{equation}}
	
\def\x{x}
\def\z{z}

\def\hs{\hskip 1pt}

\allowdisplaybreaks

\usepackage{tocloft}

\begin{document}

\newgeometry{top=2cm, bottom=2cm, left=2cm, right=2cm}

\begin{titlepage}
	\setcounter{page}{1} \baselineskip=15.5pt 
	\thispagestyle{empty}

	\begin{center}
        {\fontsize{19.}{18} \bf Bootstrapping Weakly Broken Gauge Theories in (A)dS}
	\end{center}

	\vskip 20pt
	\begin{center}
		\noindent
		{\fontsize{14}{18}\selectfont 
				Daniel Baumann\hs$^{1,2}$, Kurt Hinterbichler\hs$^{3}$,\\[8pt] Callum R.~T.~Jones\hs$^{4}$   and Nathan Meurrens\hs$^{2,5}$}
	\end{center}

	\begin{center}
    \vskip8pt
		\textit{$^1$ DAMTP, University of Cambridge,
Cambridge, CB3 0WA, UK}
			
		\vskip8pt
		\textit{$^2$  Leung Center for Cosmology and Particle Astrophysics,
				Taipei 10617, Taiwan}

        \vskip8pt
		\textit{$^3$  CERCA, Department of Physics,
				Case Western Reserve University, Cleveland, OH 44106, USA}
        
		\vskip8pt
		\textit{$^4$  Department of Physics, University of Arizona, Tucson, Arizona 85721, USA}

            \vskip8pt
		\textit{$^5$ Institute of Physics, University of Amsterdam, Amsterdam, 1098 XH, The Netherlands}
	\end{center}

\vspace{0.4cm}
\begin{center}{\bf Abstract}
\end{center}
\noindent	
We develop a bootstrap approach for weakly broken gauge theories in (anti-)de Sitter space, focusing on cases in which the conservation of the dual boundary current is broken by a double-trace operator. Integrating the corresponding Ward identity, we derive pseudo-charge conservation identities that contain fixed nonlocal contributions as integrals of three-point functions. These identities impose consistency conditions on the space of allowed theories. As an illustrative example, we study Yang--Mills theory in AdS with symmetry-breaking boundary conditions for charged matter and show that the pseudo-charge conservation identities constrain the mixing between the two quantization sectors. We then apply the framework to the Higgs mechanism for gravity in two AdS spaces with a common boundary, coupled so that the corresponding stress tensors are not separately conserved. In this setting, the ordinary charge conservation identities recover the factorization into two independent CFTs, whereas the pseudo-charge conservation identities hold for arbitrary values of the symmetry-breaking parameter. Finally, for conformal gravity in de Sitter space, we constrain the interactions between the graviton and the partially massless spin-2 field, both in the minimal theory and in the presence of additional scalar or vector matter, finding agreement with the predictions of the corresponding bulk theories.
\end{titlepage}
\restoregeometry

\clearpage
\setcounter{page}{2}
\makeatletter
\def\ps@plain{
  \renewcommand\@oddfoot{\hfill\thepage\hfill}
  \renewcommand\@evenfoot{\hfill\thepage\hfill}
}
\pagestyle{plain}
\makeatother

\setcounter{tocdepth}{2}

\begingroup
\renewcommand{\baselinestretch}{1.0}\normalsize
\tableofcontents
\endgroup

\linespread{1.1}

\newpage
\section{Introduction}\label{sec: intro}
Gauge symmetry is a central organizing principle of quantum field theory. In Minkowski spacetime, it sharply constrains the interactions of massless particles and the allowed long-range forces they can mediate~\cite{Weinberg:1965nx, Benincasa:2007xk}. A similar rigidity is present in anti-de Sitter (AdS) and de Sitter (dS) space, where bulk gauge symmetries can be studied through the algebra of boundary currents. Massless spin-$s$ fields are associated with conserved spin-$s$ currents, while partially massless (PM) fields correspond to partially conserved currents. The integrated Ward identities of these currents act directly on boundary correlators and therefore provide a powerful way of constraining the bulk interactions without assuming a particular Lagrangian description.

\vskip 4pt
In~\cite{Baumann:2025tkm}, the boundary current algebra of partially conserved currents was used to constrain the interactions of PM fields in (A)dS, building on earlier work by Maldacena and Zhiboedov~\cite{Maldacena:2011jn}. Each (partially) conserved current $J$ has an associated conserved charge $Q$. Assuming that the action of the charge on any operator can be written as an expansion in local operators, $[Q,\cO_i] = \sum_j a_{ij} \cO_j$, and that the vacuum is invariant under the symmetry, leads to a set of ``charge conservation identities.'' For example, $\la [Q,\cO_1 \cO_2 \cO_3]\ra=0$ implies the sum rule
\begin{equation}\label{equ:CC}
\sum_j \Big[a_{1j} \la \cO_j \cO_2 \cO_3\ra + a_{2j} \la \cO_1 \cO_j \cO_3\ra + a_{3j} \la \cO_1 \cO_2 \cO_j\ra\Big] = 0\,.
\end{equation}
Expanding each three-point function in conformally invariant tensor structures, $\la \cO_i \cO_j \cO_k \ra = \sum_n n_{ijk}^{(n)} F^{(n)}_{ijk}$, then leads to a constraint on the coefficients $a_{ij}$ and $n_{ijk}^{(n)}$. For the spectra studied in~\cite{Baumann:2025tkm}, these constraints led to sharp obstructions: in four-dimensional de Sitter space, PM fields of spin 2 or 3 cannot be coupled consistently to gravity without additional massive fields that violate unitarity, while in higher dimensions a consistent coupling requires additional PM fields and may ultimately point toward an infinite tower of higher-spin states.

\vskip 4pt
In this paper, we develop the corresponding bootstrap when the bulk gauge symmetry is weakly broken. On the boundary, this breaking manifests itself as a weak violation of current conservation. Schematically, we have\footnote{Equations of the form~\eqref{equ:broken} were recently derived for late-time de Sitter correlators in scalar QED, Yang--Mills theory, and gravity, where boundary currents and the stress tensor recombine with double-trace composites built from shadow-paired boundary modes and consequently acquire anomalous dimensions~\cite{Sleight:2025dmt}.} 
\begin{equation}\label{equ:broken}
\partial^{s-t} X_{(s,t)} = g\,\mathcal O_i\mathcal O_j\,,
\end{equation}
where the right-hand side is a double-trace operator and the small parameter $g$ controls the strength of the symmetry breaking. When $J^{(s,t)}$ is a spin-$s$, depth-$t$ current, the left-hand side contains $s-t$ divergences. The consequences of this symmetry breaking can be captured by defining a \textit{pseudo-charge} $\pQ$, with the charge conservation identities like \eqref{equ:CC} replaced by the corresponding ``pseudo-charge conservation identities''\,\cite{Maldacena:2012sf}. 

\vskip 4pt
An important feature of the pseudo-charge is that it receives a {\it nonlocal} contribution arising from the integrated double-trace operator. For example, in the case shown in (\ref{equ:broken}), the action of the pseudo-charge is
\begin{equation}
\begin{aligned}
    \big[\pQ, \cO_i(x)\big] &= \big[Q, \cO_i(x)\big] - g\, \wt\cO_j(x)\,,
\end{aligned}
\label{eq: pseudo-charge action intro}
\end{equation}
and similarly for $[\pQ,\cO_j(x)]$,
where the nonlocal transform is
\begin{align}
    \wt\cO_j(x) \equiv \int \d^d x'\, \la \cO_i(x')\hs\cO_i(x)\ra\,\cO_j(x')\,.
\label{eq:nonlocal-transform general}
\end{align}
We see that the local part of the pseudo-charge action contains the usual unknown current-algebra coefficients $a_{ij}$, whereas the coefficient of the nonlocal term is fixed completely by the non-conservation equation (\ref{equ:broken}). The resulting pseudo-charge conservation identities mix ordinary three-point functions with convolution integrals of three-point functions. The transform in \eqref{eq:nonlocal-transform general} is generally not a conformal shadow transform, so non-conformal terms must cancel nontrivially between the local and nonlocal parts of the identity. 

\vskip 4pt
We first study symmetry breaking by boundary conditions in anti-de Sitter space. In pure {\it Yang--Mills theory}, ordinary charge conservation reproduces the Jacobi identity for the structure constants. After coupling to scalar matter, it also requires the scalar couplings to furnish a representation of the same Lie algebra. We then impose different boundary conditions on the scalar components. Generators that mix the two quantization sectors are broken, and the associated currents become weakly non-conserved. The pseudo-charge identities nevertheless reconstruct the full non-Abelian representation on the scalar fields, including the generators responsible for the breaking. In this way, the boundary correlators recover not only the unbroken subalgebra, but also its embedding into the complete bulk gauge algebra.

\vskip 4pt
We also apply the formalism to a model of {\it massive bi-gravity} in $\mathrm{AdS}\times\mathrm{AdS}$, consisting of two AdS spaces that share a common conformal boundary~\cite{Aharony:2006hz, Kiritsis:2006hy}. It is well known that, in flat space, a theory cannot contain multiple massless spin-2 fields (gravitons) with nontrivial cross-couplings~\cite{Weinberg:1965nx,Boulanger:2000rq}. The corresponding statement in AdS is that the boundary theory containing multiple stress tensors must factorize into decoupled CFTs~\cite{Maldacena:2011jn}. Any cross-couplings between the boundary CFTs will break conservation of the additional stress tensors. This is the holographic counterpart of the statement that transparent boundary conditions in AdS$_4$ generate a graviton mass~\cite{Porrati:2001db,Karch:2000ct}.\footnote{This mechanism is also closely related to double-holographic constructions of entanglement islands~\cite{Almheiri:2019hni}. In the known higher-dimensional examples, coupling a gravitating system to an auxiliary bath is accompanied by a massive graviton, and the graviton mass can be essential for the island contribution~\cite{Geng:2020qvw}.} An explicit realization of this phenomenon was studied in~\cite{Aharony:2006hz, Kiritsis:2006hy}. In this case, the action of the pseudo-charge can be computed directly in conformal perturbation theory. The model therefore serves as a controlled test of the formalism. Applying the bootstrap approach to this model, we find perfect agreement with the perturbative computation carried out in Appendix~\ref{sec: CPT of AdSxAdS}.

\vskip 4pt
Finally, we turn to {\it conformal gravity} in four-dimensional de Sitter space. Unlike in AdS, there is no analogous freedom to impose independent boundary conditions at the future boundary: the initial state fixes the late-time behavior of the fields. Nevertheless, interactions break the bulk gauge symmetry. The linearized spectrum of conformal gravity contains a massless graviton and a partially massless spin-2  field \cite{Maldacena:2011mk}, dual to a conserved stress tensor $T_{\mu\nu}$ and a partially conserved current $X_{\mu\nu}$. Nonlinear interactions deform the conservation equation for the PM current to
\begin{equation}\label{equ:CG}
\partial_\mu\partial_\nu X^{\mu\nu} = g\hs X_{\mu\nu}X^{\mu\nu}\,.
\end{equation}
Note that the conservation of the PM current is broken by a double-trace deformation of the current itself. In the bulk, this corresponds to a form of {\it self-Higgsing} of the PM gauge symmetry. For the minimal spectrum $\{T,X\}$, pseudo-charge conservation sets the $\la XTT\ra$ correlator to zero and fixes all remaining parity-even three-point functions up to the overall gravitational normalization, reproducing precisely the cubic data of conformal gravity (see Appendix~\ref{sec: Details on CG}). Coupling conformal gravity to scalar and vector matter opens additional symmetry-breaking channels with breaking parameters $g'$ and $g''$. We find that the size of $g'$ and $g''$ is fixed in terms of $g$, again in perfect agreement with the bulk analysis.

\paragraph{Outline}
The remainder of the paper is organized as follows. In Section~\ref{sec:review}, we review the boundary description of gauge fields in (A)dS and derive the pseudo-charge conservation identities for weakly broken currents. In Section~\ref{sec: YM in AdS}, we apply this formalism to Yang--Mills theory in AdS, first with gauge-invariant boundary conditions and then with boundary conditions that break part of the gauge symmetry. In Section~\ref{sec: Bi-Gravity in AdSxAdS}, we study the ${\rm AdS}\times{\rm AdS}$ model of massive gravity, and show that ordinary charge conservation forces the boundary theory to be a product of two CFTs; we also confirm that the pseudo-charge conservation identities are satisfied for any value of the breaking parameter. In Section~\ref{sec: Higgsing in CG}, we analyze conformal gravity in dS, both in the pure spin-two sector and after coupling to scalar and vector matter. Finally, our conclusions are stated in Section~\ref{sec: conclusions}.

\vskip 4pt
The appendices contain additional technical material: In Appendix~\ref{sec: evaluating tensor integrals}, we develop tools for evaluating the tensor integrals that arise from the nonlocal transforms in the pseudo-charge conservation identities. In Appendix~\ref{sec:semianalytic}, we give a worked example in which the conservation of a spin-4, depth-0 field holds up to semi-analytic terms. In Appendix~\ref{app: Semi-Local Terms}, we show how differential regularization determines the semi-local terms needed to complete conformal correlators. In Appendix~\ref{sec: CPT of AdSxAdS}, we derive the pseudo-charge action of the ${\rm AdS}\times{\rm AdS}$ model directly in conformal perturbation theory. Finally, in Appendix~\ref{sec: Details on CG}, we present the bulk analysis of conformal gravity.

\paragraph{Notation}
Throughout the paper, we use natural units, $\hbar=c\equiv 1$, and the mostly-plus convention for the bulk metric. We work in $D\geq 4$ bulk spacetime dimensions and write $d=D-1$ for the boundary dimension. Bulk coordinates are denoted by $X^M$, with $M=0,1,\ldots,d$, and boundary coordinates by $x^\mu$, with $\mu=0,1,\ldots,d-1$ (for AdS) and $\mu=1,\ldots,d$ (for dS). 
Embedding-space coordinates and polarization vectors are written as $P^A$ and $Z^A$, with $A=0,1,\ldots,d+1$. 

\vskip 4pt
We represent symmetric-traceless operators in index-free notation as
\begin{equation}\label{equ:index-free}
\cO^{(s)}(\x,\z) \equiv \cO_{\mu_1\ldots\mu_s}(\x)\, z^{\mu_1}\cdots z^{\mu_s}\,,
\end{equation}
where $z^\mu$ is an auxiliary null vector. Indices can be restored by acting with the Thomas--Todorov operator~\cite{thomas,Dobrev:1975ru,tractors}
\begin{equation}\label{eq: Todorov}
D^{\mu}_z \equiv \left(\frac{d}{2}-1 + \z \cdot \frac{\partial}{\partial \z}\right)\frac{\partial}{\partial z_{\mu}}-\frac{1}{2}\hs z^{\mu} \frac{\partial^2 }{\partial \z \cdot \partial \z}\,,
\end{equation}
which automatically projects onto the symmetric-traceless part. 
 In embedding space, conformal correlators can then be written in terms of the following building blocks \cite{Costa:2011mg}:
\begin{equation}
\begin{aligned}
P_{ij}&\equiv P_i\cdot P_j\,,\\[2pt]
H_{ij}&\equiv -2\Big[(P_i\cdot P_j)(Z_i\cdot Z_j)-(P_i\cdot Z_j)(P_j\cdot Z_i)\Big]\,,\\[2pt]
V_{i,jk}&\equiv \frac{(Z_i\cdot P_j)(P_i\cdot P_k)-(Z_i\cdot P_k)(P_i\cdot P_j)}{P_j\cdot P_k}\,.
\end{aligned}
\label{equ:conformal-structures}
\end{equation}
We will write $V_{i,jk} \equiv V_i$ as a shorthand, where the indices $j,k$ are implicit, assuming cyclic ordering. On the Poincar\'e section, we have $-2P_{ij}=x_{ij}^2 \equiv (x_i -x_j)^2$. 

\vskip 4pt
We will often use the shorthand $\cO_i \equiv \cO_i(x_i)$, where the subscript denotes both the position of the operator and possibly its flavor. Double brackets $\la\!\la\cO_1\cdots\cO_n\ra\!\ra$ denote the kinematic conformal structure with its overall coefficient and internal-symmetry tensor stripped off. Two structures will appear frequently: the two-point function of a spin-$s$ operator $Y$ with scaling dimension $\Delta_Y$, and its three-point function with two scalar operators $\cO_i$ and $\cO_j$:
\begin{align}\label{eq: YY structure}
\la\!\la Y Y\ra\!\ra &\equiv \frac{H_{12}^{\hs s}}{(-2\hs P_{12})^{\Delta_Y+s}}\,,\\[4pt] \label{eq: YOO strcuture}
\la\!\la Y \cO_i\hs \cO_j\ra\!\ra &\equiv \frac{V_{1}^{\hs s}}{(-2P_{12})^{\tfrac{1}{2}(\Delta_Y+\Delta_i-\Delta_j+s)} (-2P_{23})^{\tfrac{1}{2}(\Delta_i+\Delta_j-\Delta_Y-s)} (-2P_{31})^{\tfrac{1}{2}(\Delta_Y+\Delta_j-\Delta_i+s)}}\,.
\end{align}
Structures involving more than one spinning operator are built from the same ingredients $P_{ij}$, $H_{ij}$ and $V_{i,jk}$, and will be introduced when needed.

\section{Background and Strategy}
\label{sec:review}
We will begin by introducing the cast of characters of this work. In Section~\ref{sec: Holographic Dictionary}, we review how boundary conditions in (anti-)de Sitter space determine the dual operator content, and how a bulk gauge symmetry implies the conservation of a boundary current. We also present the Ward identities obeyed by the boundary correlators. In Section~\ref{sec: Weakly Broken Gauge Symmetry}, we allow this symmetry to be weakly broken, so that the conservation of the associated boundary current is broken by a double-trace operator. Finally, in Section~\ref{sec: pseudo}, we demonstrate that the charge conservation identities continue to hold in a modified form, with the charge replaced by a pseudo-charge whose action is partly nonlocal.

\subsection{Holographic Dictionary}
\label{sec: Holographic Dictionary}
For most of this section, we will work in Euclidean ${\rm AdS}_{d+1}$, which in Poincar\'e coordinates has the line element
\begin{equation}\label{eq: AdS metric}
\d s^2 = L^2\, \frac{\d u^2 + \d \x^2}{u^2}\,,
\end{equation}
with the conformal boundary located at $u \to 0$. We set the AdS radius to unity, $L\equiv 1$. The case of de Sitter space is related by analytic continuation and will be discussed at the end of this section. Readers familiar with this material may skip ahead to Section~\ref{sec: Weakly Broken Gauge Symmetry}.

\subsubsection{Boundary Operators}
Bosonic spin-$s$ fields in the bulk are described by symmetric tensor fields $A_{M_1 \cdots M_s}$, which on-shell are transverse ($\nabla^M A_{M N_2\cdots N_s} = 0$) and traceless ($A^M_{\ \ M N_3\cdots N_{s}} = 0$), and satisfy 
\begin{equation}
    \Big(\square + \big[\hs d - (s-1)(s+d-4)\big] - m^2 \Big) A_{M_1\cdots M_s} = 0\,.
\label{eq: bulk eom}
\end{equation}
Near the boundary, the components of the field along the boundary directions have two characteristic fall-offs,
\begin{equation}\label{eq: falloffs}
A_{\mu_1\cdots\mu_s}(u,\x) \ \xrightarrow{\ u \hs \to\hs 0\ }\ \alpha_{\mu_1\cdots\mu_s}(\x)\, u^{\hs\Delta_- - s} + \beta_{\mu_1\cdots\mu_s}(\x)\, u^{\hs\Delta_+ - s}\,,
\end{equation}
where the exponents are determined by the mass and spin of the field,
\begin{equation}\label{eq: Delta pm}
\Delta_\pm = \frac{d}{2} \pm \sqrt{\left(\frac{d}{2}+s-2\right)^{\!2} + m^2}\,.
\end{equation}
Scalar fields obey a different relation, which coincides with \eqref{eq: Delta pm} evaluated at $s=2$. The AdS isometries act as conformal transformations on the boundary, under which $\alpha_{\mu_1\cdots\mu_s}$ and $\beta_{\mu_1\cdots\mu_s}$ transform as spin-$s$ conformal primaries of weight $\Delta_-$ and $\Delta_+$, respectively.

\vskip 4pt
In the standard holographic dictionary~\cite{Witten:1998qj}, the leading coefficient $\alpha$ is interpreted as the source of a spin-$s$ primary operator $\cO_{\mu_1\cdots\mu_s}$ of dimension $\Delta_+$ in the partition function of the dual conformal field theory (CFT),
\begin{equation}\label{eq: AdS/CFT}
    Z_{\rm AdS}[\alpha] = \Big\langle \exp\left( \int \d^d x \, \alpha_{\mu_1\cdots\mu_s}\hs \cO^{\mu_1\cdots\mu_s} \right)\Big\rangle_{\rm CFT}\,,
\end{equation}
while the subleading coefficient $\beta$ determines the one-point function, $\beta \propto \la \cO_{\mu_1\cdots\mu_s} \ra$. More generally, each bulk field $A^{(i)}$ contributes its own source $\alpha_i$ to \eqref{eq: AdS/CFT}, and boundary correlation functions of the dual operators $\cO_i$ are obtained by functional differentiation with respect to the corresponding sources,
\begin{equation}\label{eq: correlators from Z}
    \la \cO_{1,\hs\mu_1\cdots\mu_{s_1}}(\x_1) \cdots\hs \cO_{n,\hs\nu_1\cdots\nu_{s_n}}(\x_n) \ra = \frac{\delta^{\hs n} \log Z_{\rm AdS}[\alpha_i]}{\delta \alpha_1^{\mu_1\cdots\mu_{s_1}}(\x_1) \cdots\, \delta \alpha_n^{\nu_1\cdots\nu_{s_n}}(\x_n)}\bigg|_{\alpha_i\hs=\hs0}\,,
\end{equation}
where the logarithm selects the connected correlators. 

\vskip 4pt
Throughout, we take the dual CFT to be a large-$N$ theory, dual to a weakly coupled, semiclassical bulk. The parameter $N$ counts the boundary degrees of freedom and is fixed by the bulk couplings, with the central charge scaling as $N^2$. Normalizing the single-trace operators to have unit two-point functions up to a sign, their connected $n$-point functions scale as $\la \cO_1 \cdots \cO_n \ra \sim N^{2-n}$. As a result, correlators factorize into sums of products of two-point functions, up to $1/N$-suppressed connected pieces. We will use this factorization property repeatedly.

\subsubsection{Mixed Boundary Conditions}
The dictionary described so far, with a source $\alpha$ and one-point function $\beta$, corresponds to \textit{Dirichlet} boundary conditions, in which the coefficient of the leading fall-off in \eqref{eq: falloffs} is held fixed. This is often called \textit{standard quantization}. However, when the mass of the field lies in the Klebanov--Witten window~\cite{Breitenlohner:1982jf, Klebanov:1999tb} (quoted here for scalars), 
\begin{equation}\label{eq: quantization window}
    -\frac{d^2}{4} \,\le\, m^2 \,<\, -\frac{d^2}{4} + 1\,,
\end{equation}
\textit{both} fall-offs are normalizable, and it is equally consistent to impose \textit{Neumann} boundary conditions, holding $\beta$ fixed instead. In this \textit{alternate quantization}~\cite{Klebanov:1999tb}, the roles of source and response are exchanged: $\beta$ sources an operator of dimension $\Delta_-$, and the generating functionals of the two theories are related by a Legendre transform. A recurring example in this paper is a conformally coupled scalar in ${\rm AdS}_4$, with $m^2 = -2$: standard quantization yields an operator $\cO_2$ of dimension $\Delta_+ = 2$, while alternate quantization yields an operator $\cO_1$ of dimension $\Delta_-=1$.\footnote{Analogous choices exist for spinning fields. For instance, both fall-offs of a Maxwell field in ${\rm AdS}_4$ are normalizable: Dirichlet boundary conditions give a boundary theory with a conserved current $J^\mu$, while Neumann boundary conditions gauge the corresponding boundary global symmetry~\cite{Marolf:2006nd}. Unless stated otherwise, we will impose Dirichlet boundary conditions on the gauge fields, so that the dual currents exist as conserved operators in the boundary theory.}

\vskip 4pt
Importantly, for a scalar in the window \eqref{eq: quantization window}, we may also impose the \textit{mixed} boundary condition
\begin{equation}\label{eq: mixed bc}
    \beta(\x) = W'\big(\alpha(\x)\big)\,,
\end{equation}
for some function $W$. In the boundary CFT, this corresponds to deforming the theory with alternate quantization by a \textit{multi-trace} operator~\cite{Witten:2001ua, Berkooz:2002ug}
\begin{equation}
    S \ \to \ S + \int \d^d x \ W\big(\cO(\x)\big)\,.
\label{eq: multitrace deformation}
\end{equation}
The simplest nontrivial case is the \textit{double-trace} deformation $W = \frac{1}{2}\hs g \hs \cO^2$, corresponding to the linear boundary condition $\beta = g\hs\alpha$. For an operator of dimension $\Delta_-$, the deformation has dimension $2\Delta_- < d$ and is relevant: it triggers an RG flow from the theory with alternate quantization in the UV to the theory with standard quantization in the IR, with the dimension of $\cO$ flowing from $\Delta_-$ to $\Delta_+$~\cite{Witten:2001ua, Gubser:2002vv}.

\vskip 4pt
When several bulk fields are present, boundary conditions can also \textit{mix} different fields. Given two scalar operators $\cO_1$ and $\cO_2$, the deformation $g \int \d^d x\, \cO_1 \cO_2$ is marginal precisely when
\begin{equation}
    \Delta_1 + \Delta_2 = d\,,
\label{eq: marginality}
\end{equation}
which is automatic when $\cO_1$ and $\cO_2$ are dual to bulk fields of equal mass with opposite choices of quantization. At large $N$, where the boundary correlators exhibit large-$N$ factorization, the effects of such deformations are computable in conformal perturbation theory.

\subsubsection{Gauge Fields and Conserved Currents}
Our main focus will be on theories containing gauge fields in the bulk, whose boundary duals are conserved currents. Gauge symmetry arises only at special values of the bulk mass, where the corresponding representation becomes shortened. In flat spacetime, this shortening leads to the familiar massless fields. In (A)dS spacetime, however, the richer representation theory~\cite{Sun:2021thf, Hinterbichler:2026xqf} also allows for partially massless fields, which possess an intermediate gauge symmetry and propagate fewer degrees of freedom than a generic massive field, but more than a massless field.

\paragraph{Massless fields} We begin with the massless case. For a spin-$s$ field, gauge invariance emerges for $m^2=0$. At this value, the field transforms as
\begin{equation}
    \delta_\xi A_{M_1\cdots M_s} = \nabla_{(M_1} \xi_{M_2 \cdots M_s)}\,,
\label{eq: massless gauge transformation}
\end{equation}
with a symmetric, traceless, rank-$(s-1)$ gauge parameter $\xi_{M_2 \cdots M_s}$. Choosing the gauge parameter to have the same leading fall-off as the field, the boundary source inherits the transformation 
\begin{equation}
    \delta_\xi\hs \alpha^{(s)} = (\z\cdot\partial)\hs \xi^{(s-1)}\,,
\label{eq: boundary gauge transformation}
\end{equation}
where $\xi^{(s-1)} \equiv z^{\mu_2} \cdots z^{\mu_s} \xi_{\mu_2 \cdots \mu_s}$ in index-free notation. Invariance of the source term in \eqref{eq: AdS/CFT} under \eqref{eq: boundary gauge transformation} then requires the dual operator to be a conserved current,
\begin{equation}
    (\partial \cdot D_z)\hs J^{(s)} = 0\,, \qquad \Delta_J = d - 2 + s\,,
\label{eq: exact conservation}
\end{equation}
where $J^{(s)} \equiv z^{\mu_1} \cdots z^{\mu_s} J_{\mu_1 \cdots \mu_s}$ and $D_z$ is the Thomas--Todorov operator defined in~\eqref{eq: Todorov}. Familiar examples are the massless vector, dual to a spin-1 current $J^\mu$ of dimension $d-1$, and the graviton, dual to the stress tensor $T^{\mu\nu}$ of dimension $d$. In this way, bulk gauge symmetries manifest themselves as global symmetries of the boundary theory.\footnote{The correspondence between bulk gauge fields and boundary conserved currents requires that both the bulk action \textit{and} the boundary conditions respect the gauge symmetry. As we describe in Section~\ref{sec: Weakly Broken Gauge Symmetry}, relaxing the second requirement is precisely the subject of this paper.} 

\paragraph{Partially massless fields} In (A)dS, massless fields are not the only special points in the representation theory. For the discrete set of masses\hs\footnote{We quote masses in AdS units; in de Sitter, the sign is reversed, $m^2_{\text{dS}}L^2 = (s-t-1)(s+t+d-3)$, as in~\cite{Baumann:2025tkm}. This is the convention used in Section~\ref{sec: Higgsing in CG} and Appendix~\ref{sec: Details on CG}.}
\begin{equation}
    m^2 = -(s-t-1)(s+t+d-3)\,, \qquad t \in \{0,1,\ldots,s-1\}\,,
\label{eq: PM masses}
\end{equation}
the field develops a smaller gauge invariance, with a rank-$t$ gauge parameter and $s-t$ derivatives,
\begin{equation}
    \delta_\xi A_{M_1\cdots M_s} = \nabla_{(M_{t+1}}\cdots \nabla_{M_s}\hs \xi_{M_1\cdots M_t)} + \text{trace terms}\,.
\label{eq: PM gauge transformation}
\end{equation}
This symmetry removes the helicity components $0,\pm1,\ldots,\pm t$. The resulting fields are known as \textit{partially massless} fields~\cite{Deser:1983mm, Deser:2001pe} and are labeled by their \textit{depth}~$t$, with $t=s-1$ corresponding to the massless case. Repeating the argument that led to \eqref{eq: exact conservation}, the boundary source now shifts as $\delta_\xi \hs\alpha^{(s)} = (\z\cdot\partial)^{s-t}\hs\xi^{(t)}$, so the dual operator $X_{\mu_1\cdots\mu_s}$ is a \textit{partially conserved current}~\cite{Dolan:2001ih},
\begin{equation}
    (\partial\cdot D_z)^{s-t}\, X^{(s)} = 0\,, \qquad \Delta_X = d - 1 + t\,.
\label{eq: partial conservation}
\end{equation}
Note that, for $t<s-1$, $\Delta_X$ lies \textit{below} the unitarity bound, reflecting the fact that PM fields are non-unitary in AdS; in de Sitter space, by contrast, they furnish unitary (exceptional series) representations~\cite{Higuchi:1986py}. Throughout this paper, we reserve the letter $X$ for a spin-2 depth-0 partially conserved current, while $J$ and $T$ denote exactly conserved spin-1 and spin-2 currents.

\subsubsection{Conservation Identities}
\label{sec: conservation identities}
The existence of a conserved current imposes \textit{Ward identities} on corresponding boundary correlation functions. To derive these identities, one couples the current to a background source and requires invariance under the gauge transformations \eqref{eq: boundary gauge transformation}. It follows that the divergence of a correlation function containing a current insertion is supported only at coincident points, where the current approaches one of the charged operators. For a spin-1 current and a collection of charged scalar operators, the resulting Ward identity is
\begin{equation}
\partial_\mu \la J^\mu(x)\, \cO_1(x_1) \cdots \cO_n(x_n)\ra=-\sum_{i=1}^{n}\delta^{(d)}(x-x_i)\,q_i\,\la \cO_1(x_1) \cdots \cO_n(x_n)\ra\,,
\label{eq: J Ward identity}
\end{equation}
where $q_i$ denotes the charge of the operator $\cO_i$. We note that the right-hand side is a sum of contact terms localized at the operator insertions. For the stress tensor, the analogous relation is the translation Ward identity, in which the contact terms generate translations of the inserted operators
\begin{equation}
    \partial_\mu \la T^{\mu\nu}(x)\, \cO_1(x_1)\cdots \cO_n(x_n)\ra = -\sum_{i = 1}^{n}\hs\delta^{(d)}(x-x_i)\, \partial^{\nu}_{x_i} \la \cO_1(x_1) \cdots \cO_n(x_n)\ra\,.
\label{eq: T Ward identity}
\end{equation}
Partially conserved currents obey similar identities, with the single divergence replaced by the multiple divergences appearing in \eqref{eq: partial conservation}: the correlator $(\partial\cdot D_z)^{s-t}\la X^{(s)}\hs \cO_1 \cdots \cO_n\ra$ is again a sum of contact terms.

\vskip 4pt
The contact terms in \eqref{eq: J Ward identity} and \eqref{eq: T Ward identity} deserve a comment. When a correlator is too singular in an OPE limit to be Fourier transformed, its definition must be supplemented by semi-local terms, fixed by demanding consistency with the Ward identities. This can be done systematically by \textit{differential regularization}~\cite{Osborn_1994}, and cases relevant to this paper are discussed in Appendix~\ref{app: Semi-Local Terms}. While such terms are invisible at separated points, they will contribute to the integral transforms that appear once the conservation laws are weakly broken, and we will have to track them carefully. We defer this discussion to Section~\ref{sec: pseudo}.
\begin{figure}[t!]
	\centering
    \includegraphics[scale=1.0]{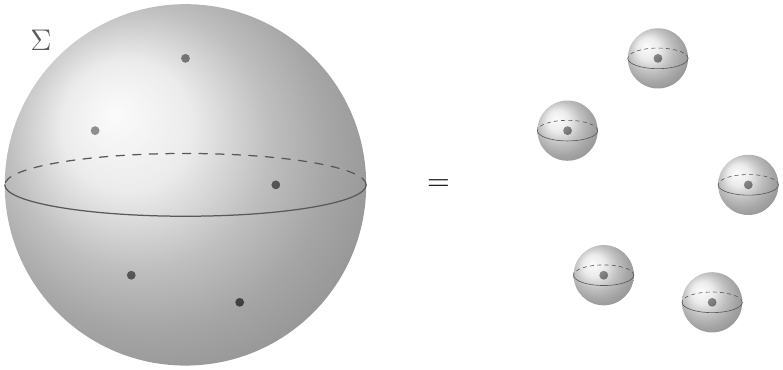}
	\caption{Insertion of a charge operator $Q[\Sigma]$ on a surface $\Sigma$ linking all local operator insertions. On the right-hand side, the surface is deformed to act locally on each operator $[Q,\cO_i]$. The equality of the two pictures leads to the charge conservation identity (\ref{eq: charge conservation identity}).}
	\label{fig: spheres}
\end{figure}

\vskip 4pt
Conserved and partially conserved currents can be integrated to form charges. For an exactly conserved current, the charge is the familiar surface operator $Q[\Sigma] = -\oint_\Sigma \d \Sigma_\mu\hs J^\mu$, while a depth-$t$ partially conserved current gives rise to the charge~\cite{Baumann:2025tkm},\footnote{This is not the only charge that can be built from a (partially) conserved current. More generally, one contracts $X_{\mu_1\cdots\mu_s}$ with a Killing-like tensor $\zeta^{(t)} \equiv \z^{\mu_1}\cdots\z^{\mu_t}\hs\zeta_{\mu_1\cdots\mu_t}(\x)$ satisfying $(\z\cdot\partial)^{s-t}\hs \zeta^{(t)}=0$, and distributes the remaining $s-t-1$ derivatives between $\zeta$ and $X$ with alternating signs, so as to obtain an exactly conserved current whose flux through $\Sigma$ defines a charge $Q_\zeta[\Sigma]$ (see~\cite{Baumann:2025tkm} for explicit expressions). The charge \eqref{eq: PM charge} corresponds to the constant solutions for $\zeta$, which is singled out by the fact that it commutes with the momentum generator, $[Q,P^\mu]=0$. Whether the identities associated with the position-dependent charges lead to further constraints is not known and we will not consider them here.}
\begin{equation}
    Q^{\mu_1\cdots\mu_t}[\Sigma] \equiv- \oint_\Sigma \d\Sigma_{\nu_1}\, \partial_{\nu_2} \cdots \partial_{\nu_{s-t}}\hs X^{\mu_1\cdots\mu_t\hs \nu_1 \nu_2 \cdots\hs \nu_{s-t}}\,,
\label{eq: PM charge}
\end{equation}
which is topological by virtue of \eqref{eq: partial conservation}. We may insert the charge into a correlation function by choosing the surface $\Sigma$ to enclose all operator insertions. If the symmetry is unbroken, the surface can then be deformed to infinity, where the charge annihilates the vacuum, $\la [Q, \cO_1 \cdots \cO_n] \ra=0$. Alternatively, $\Sigma$ may be deformed into a collection of small surfaces, each enclosing a single operator insertion. This yields the \textit{charge conservation identities}~\cite{Maldacena:2011jn} (see Figure~\ref{fig: spheres}):
\begin{equation} 
    \la [Q, \cO_1 \cdots \cO_n] \ra = \sum_{i=1}^{n} \la \cO_1 \cdots [Q, \cO_i] \cdots \cO_n\ra = 0\,,
\label{eq: charge conservation identity}
\end{equation}
where $[Q,\cO_i]$ denotes a charge insertion on a small surface enclosing only the operator $\cO_i$. These are integrated versions of the Ward identities and are therefore insensitive to contact terms: they enforce the consistency of the conservation laws at separated points.

\vskip 4pt
To evaluate constraints like \eqref{eq: charge conservation identity}, one needs the action of the charge on the local operators of the theory, which we will refer to as the \textit{current algebra}. On general grounds, this is a finite sum of local operators, which we schematically write as
\begin{equation}
    [Q, \cO_i(\x)] = \sum_j a_{ij}\, \partial^{\hs n_{ij}} \cO_j(\x)\,,
\label{eq: current algebra ansatz}
\end{equation}
where the spectrum of operators that can appear on the right-hand side is fixed by matching spins and dimensions (see~\cite{Baumann:2025tkm} for a detailed discussion). Substituted into \eqref{eq: charge conservation identity} for $n=3$, the current algebra turns the charge conservation identities into {\it sum rules} on three-point functions. Although these are conditions on three-point data, they originate from the existence of a consistent \textit{four}-point function of the current with the operators $\cO_i$, and are therefore genuinely dynamical constraints. In~\cite{Maldacena:2011jn, Baumann:2025tkm}, such identities were used to constrain theories with fully and partially conserved currents; in Section~\ref{sec: pseudo}, we will generalize them to currents whose conservation is weakly broken by double-trace deformations.

\subsubsection{Comment on De Sitter Space}
Finally, let us comment on de Sitter space, where PM fields are unitary and potentially of direct cosmological interest. In that case, the natural boundary object is not a partition function but the late-time wavefunction of the universe, $\Psi[\alpha]$, computed by the bulk path integral with Bunch--Davies conditions in the far past and the field profile $\alpha(\x)$ fixed at the future boundary. The wavefunction coefficients obtained by expanding $\log \Psi$ in powers of $\alpha$ have the same kinematic properties as correlation functions of a Euclidean CFT, with operator dimensions $\Delta_+$, and are related to their AdS counterparts by analytic continuation~\cite{Maldacena:2002vr,Harlow:2011ke,Anninos:2014lwa}. All of the preceding kinematic statements, including the conservation conditions \eqref{eq: exact conservation} and \eqref{eq: partial conservation}, the Ward identities, and the charge conservation identities, therefore apply without modification to wavefunction coefficients. 

\vskip 4pt
There is, however, one important structural difference: in AdS, the boundary conditions are part of the definition of the theory and may be chosen at will; in dS, the ``boundary condition" is related to the choice of vacuum state in the far past, and one does not get to impose gauge-invariant conditions at the late-time boundary~\cite{Baumann:2025tkm}. As we will see, this means that in de Sitter space the breaking of current conservation can be unavoidable, rather than optional (see Section~\ref{sec: Higgsing in CG}).

\vskip 4pt
A further distinction arises for physical late-time correlation functions, which are obtained from the wavefunction by applying the Born rule and therefore receive contributions from both asymptotic modes of each bulk field. In the free theory, the corresponding boundary operators have complementary dimensions and form shadow pairs. As shown in~\cite{Sleight:2025dmt}, interactions can then induce multiplet recombination of boundary currents and the stress tensor with double-trace composites containing both members of a shadow pair. The currents consequently acquire anomalous dimensions and become weakly non-conserved. The resulting non-conservation equations have precisely the double-trace form studied below.

\subsection{Weakly Broken Gauge Symmetry}
\label{sec: Weakly Broken Gauge Symmetry}
As described above, the usual holographic dictionary (\ref{eq: correlators from Z}) maps bulk interactions in (A)dS$_{d+1}$ to conformally invariant correlation functions of dual operators. For massive spin-$s$ fields, with dual spin-$s$ operators in long conformal multiplets, the set of available three-point conformal correlators is in one-to-one correspondence with the set of independent bulk interactions. However, as we describe below, the holographic map for (partially) massless fields can sometimes be more complicated.

\vskip 4pt
For a partially massless gauge field $A^{M_1\ldots M_s}$ of spin $s$ and depth $t$, we expect a dual boundary operator $X_{(s,t)}^{\mu_1 \ldots \mu_s}$ in a short conformal multiplet, with the shortening condition given by the conservation condition (\ref{eq: partial conservation}) \cite{Dolan:2001ih}. This implicitly assumes that we have chosen gauge-invariant boundary conditions for the bulk field $A^{M_1...M_s}$ and any other fields with which it interacts. The boundary three-point functions must then satisfy the following condition at separated points
\begin{equation}
    (\partial_{x_1} \cdot D_{z_1})^{s-t} \langle X_{(s,t)}(x_1) \mathcal{O}_i(x_2)\mathcal{O}_j(x_3)\rangle = 0\,.
\label{fullconservation3pt}
\end{equation}
As discussed in \cite{Baumann:2025tkm}, sometimes such symmetry-preserving boundary conditions do not exist, so that there is no longer a one-to-one correspondence between gauge-invariant bulk interactions and conserved boundary correlators. When a model with gauge-invariant local interactions is subject to symmetry-breaking boundary conditions, conservation of the dual operators is broken by a double-trace operator \cite{Porrati:2001db,Porrati:2024zvi}. Schematically, we write this as\hs\footnote{The composite operator is defined by $:\!\cO_i \cO_j\!:\!(x) \equiv \lim_{y \to x} \big[\hs \cO_i(x)\hs\cO_j(y) - \la \cO_i(x)\hs\cO_j(y)\ra \hs\big]$, where the subtraction removes the identity channel of the OPE, which is the only singular channel whose coefficient is not $1/N$-suppressed.} 
\begin{equation}
    (\partial \cdot D_z)^{s-t}\, X_{(s,t)} = g :\!\partial^{\hs k}\cO_i \partial^{\hs l}\cO_j \!:\,,
\label{eq: generic non-conservation}
\end{equation}
where the number of derivatives on the right-hand side is fixed by matching scaling dimensions. The parameter $g$ is either an independent small coupling or suppressed in $1/N$, justifying a perturbative expansion.

\vskip 4pt
The weakly broken conservation law \eqref{eq: generic non-conservation} leads to a modification of the conservation condition~\eqref{fullconservation3pt}:
\begin{equation}
\begin{aligned}
    (\partial_{x_1} \cdot D_{z_1})^{s-t} \langle X_{(s,t)}(x_1) \mathcal{O}_i(x_2)\mathcal{O}_j(x_3)\rangle &= g\,\langle\, :\! \partial^{\hs k}\mathcal{O}_i(x_1) \partial^{\hs l}\mathcal{O}_j(x_1)\!: \mathcal{O}_i(x_2)\mathcal{O}_j(x_3) \rangle \\[4pt]
    &= g\, \langle \partial^{\hs k}\mathcal{O}_i(x) \mathcal{O}_i(x_1)\rangle \langle \partial^{\hs l}\mathcal{O}_j(x) \mathcal{O}_j(x_2)\rangle + \cdots\,,
\end{aligned}
\label{semiconservation3pt}
\end{equation}
where in the second line we are neglecting subleading contributions, corresponding to the tree-level truncation in the bulk. Since the operator $X_{(s,t)}$ is assumed to have dimension $\Delta_X = d-1+t$, the descendant operator $(\partial\cdot D_{z})^{s-t} X_{(s,t)}$ is itself a conformal primary \cite{Zhiboedov:2012bm}. However, the right-hand side of \eqref{semiconservation3pt} does not have the structure of a generic conformal three-point function of primary operators, since it has singularities in the OPE limits $x_{12}\rightarrow 0$ and $x_{13}\rightarrow 0$, but is an \textit{analytic} function at $x_{23} = 0$. We will refer to this kind of structure as \textit{semi-analyticity}. 

\vskip 4pt
The classification of all independent, locally gauge-invariant cubic bulk interactions in (A)dS therefore requires a generalization of the standard procedure. In addition to the familiar fully conserved three-point functions satisfying (\ref{fullconservation3pt}), one must allow for \textit{anomalous} structures whose non-conservation is entirely accounted for by the double-trace breaking in (\ref{semiconservation3pt}). In the following, we will illustrate this in an explicit example.

\paragraph{Example} In $d$ dimensions, a general ansatz for the three-point correlator of a spin-2 depth-0 current $X_{(2,0)}$ is
\begin{equation}
    \langle X_{(2,0)} X_{(2,0)} X_{(2,0)}\rangle = \frac{\sum_{n=1}^5 c_n G^{(2,0)}_n}{\left(-2P_{12}\right)^{\frac{d+1}{2}}\left(-2P_{23}\right)^{\frac{d+1}{2}}\left(-2P_{31}\right)^{\frac{d+1}{2}}}\,,
\label{XXXPM2ansatz}
\end{equation}
with
\begin{equation}
    G^{(2,0)}_n \equiv 
    \begin{pmatrix}
    V_1^2 H_{23}^2+\text{cyclic} \\
    V_1 V_2 H_{23} H_{31}+\text{cyclic} \\
    V_2 V_3 V_1^2 H_{23}+\text{cyclic} \\
    V_1^2 V_2^2 V_3^2 \\
    H_{12}H_{23}H_{31}
    \end{pmatrix} ,
\end{equation}
where the structures $P_{ij}$, $H_{ij}$ and $V_i$ were defined in \eqref{equ:conformal-structures}.
We begin with the physically relevant case of $d=3$; in this case, one of the tensor structures becomes evanescent due to a vanishing Gram determinant, and so, without loss of generality, we restrict to $n=1,\ldots,4$.\footnote{The counting of independent three-point structures in general $d$, and the Gram relations responsible for such degenerations, are treated systematically in~\cite{Borovik:2026zgl}.} Imposing conservation up to semi-analytic terms, we get
\begin{equation}
    (\partial_{x_1}\cdot D_{z_1})^2\langle X_{(2,0)} X_{(2,0)} X_{(2,0)}\rangle = \frac{b_1 H_{23}^2+b_2 V_2 V_3 H_{23}+b_3 V_2^2 V_3^2}{(-2P_{12})^2 (-2P_{23})^2 (-2P_{31})^2}\,,
\label{eq: sa example1}
\end{equation}
where
\begin{align}
    b_1 &= 5 c_1-\frac{5}{2} c_2+c_3\,,\\
    b_2 &= 48 c_1-12 c_2-6 c_3+4c_4 \,, \\[4pt]
    b_3 &= 72 c_1-18 c_2-18c_3+9c_4\,.
\end{align}
As expected, the right-hand side of \eqref{eq: sa example1} is a linear combination of the three independent tensor structures for a conformal three-point function $\langle \mathcal{O} X_{(2,0)} X_{(2,0)}\rangle$, where $\mathcal{O}$ is the spin-zero descendant of $X$ with conformal dimension $\Delta_{\mathcal{O}}=4$. Full conservation would require the coefficient of each structure to vanish separately, which yields three equations for four unknowns and hence a one-dimensional solution space. To impose the weaker condition of conservation up to semi-analytic terms, we instead observe that, in this example, one linear combination of the three structures is \textit{regular} in the limit $x_{23}\to 0$:
\begin{equation}
    \frac{\left(H_{23}+2V_2 V_3\right)^2}{(-2P_{12})^2 (-2P_{23})^2 (-2P_{31})^2} = \frac{1}{|x_{12}|^4 |x_{13}|^4} \,f(x_{ab},z_a)\,,
\label{eq: sa example1a}
\end{equation}
where $f(x_{ab},z_a)$ is an analytic function of $x_{23}^2$ since $\left(H_{23}+2V_2 V_3\right)^2 \propto |x_{23}|^4$. Requiring the right-hand side of (\ref{eq: sa example1}) to take the form (\ref{eq: sa example1a}) instead yields two independent equations for four unknowns, and hence a two-dimensional solution space. As expected, this space contains a one-dimensional subspace of fully conserved solutions, spanned by
\begin{equation}
    c_n = \big\{-1,\;6,\;20,\;60\big\}\,.
\end{equation}
Any solution that is linearly independent of this one may be chosen as a representative of the nontrivial cohomology class of anomalous interactions. One convenient choice is
\begin{equation}
    c_n = \big\{-7,\;22,\;0,\;60\big\}\,.
\end{equation}
Matching this to the double-trace breaking of the conservation law then gives
\begin{equation}
    (\partial\cdot D)^2 X_{(2,0)} \propto X_{(2,0)\mu\nu} X_{(2,0)}^{\mu\nu}\,,
\end{equation}
where the precise constant of proportionality depends on the chosen normalization of the two-point function $\langle X_{(2,0)} X_{(2,0)}\rangle $. Thus, there are two independent locally gauge-invariant, Bose-symmetric cubic interactions among partially massless spin-2 fields in (A)dS$_4$. One gives rise to a fully conserved boundary correlator, while the other is conserved only up to a double-trace breaking. This counting agrees with the bulk analysis of \cite{Goon:2018fyu}.

\vskip 4pt
It is instructive to extend this analysis to general $d$. For $d>3$, the ansatz (\ref{XXXPM2ansatz}) contains an additional independent tensor structure, so that $n=1,\dots,5$. As before, the unique linear combination that vanishes in the limit $x_{23}\to 0$ is $\left(H_{23}+2V_2 V_3\right)^2 \sim |x_{23}|^4$. However, in a general dimension, the denominator also contains the factor $(-2P_{23})^{\frac{d+1}{2}} \sim |x_{23}|^{d+1}$. For $d>3$, the zero in the numerator is insufficient to cancel the singularity arising from the denominator. Consequently, every tensor structure that is conserved up to semi-analytic terms is in fact fully conserved. Equivalently, there are no anomalous cubic self-interactions of partially massless spin-2 fields in ${\rm (A)dS}_{d+1}$ for $d>3$.\footnote{This conclusion may also be understood from the absence of suitable double-trace operators. Schematically, the non-conservation equation would take the form $\partial^2 X_{(2,0)}\sim \partial^{k_1}X_{(2,0)}\,\partial^{k_2}X_{(2,0)}$. Matching the scaling dimensions of the two sides requires $k_1+k_2=3-d$. Since the right-hand side defines a local operator only if $k_1$ and $k_2$ are non-negative integers, such a double-trace contribution can exist only for $d\leq 3$.} A further worked example, involving a spin-4 depth-0 field, is presented in Appendix~\ref{sec:semianalytic}.

\subsection{Pseudo-Charge Conservation}
\label{sec: pseudo}
In the previous subsections, we discussed the weak breaking of bulk gauge symmetries and the associated weak non-conservation of the dual boundary currents. This raises the question of whether the charge conservation identities of Section~\ref{sec: conservation identities}, which assumed exact conservation, still constrain the theory when conservation is broken. In this section, we show that they do. Despite its weak non-conservation, the current continues to generate exact identities among correlation functions. These relations take the form of charge conservation identities for a modified charge action containing a \textit{nonlocal} contribution. We refer to the resulting modified charge as a \textit{pseudo-charge}. A similar strategy was used in~\cite{Maldacena:2012sf} to constrain theories with slightly broken higher-spin symmetry.

\vskip 4pt
The relation \eqref{eq: generic non-conservation} is an operator equation and therefore holds inside correlation functions, up to contact terms. The Ward identities of Section~\ref{sec: Holographic Dictionary} are then modified to
\begin{equation}
\begin{aligned}
    (\partial \cdot D_z)^{s-t} \la X_{(s,t)}(x)\, \cO_1(x_1) \cdots \cO_n(x_n)\ra
    = &-\sum_{i=1}^{n} \delta^{(d)}(x - x_i) \la \cO_1 \cdots [Q, \cO_i] \cdots \cO_n \ra \\
    &+ g \, \la \mathcal{B}(x)\, \cO_1(x_1) \cdots \cO_n(x_n)\ra\,,
\end{aligned}
\label{eq: broken Ward identity}
\end{equation}
where $\mathcal{B}(x)$ denotes the double-trace operator appearing on the right-hand side of \eqref{eq: generic non-conservation} and $[Q,\cO_i]$ is the local action of the charge. The new feature is the final term: the divergence of a correlator with a current insertion no longer vanishes at separated points. 

\vskip 4pt
To derive the modified conservation identities, we integrate the Ward identity \eqref{eq: broken Ward identity} over all of space. The left-hand side is a total derivative and integrates to zero, while the contact terms localize onto the operator insertions. We obtain 
\begin{equation}
    0 = \sum_{i=1}^{n} \la \cO_1 \cdots [Q, \cO_i] \cdots \cO_n\ra
    - g \int \d^d x \, \la \mathcal{B}(x)\, \cO_1(x_1) \cdots \cO_n(x_n)\ra\,.
\label{eq: integrated broken identity}
\end{equation}
The first term is the familiar sum over local charge insertions appearing in \eqref{eq: charge conservation identity}, while the second term is the integrated remnant of the broken conservation law.

\vskip 4pt
The integral in \eqref{eq: integrated broken identity} simplifies at leading order in $1/N$, and can be evaluated by Wick-contracting the constituents of the double-trace operator with the external operators. Contracting $\cO_i$ inside $\mathcal{B} = \,:\!\partial^{\hs k}\cO_i \partial^{\hs l}\cO_j\!:$ with an external insertion $\cO_i(x_i)$, we get
\begin{equation}
    \la \mathcal{B}(x)\hs \cO_1 \cdots \cO_i(x_i) \cdots \cO_n \ra = \la \partial^{\hs k}\cO_i(x)\hs \cO_i(x_i)\ra\, \la \cO_1 \cdots  \partial^{\hs l}\cO_j(x) \cdots \cO_n \ra + \cdots\,,
\label{eq: largeN factorization}
\end{equation}
with one such term for each external operator that has a non-vanishing two-point function with a constituent of $\mathcal{B}$. Performing the $x$-integral therefore replaces the external operator $\cO_i(x_i)$ by its associated transformed operator
\begin{equation}
    \wt\cO_j(x_i) \equiv \int \d^d x \, \la \partial^{\hs k}\cO_i(x)\hs \cO_i(x_i)\ra \, \partial^{\hs l}\cO_j(x)\,,
\label{eq: nonlocal transform}
\end{equation}
where the tensor indices of the two-point kernel $\la \cO_i(x)\hs \cO_i(x_i)\ra$ and the operator $\cO_j(x)$ are implicitly contracted. Counting dimensions in \eqref{eq: generic non-conservation}, it is clear that the operator $\wt\cO_j$ has scaling dimension
\begin{equation}
    \Delta_{\wt\cO_j} = (2\hs\Delta_i + k) + (\Delta_j + l) - d = \Delta_i + (s - 1)\,,
\label{eq: transform dimension}
\end{equation}
which is precisely the dimension required to appear in the action of a charge of dimension $\Delta_Q = s-1$ on $\cO_i$. All terms in \eqref{eq: integrated broken identity} then organize into a charge conservation identity,
\begin{equation}
    \la [\pQ, \cO_1 \cdots \cO_n] \ra = \sum_{i=1}^{n} \la \cO_1 \cdots [\pQ, \cO_i] \cdots \cO_n\ra = 0\,,
\label{eq: pseudo-charge conservation identity}
\end{equation}
for a \textit{pseudo-charge} $\pQ$, whose action on the constituents of the double-trace operator is
\begin{equation}
\begin{aligned}
    \big[\pQ, \cO_i(x)\big] &= \big[Q, \cO_i(x)\big] - g\, \wt\cO_j(x)\,,\\[3pt]
    \big[\pQ, \cO_j(x)\big] &= \big[Q, \cO_j(x)\big] - g\, \wt\cO_i(x)\,,
\end{aligned}
\label{eq: pseudo-charge action}
\end{equation}
while $[\pQ, \cO] = [Q, \cO]$ for all operators that do not appear in $\mathcal{B}$.

\vskip 4pt
The rules for constructing the most general pseudo-charge action are therefore a simple extension of those reviewed in Section~\ref{sec: Holographic Dictionary}. The \textit{local} part $[Q,\cO_i]$ is parametrized exactly as in \eqref{eq: current algebra ansatz}, as a finite sum of primaries and descendants with matching spins and dimensions, with a priori unknown coefficients. The \textit{nonlocal} part, by contrast, contains no free parameters: each operator appearing in the double-trace operator receives a single term, given by $(-g)$ times the transform~\eqref{eq: nonlocal transform}. The action on operators that are absent from the double-trace operator receives no nonlocal correction. Finally, the reciprocity property of the current algebra continues to hold, as enforced by the two-point charge conservation constraints $\la [\pQ, \cO_i \cO_j]\ra = 0$.

\paragraph{Evaluating the constraints}
Substituted into \eqref{eq: pseudo-charge conservation identity} for $n = 3$, the pseudo-charge action turns the conservation identities into sum rules that mix ordinary three-point functions, coming from the local terms, with \textit{integrated} three-point functions, coming from the nonlocal terms of the schematic form:
\begin{equation}
    \la \cO_1(x_1)\, \wt\cO_j(x_2)\, \cO_3(x_3) \ra = \int \d^d x \, \la \partial^{\hs k}\cO_i(x)\hs\cO_i(x_2) \ra\, \la \cO_1(x_1)\, \partial^{\hs l}\cO_j(x)\, \cO_3(x_3)\ra\,.
\label{eq: integrated three-point function}
\end{equation}
Two technical points deserve emphasis. First, the integrals \eqref{eq: integrated three-point function} require care to define and evaluate: we reduce the integrals to a small set of master integrals using tensor reduction and integration-by-parts identities. When half-integer powers arise, we employ dimensional regularization as needed and extract the resulting constraints from an OPE series expansion. These methods are developed in Appendix~\ref{sec: evaluating tensor integrals}. Second, the three-point functions appearing in the integrand must be equipped with their semi-local pieces, as dictated by differential regularization: while such contact terms are invisible in the ordinary (local) contributions to the identities, the integral transform \eqref{eq: integrated three-point function} converts them into non-vanishing contributions \textit{at separated points}. Details on differential regularization can be found in Appendix~\ref{app: Semi-Local Terms}.

\paragraph{Strategy} 
We can now state the strategy of this paper compactly. Given an assumed operator spectrum and a double-trace deformation \eqref{eq: generic non-conservation}, we proceed in three steps: (\textit{i}) write down the most general pseudo-charge action, consisting of the freely parametrized local terms and the fixed nonlocal terms; (\textit{ii}) impose the (two- and) three-point pseudo-charge conservation identities~\eqref{eq: pseudo-charge conservation identity}; (\textit{iii}) solve the resulting identities for the current algebra coefficients, the three-point normalizations, and the symmetry-breaking parameters. The existence of a nontrivial solution is a necessary condition for the consistency of a weakly broken gauge theory in the bulk, and the solution itself reconstructs the couplings of that theory from boundary data. The remainder of this paper carries out this program in a series of examples.

\section{Yang--Mills in Anti-De Sitter}\label{sec: YM in AdS}
As an illustrative example, we consider a theory which in the bulk contains a set of $n$ spin-one gauge fields, $A_M^a$, where $M=0,1,2,3$ is the spacetime index and $a=1,\ldots,n$ labels the fields. In addition, we allow for a collection of real scalar fields $\phi^I$, which we take to be conformally coupled for simplicity. On the boundary, these fields are dual to conserved current operators $J_a^{\mu}$, with $\mu=0,1,2$, and scalar operators~$\cO_I$. We will bootstrap their interactions and study how the boundary conditions for the scalars can break current conservation and, consequently, the associated bulk gauge symmetry. We will see that, despite this breaking, pseudo-charge conservation still constrains the structure to be that of a bulk theory of Yang--Mills gauge fields coupled to charged matter.

\vskip 4pt
Throughout this section, we specialize to ${\rm AdS}_4/{\rm CFT}_3$ and set the AdS radius to one. A conformally coupled scalar then has $m^2=-2$ and admits both alternate and standard quantization, corresponding to operator dimensions $\Delta_-=1$ and $\Delta_+=2$, respectively.

\subsection{Unbroken Symmetry}
With symmetry-preserving boundary conditions, each bulk gauge field is dual to a conserved boundary current, $\partial_\mu J_a^{\mu}=0$, which defines the charges $Q_a\equiv-\oint_\Sigma \d \Sigma_\mu\,J_a^{\mu}$. These charges act on the current and the scalars as
\begin{align}
    \big[Q_a,J_b(x)\big] &=-f_{ab}^{\ \ c}J_c(x)\,, \label{eq: current algebra J} \\
    \big[Q_a,\cO_I(x)\big] &=(T_a)_I{}^J\cO_J(x)\,,
\label{eq: charge action on scalars}
\end{align}
where $f_{ab}^{\ \ c}$ and $(T_a)_I{}^J$ are collections of real constants. As we will see below, the charge conservation identities require $f_{ab}{}^{c}$ to satisfy the antisymmetry and Jacobi identities defining a Lie algebra, and $(T_a)_I{}^J$ to furnish a representation of that algebra.
 
\vskip 4pt
The charge-conservation identity for a correlator of three currents is 
\begin{align}
    0 &= \la[Q_a,J_bJ_cJ_d]\ra\nonumber\\[4pt]
    &= -f_{ab}^{\ \ e}\la J_eJ_cJ_d\ra -f_{ac}^{\ \ e}\la J_bJ_eJ_d\ra -f_{ad}^{\ \ e}\la J_bJ_cJ_e\ra\,.
\label{eq: charge identity JJJ}
\end{align}
The three-point function of the currents is fixed to be
\begin{equation}
    \la J_aJ_bJ_c\ra = N_{JJJ}\,f_{abc}\,\langle\!\langle JJJ\rangle\!\rangle\,, 
\label{eq: 3pt identity JJJ}
\end{equation}
where $\langle\!\langle JJJ\rangle\!\rangle$ denotes the common conformal tensor structure and $f_{abc}=f_{ab}{}^{e}\kappa_{ce}$. Here, $\kappa_{ab}$ is the constant symmetric tensor appearing in the current two-point function, $\langle J_a J_b\rangle=\kappa_{ab}\langle\!\langle JJ\rangle\!\rangle$, with $\langle\!\langle JJ\rangle\!\rangle$ the standard spin-one two-point structure. We assume that $\kappa_{ab}$ is non-degenerate. The conformal tensor structure $\langle\!\langle JJJ\rangle\!\rangle$ is completely antisymmetric under interchange, so $f_{abc}$ is fully antisymmetric.\footnote{This can also be derived from the two-point charge-conservation identity, $\la[Q_a,J_bJ_c]\ra = 0$, which implies $f_{abc}=-f_{acb}$. Combined with the antisymmetry of $f_{ab}^{\ \ c}$ in its first two indices, which follows from $[Q_a,Q_b]=-f_{ab}^{\ \ c}\hs Q_c$, this implies that $f_{abc}$ is fully antisymmetric.} Substituting \eqref{eq: 3pt identity JJJ} into \eqref{eq: charge identity JJJ} gives
\begin{equation}
    \boxed{f_{ab}{}^{e}f_{ecd} +f_{ac}{}^{e}f_{bed} +f_{ad}{}^{e}f_{bce} =0}\ ,
\label{eq: Jacobi from charge conservation}
\end{equation}
which is indeed the {\it Jacobi identity}. Thus, the closure of the bulk gauge algebra is encoded in the charge conservation identities obeyed by the boundary current correlators.

\vskip 4pt
Next, we consider the charge-conservation identity for a correlator of one current and two scalars:
\begin{align}
    0 &= \la[Q_a,J_b\cO_I\cO_J]\ra\nonumber\\[4pt]
    & = -f_{ab}{}^{ c}\la J_c\cO_I\cO_J\ra +(T_a)_I{}^K\la J_b\cO_K\cO_J\ra +(T_a)_J{}^K\la J_b\cO_I\cO_K\ra\,.
\label{eq: charge identity JOO}
\end{align}
The Ward identity associated with $J_a$ fixes the current--scalar--scalar correlator to be of the form
\begin{equation}
    \la J_a\cO_I\cO_J\ra = \frac{(T_a)_{IJ}}{4\pi} \langle\!\langle J\cO\cO\rangle\!\rangle\,,
\label{eq: JOO group structure}
\end{equation}
where $(T_a)_{IJ}=(T_a)_{I}{}^{K} \kappa_{KJ}$. Here, $\kappa_{IJ}$ is the constant symmetric tensor appearing in the scalar two-point function, $\langle \mathcal O_I \mathcal O_J\rangle = \kappa_{IJ}\langle\!\langle \mathcal O\mathcal O\rangle\!\rangle$, with $\langle\!\langle \mathcal O\mathcal O\rangle\!\rangle$ the standard spin-zero two-point structure. We assume that $\kappa_{IJ}$ is non-degenerate. Substituting this into (\ref{eq: charge identity JOO}), and using the antisymmetry of the generators $(T_a)_{IJ}$,\footnote{This can be derived from the two-point charge-conservation constraint $\la [Q_a,\cO_I \cO_J] \ra =0$.} the common conformal structure factors out and we get the following matrix equation for $(T_a)_{I}{}^{J}$,
\begin{equation}
    \boxed{[T_a,T_b]=f_{ab}{}^{c}\hs T_c}\ .
\label{eq: scalar representation algebra}
\end{equation}
Charge conservation therefore implies that the matrices appearing in the three-point functions $\la J_a\cO_I\cO_J\ra$ furnish a representation of the same Lie algebra as the currents. Together with \eqref{eq: Jacobi from charge conservation}, this reconstructs both the gauge algebra and its action on matter directly from the boundary correlators.

\vskip 4pt
Note that we have made no assumption about the positive-definiteness of the vector and scalar two-point functions, encoded in $\kappa_{ab}$ and $\kappa_{IJ}$, respectively; we required only that they be non-degenerate. The argument therefore goes through unchanged in the non-unitary case, where some eigenvalues of these two-point matrices may be negative. All of this is consistent with the bulk Yang--Mills action
\begin{equation}
    S = \int \d^4x\,\sqrt{-g}\left[ -\frac{1}{4}\,\kappa_{ab}\, F^a_{MN}F^{b\hs MN} -\frac{1}{2}\,\kappa_{IJ}\Big( (D_M\phi)^I(D^M\phi)^J +m^2\phi^I\phi^J\Big) \right] ,
\label{eq: scalar YM action}
\end{equation}
where $F^a_{MN} \equiv \partial_M A^a_N-\partial_N A^a_M +f_{bc}{}^{a}A^b_M A^c_N$ and $(D_M\phi)^I \equiv \nabla_M\phi^I-A^a_M(T_a)_{J}{}^{I}\phi^J$ are the non-Abelian field strength and gauge-covariant derivative, with $f_{bc}{}^{a}$ the structure constants and $(T_a)_{J}{}^{I}$ the generators in the scalar representation. In this action, $\kappa_{ab}$ and $\kappa_{IJ}$ play the role of (possibly indefinite) metrics on the gauge algebra and on the scalar representation space, and the two-point functions of the boundary operators inherit their signatures directly.

\subsection{Symmetry Breaking}
Next, we impose boundary conditions on the scalar fields that break part of the bulk gauge symmetry.\footnote{Boundary conditions for gauge theories in AdS have been studied extensively; see e.g.~\cite{Aharony:2012jf, Ankur:2023lum, Ankur:2026ylr, Ciccone:2025dqx, DiPietro:2025ozw}.} Conservation of the associated boundary currents is then weakly broken. We show that the resulting pseudo-charge conservation identities continue to constrain the theory and, in fact, reconstruct the representation of the full bulk gauge algebra on the scalar fields.

\vskip 4pt
For a conformally coupled scalar in ${\rm AdS}_4$, with $m^2=-2$, both asymptotic fall-offs are normalizable and correspond to boundary operators of dimensions $\Delta_-=1$ and $\Delta_+=2$. We may therefore decompose the scalar representation space as
\begin{equation}
    \mathcal{V}=\mathcal{V}_1\oplus \mathcal{V}_2
\end{equation}
and impose alternate quantization on the fields in $\mathcal{V}_1$ and standard quantization on those in~$\mathcal{V}_2$. We denote the corresponding boundary operators by $\cO_{1I}$ and $\cO_{2\bar I}$, with $\Delta(\cO_{1I})=1$ and $\Delta(\cO_{2\bar I})=2$, respectively. Furthermore, we will now assume that the kinetic terms do not mix across these two subsectors, so that the scalar kinetic matrix is ${\rm diag}(\kappa_{IJ},\kappa_{\bar I \bar J})$, and we use $\kappa_{IJ}$ and $\kappa_{\bar I \bar J}$ and their inverses to raise and lower unbarred and barred indices, respectively.

\vskip 4pt
Because the fields in $\mathcal{V}_1$ and $\mathcal{V}_2$ obey different boundary conditions, only gauge transformations that act separately within the two sectors preserve the boundary conditions. A generator that mixes $\mathcal{V}_1$ and $\mathcal{V}_2$ changes the quantization of the scalar fields and is therefore broken. At leading order in the breaking, the corresponding boundary current obeys
\begin{equation}
    \partial_\mu J_a^{\mu} = g_{a}^{I\bar J}\,\cO_{1I}\cO_{2\bar J}\,,
\label{eq: Scalar YM double trace deformation}
\end{equation}
where the coincident product is understood as a renormalized double-trace operator. Since $\Delta_1+\Delta_2=3$, the coefficients $g_{a}^{I\bar J}$ are dimensionless. A linear combination $v^aJ_a^{\mu}$ remains conserved precisely when $v^a g_{a}^{I\bar J}=0$. The kernel of $g_{a}^{I\bar J}$ therefore identifies the unbroken gauge subalgebra.

\vskip 4pt
The broken conservation equation modifies the action of the charges on the scalar operators. In addition to the local transformations within $\mathcal{V}_1$ and $\mathcal{V}_2$, the pseudo-charges contain nonlocal terms that exchange the two quantization sectors:
\begin{equation}
\begin{aligned}
    [\pQ_a,\cO_{1I}] &=(T_a)_I{}^J\cO_{1J} - (g_a)_I{}^{\bar J}\widetilde\cO_{2\bar J}\,, \\[4pt]
    [\pQ_a,\cO_{2\bar I}] &=(\bar T_a)_{\bar I}{}^{\bar J}\cO_{2\bar J} -(g_a)^J{}_{\bar I}\widetilde\cO_{1J}\,.
\end{aligned}
\label{eq: Scalar YM current algebra}
\end{equation}
$T_a$ and $\bar T_a$ describe the action within the two quantization sectors, whereas $g_a$ measures the mixing between them. The transformed operators appearing in (\ref{eq: Scalar YM current algebra}) are defined by
\begin{equation}
    \widetilde{\cO}_{1I}(x) \equiv\int\d^3x'\,\frac{\cO_{1I}(x')}{|x'-x|^4}\,,\qquad \widetilde{\cO}_{2\bar I}(x) \equiv\int\d^3x'\,\frac{\cO_{2\bar I}(x')}{|x'-x|^2}\,.
\end{equation}
The operators $\widetilde{\cO}_{1I}$ and $\widetilde{\cO}_{2\bar I}$ have
dimensions two and one, respectively. 

\vskip 4pt
We next consider the three-point pseudo-charge identities involving one current and two scalars. There are three possibilities:
\begin{align}
    0 & =\la[\pQ_a,J_b\mathcal O_{1I}\mathcal O_{1J}]\ra\,,\\
    0 & =\la[\pQ_a,J_b\mathcal O_{2\bar I}\mathcal O_{2\bar J}]\ra\,,\\
    0 & =\la[\pQ_a,J_b\mathcal O_{1I}\mathcal O_{2\bar J}]\ra\,.
\end{align}
To evaluate these constraints, we require the following three-point functions:\footnote{The shadow transforms can be evaluated using the techniques described in Appendix~\ref{sec: evaluating tensor integrals}. The correlators in this section do not require the semi-local terms discussed in Appendix~\ref{app: Semi-Local Terms}.}
\begin{equation}
\begin{aligned}
    \la J_a\mathcal O_{1I}\mathcal O_{1J}\ra
    &=\frac{T_{a\, IJ}}{4\pi}\,\langle\!\langle J\mathcal O_1\mathcal O_1\rangle\!\rangle\,,
    &\quad
    \la J_a\mathcal O_{2\bar I}\mathcal O_{2\bar J}\ra
    &=\frac{\bar T_{a\, \bar I\bar J}}{4\pi}\,\langle\!\langle J\mathcal O_2\mathcal O_2\rangle\!\rangle\,,\\[4pt]
    \la J_a\widetilde{\mathcal O}_{1I}\mathcal O_{1J}\ra
    &=-T_{a\, IJ}\,\langle\!\langle J\mathcal O_2\mathcal O_1\rangle\!\rangle\,,
    &
    \la J_a\mathcal O_{1I}\widetilde{\mathcal O}_{1J}\ra
    &=-T_{a\, IJ}\,\langle\!\langle J\mathcal O_1\mathcal O_2\rangle\!\rangle\,,\\[4pt]
    \la J_a\widetilde{\mathcal O}_{2\bar I}\mathcal O_{2\bar J}\ra
    &=+\bar T_{a\, \bar I\bar J}\,\langle\!\langle J\mathcal O_1\mathcal O_2\rangle\!\rangle\,,
    &
    \la J_a\mathcal O_{2\bar I}\widetilde{\mathcal O}_{2\bar J}\ra
    &=+\bar T_{a\, \bar I\bar J}\,\langle\!\langle J\mathcal O_2\mathcal O_1\rangle\!\rangle\,,\\[4pt]
    \la J_a\mathcal O_{1I}\mathcal O_{2\bar J}\ra
    &=-g_{a\, I\bar J}\,\langle\!\langle J\mathcal O_1\mathcal O_2\rangle\!\rangle\,,
    &
    \la J_a\widetilde{\mathcal O}_{1I}\mathcal O_{2\bar J}\ra
    &=-\la J_a\mathcal O_{2\bar J}\widetilde{\mathcal O}_{1I}\ra
      =\frac{\pi^3}{2}\,g_{a\, I\bar J}\,\langle\!\langle J\mathcal O_2\mathcal O_2\rangle\!\rangle\,,\\[4pt]
    \la J_a\mathcal O_{2\bar I}\mathcal O_{1J}\ra
    &=+g_{a\, J\bar I}\,\langle\!\langle J\mathcal O_2\mathcal O_1\rangle\!\rangle\,,
    &
    \la J_a\widetilde{\mathcal O}_{2\bar I}\mathcal O_{1J}\ra
    &=-\la J_a\mathcal O_{1J}\widetilde{\mathcal O}_{2\bar I}\ra
      =\frac{\pi^3}{2}\,g_{a\, J\bar I}\,\langle\!\langle J\mathcal O_1\mathcal O_1\rangle\!\rangle\,,
\end{aligned}
\label{eq: JOOs}
\end{equation}
where $\langle\!\langle J\mathcal O_i\mathcal O_j\rangle\!\rangle$ denote the conformal structures defined in~\eqref{eq: YOO strcuture}. It is convenient to introduce the rescaled coefficients
\begin{equation}
    G_{a}^{I\bar J}\equiv \sqrt{2}\hs\pi^2 \hs g_{a}^{I\bar J}\,.
\end{equation}
The resulting constraints are then the following matrix equations for the matrices $(T_a)_I^{\ J}$, $(\bar T_a)_{\bar I}^{\ \bar J}$, $(G_a)_I^{\ \bar J}$, and $(G_a^T)_{\bar J}^{\ I}\equiv (G_a)^I_{\ \bar J}$,
\begin{align}
    [T_a,T_b]-G_a(G_b)^T+G_b(G_a)^T &=f_{ab}^{\ \ c}T_c\,,\label{eq: doubled closure 11}\\[4pt]
    [\bar T_a,\bar T_b]-(G_a)^TG_b+(G_b)^TG_a &=f_{ab}^{\ \ c}\bar T_c\,,\label{eq: doubled closure 22}\\[4pt]
    T_aG_b-T_bG_a+G_a\bar T_b-G_b\bar T_a &=f_{ab}^{\ \ c}G_c\,,\label{eq: doubled closure 12}
\end{align}
which can also be written as
\begin{equation}
   \boxed{ [\mathbb T_a,\mathbb T_b]=f_{ab}^{\ \ c}\hs \mathbb T_c\,, \quad {\rm with} \quad \mathbb T_a\equiv \begin{pmatrix}
        T_a&-G_a\\
        (G_a)^T&\bar T_a
    \end{pmatrix}}\ .
\label{eq: real doubled closure}
\end{equation}
Since $T_a$ and $\bar T_a$ are antisymmetric when all the indices are lowered, the same is true of the matrices $\mathbb T_a$. They therefore furnish an orthogonal representation of the gauge algebra on the space $\mathcal{V}_1\oplus \mathcal{V}_2$. The diagonal blocks generate transformations that preserve the two quantization sectors separately, whereas the off-diagonal blocks mix the two sectors and are precisely the source of current non-conservation. Pseudo-charge conservation has therefore reconstructed the full bulk representation from the boundary data.

\paragraph{Example 1}
As a simple example, consider three scalars, with standard canonical kinetic terms, in the adjoint representation of $SU(2)$, and impose alternate quantization on the first scalar and standard quantization on the other two.

\vskip 4pt
The adjoint generators may be written as $(\mathbb T_a^{\rm Adj})_{bc}=-\epsilon_{abc}$, so that $[\mathbb T^{\rm Adj}_a,\mathbb T^{\rm Adj}_b] =\epsilon_{abc}\hs \mathbb T^{\rm Adj}_c$. Explicitly, we have
\begin{equation}
    \mathbb T^{\rm Adj}_1= \begin{pmatrix}
        0&0&0\\
        0&0&-1\\
        0&1&0
    \end{pmatrix}, \quad
    \mathbb T^{\rm Adj}_2= \begin{pmatrix}
        0&0&1\\
        0&0&0\\
        -1&0&0
    \end{pmatrix},\quad \mathbb T^{\rm Adj}_3= \begin{pmatrix}
        0&-1&0\\
        1&0&0\\
        0&0&0
    \end{pmatrix}.
\label{eq: Adj rep generators}
\end{equation}
Comparing this with \eqref{eq: real doubled closure}, we find
\begin{equation}
\begin{aligned}
    T_1&=T_2=T_3=0\,, &\quad \bar T_2&=\bar T_3=0_{2\times 2}\,, & G_1&=0_{1\times 2}\,,\\[3pt]
    \bar T_1& = \begin{pmatrix}
        0&-1\\
        1&0
    \end{pmatrix}, & G_2&=(0,-1)\,, & G_3&=(1,0)\,.
\end{aligned}
\label{eq: T G in adjoint rep}
\end{equation}
The generator $\mathbb T_1^{\rm Adj}$ rotates only the two standard-quantized scalars and therefore preserves the boundary conditions. By contrast, $\mathbb T_2^{\rm Adj}$ and $\mathbb T_3^{\rm Adj}$ mix the alternate-quantized scalar with the standard-quantized scalars. Accordingly, $G_1=0$ and $J_{1}^{\mu}$ remains conserved, whereas $G_2$ and $G_3$ are nonzero and the currents $J_{2}^{\mu}$ and $J_{3}^{\mu}$ are weakly non-conserved. The mixed boundary conditions therefore break $SU(2)$ to the $U(1)$ subgroup generated by $\mathbb T_1^{\rm Adj}$.

\paragraph{Example 2}
As a second example, suppose that the two quantization sectors have equal multiplicity $k$, as well as canonical kinetic matrices. We can then take $I,\bar I=1,\ldots,k$ and combine the corresponding real bulk fields into complex scalars,
\begin{equation}
    \Phi_i\equiv\frac{1}{\sqrt{2}}\big(\phi_{1i}+i\phi_{2i}\big)\,.
\end{equation}
In this case, $T_a$, $\bar T_a$, and $G_a$ are all $k\times k$ matrices. A real transformation of $(\phi_1,\phi_2)$ arises from a complex-linear transformation of $\Phi$ if and only if it preserves the complex structure
\begin{equation}
    \mathbb J= \begin{pmatrix}
        0_{k\times k}&- 1_{k\times k}\\
        1_{k \times k}&0_{k \times k}
    \end{pmatrix}, \qquad \mathbb J^2=- 1_{2k \times 2k}\,.
\end{equation}
Thus, the real representation on $\mathcal{V}_1\oplus \mathcal{V}_2$ comes from a complex representation precisely when $[\mathbb T_a,\mathbb J]=0$. Substituting \eqref{eq: real doubled closure} gives $T_a=\bar T_a$ and $(G_a)^T=G_a$, so that
\begin{equation}
    \mathbb T_a = \begin{pmatrix}
        T_a&-G_a\\
        G_a&T_a
    \end{pmatrix}.
\end{equation}
For infinitesimal real parameters $\alpha^a$, the transformation of the complex scalar is
\begin{equation}
    \delta\Phi_i=i\alpha^a {\sf T}_{a\, ij}\Phi_j\,, \qquad {\sf T}_a=G_a-iT_a\,.
\end{equation}
Since $G_a$ is symmetric and $T_a$ is antisymmetric, ${\sf T}_a$ is Hermitian. With the generator convention used in \eqref{eq: real doubled closure}, the closure condition is equivalent to
\begin{equation}
    [{\sf T}_a,{\sf T}_b]=-if_{ab}{}^c\,{\sf T}_c\,.
\label{eq: Lie algebra}
\end{equation}
Hence, imposing that the real bulk representation admits a gauge-invariant complex structure, the pseudo-charge constraints reproduce the Lie algebra of the complex scalar representation. 

\vskip 4pt
These examples serve to illustrate the constraining power of the pseudo-charge conservation identities on interactions involving weakly broken current conservation. Beginning with a generic parametrization of double-trace breaking (\ref{eq: Scalar YM double trace deformation}), satisfying the constraints restricts the \textit{a priori} unknown interaction coefficients to take the familiar form corresponding to a representation of a Lie algebra (\ref{eq: real doubled closure}). This is of course the same restriction found in the unbroken case (\ref{eq: scalar representation algebra}). Physically, the constraints are re-discovering the fact that these are the same bulk theories, Yang--Mills coupled to scalar matter, albeit with different boundary conditions. In the following sections, we will apply these same constraints in cases where we do not necessarily have access to an unbroken phase of the theory; the constraints then provide a powerful way to discover and constrain genuinely new gauge-invariant bulk interactions.

\section{Bi-Gravity in ${\rm AdS} \times {\rm AdS}$}
\label{sec: Bi-Gravity in AdSxAdS}
It is well known that, in flat space, consistent local interactions do not allow multiple massless spin-2 fields (gravitons) with nontrivial cross-couplings \cite{Boulanger:2000rq}. Instead, each graviton must live in an independent completely decoupled sector. In other words, there is no non-Abelian version of gravity. In this section, we will bootstrap the same statement in anti-de Sitter space. We will first show that a theory with two conserved stress tensors must factorize into a product of two decoupled CFTs. (A brief version of the argument has appeared before in~\cite{Maldacena:2011jn}.) We then consider the breaking of one of the stress tensors, corresponding to the bulk graviton becoming massive. We show that this couples the two CFTs, solving the relevant pseudo-charge conservation identities for any value of the breaking parameters. Finally, we demonstrate that this boundary bootstrap precisely reproduces the structure of a model studied by Aharony, Clark and Karch~\cite{Aharony:2006hz}, as well as Kiritsis~\cite{Kiritsis:2006hy}. Their theory describes two AdS spaces sharing a common conformal boundary, with a coupling that breaks conservation of the respective stress tensors. In Appendix~\ref{sec: CPT of AdSxAdS}, we explicitly compute the correlators of this model in conformal perturbation theory, thereby verifying the results of our boundary bootstrap.

\subsection{Two Gravitons in Anti-De Sitter}
\label{ssec:2gravitons}
Suppose that a CFT contains two symmetric, traceless, and conserved spin-two operators. Let $T_{\mu\nu}$ be the physical stress tensor, so that its charge $P^\mu$ generates translations on every local operator, and choose the second stress tensor $S_{\mu\nu}$ to be orthogonal to it:
\begin{equation}
  \partial^\mu T_{\mu\nu}=0\,, \qquad \partial^\mu S_{\mu\nu}=0\,, \qquad \langle T S \rangle=0\,.
\label{eq:two-conserved-tensors}
\end{equation}
We denote the charge associated with $S_{\mu \nu}$ by $Q^\mu$. The two-point functions of the two stress tensors are $\langle TT \rangle = c_T \la\! \la TT\ra \!\ra$ and $\langle SS \rangle = c_S \la\! \la TT\ra \!\ra$, where $\la\! \la TT\ra \!\ra$ is a fixed unit-normalized conformal structure defined in \eqref{eq: YY structure}. Reflection positivity demands that $c_T>0$ and $c_S>0$. There is still a simultaneous normalization freedom $S_{\mu \nu}\to\lambda S_{\mu \nu}$ and $Q^\mu \to\lambda Q^\mu$. We use it below to set $c_S=c_T$. This choice is possible precisely because both central charges are positive.

\vskip 4pt
The ordinary translation charge $P^\mu$ acts in the usual way,
\begin{equation}
\begin{aligned}
  [P^\mu,T_{\alpha \beta}] &=\partial^\mu T_{\alpha \beta}\,, \\
  [P^\mu, S_{\alpha \beta}] &=\partial^\mu S_{\alpha \beta}\,.
\end{aligned}
\label{eq:P-action}
\end{equation}
The action of the second charge $Q^\mu$ is fixed by conformal covariance. We normalize the coefficient of $\partial^\mu S_{\alpha \beta}$ in $[Q^\mu, T_{\alpha\beta}]$ to unity, as required when $Q^\mu$ is the charge obtained by integrating $S_{\mu\nu}$. Before using the normalization freedom above, the two-point identity $\langle[Q,TS]\rangle=0$ fixes the coefficient of $\partial^\mu T_{\alpha \beta}$ in $[Q^\mu,S_{\alpha \beta}]$ to $c_S/c_T$. It is therefore equal to one in our conventions, and the most general charge action is 
\begin{equation}
\begin{aligned}
    [Q^\mu, T_{\alpha \beta}] & = \partial^\mu S_{\alpha \beta}\,,\\
    [Q^\mu, S_{\alpha \beta}] & = \partial^\mu T_{\alpha \beta}+a\,\partial^\mu S_{\alpha \beta}\,,
\end{aligned}
\label{eq:Q-action}
\end{equation}
for a real number $a$. In principle, $ [Q^\mu, T_{\alpha \beta}]$ could have a contribution from a spin-1 current, but the three-point identity $\langle [Q, TTS] \rangle =0$ sets its coefficient to zero~\cite{Maldacena:2011jn}. Defining $\mathcal{T}\equiv (T, S)$, we can also write (\ref{eq:Q-action}) as
\begin{equation}
    [Q^\mu, \mathcal{T}_i] =\partial^\mu (\mathcal{A}\,\mathcal{T})_i\,, \quad  \mathcal{A} \equiv \begin{pmatrix}0&1\\[2pt]1& a\end{pmatrix}.
\label{eq:ST-action}
\end{equation}
The matrix $\mathcal A$ is real and symmetric, with eigenvalues $\lambda_{1,2}=\tfrac12(a\pm\sqrt{a^2+4})$. It is useful to project the physical stress tensor onto the corresponding eigenspaces. We therefore define
\begin{equation}
\begin{aligned}
    T^{(1)}_{\mu\nu} & = \frac{1}{2}\left(1-\frac{a}{\sqrt{a^2+4}}\right)T_{\mu\nu} +\frac{1}{\sqrt{a^2+4}}\, S_{\mu\nu}\,,\\
    T^{(2)}_{\mu\nu} & = \frac{1}{2}\left(1+\frac{a}{\sqrt{a^2+4}}\right)T_{\mu\nu}-\frac{1}{\sqrt{a^2+4}}\, S_{\mu\nu}\,.
\end{aligned}
\label{eq:diagonal-stress-tensors}
\end{equation}
These operators obey
\begin{equation}
    [Q^\mu,T^{(i)}_{\alpha\beta}] = \lambda_i\,\partial^\mu T^{(i)}_{\alpha\beta}\,.
\end{equation}
They are conserved, orthogonal, and add up to the physical stress tensor, $T_{\mu \nu}=T^{(1)}_{\mu \nu}+T^{(2)}_{\mu \nu}$. Their two-point coefficients are
\begin{equation}
    c_{1,2}=\frac{c_T}{2}\left(1\mp\frac{a}{\sqrt{a^2+4}}\right),
\end{equation}
and are both positive. The same linear combinations of $P^\mu$ and $Q^\mu$
define the charges
\begin{equation}
\begin{aligned}
    P^{(1)}_\mu &=\frac{1}{2}\left(1-\frac{a}{\sqrt{a^2+4}}\right)P_\mu +\frac{1}{\sqrt{a^2+4}}Q_\mu\,,\\
    P^{(2)}_\mu &=\frac{1}{2}\left(1+\frac{a}{\sqrt{a^2+4}}\right)P_\mu -\frac{1}{\sqrt{a^2+4}}Q_\mu\,.
\end{aligned}
\end{equation}
Equations~\eqref{eq:P-action} and \eqref{eq:Q-action} then give
\begin{equation}
    [P^{(i)}_\mu,T^{(j)}_{\alpha \beta}] =\delta^{ij}\partial_\mu T^{(j)}_{\alpha \beta}\,, \qquad [P^{(1)}_\mu, P^{(2)}_\nu]=0\,,
\label{eq:split-translations}
\end{equation}
with $P_\mu=P^{(1)}_\mu+P^{(2)}_\mu$.
Hence, each $P^{(i)}_{\mu \nu}$ generates translations in one sector and annihilates the other. The Ward identities consequently organize the local operator algebra into two mutually commuting subalgebras: operators in the first are translated only by $P^{(1)}_\mu$, while operators in the second are translated only by $P^{(2)}_\mu$. Operators carrying both quantum numbers are products of operators from the two subalgebras, so that mixed connected correlators vanish and the Hilbert space factorizes~\cite{Maldacena:2011jn}. The two conserved spin-two operators are simply the stress tensors of the two factors, corresponding to two decoupled CFTs. In the bulk, they describe two decoupled massless gravitons; any nontrivial coupling between the two sectors must therefore break one of the two conservation laws (or abandon unitarity).

\subsection{Massive Gravity in ${\rm AdS} \times {\rm AdS}$}
\label{ssec:ACKK}

In the following, we briefly review the model of massive gravity first studied by Aharony, Clark, Karch and Kiritsis (ACKK)~\cite{Aharony:2006hz, Kiritsis:2006hy}. We start from the product theory ${\rm CFT}_1\times {\rm CFT}_2$, but couple the two CFTs through a double-trace deformation involving two scalar operators. Before the deformation, the bulk dual consists of two disjoint AdS spaces, each containing a massless graviton. The two spaces share a common conformal boundary, and the interaction between the two sectors is implemented through correlated boundary conditions for the bulk scalars dual to $\cO_1$ and $\cO_2$. The bulk theories therefore communicate only through their boundary conditions.

\vskip 4pt
The boundary action is
\begin{equation}
    S=S_1+S_2+g\int \d^d x\,\cO_1(x)\cO_2(x)\,, 
\label{eq:double-trace-deformation}
\end{equation}
where $g$ is a constant parameter and $\cO_i$ is a scalar primary in ${\rm CFT}_i$, with conformal dimension~$\Delta_i$. Following~\cite{Aharony:2006hz, Kiritsis:2006hy}, we choose $\Delta_1+\Delta_2=d$, so that the deformation is exactly marginal. Each CFT contains a conserved stress tensor $T^{(i)}_{\mu\nu}$, whose two-point function is $\la T^{(i)} T^{(i)}\ra = c_i\hs \la\!\la TT \ra\!\ra$.

\vskip 4pt
The coupling breaks the two independent gauge symmetries to a diagonal subgroup. To leading order in $g$, the individual stress tensors obey
\begin{align}\label{eq: non-conservation eq T1}
    \partial^\mu T^{(1)}_{\mu\nu} &=g\hs (\partial_\nu\mathcal{O}_1)\mathcal{O}_2\,,\\ \label{eq: non-conservation eq T2}
    \partial^\mu T^{(2)}_{\mu\nu} &=g\hs \mathcal{O}_1(\partial_\nu\mathcal{O}_2)\,.
\end{align}
Neither stress tensor is therefore separately conserved. However, the total stress tensor, including the contribution associated with the deformation (\ref{eq:double-trace-deformation}), remains conserved:
\begin{equation}
    \partial^\mu T_{\mu\nu} = 0 \,, \quad T_{\mu\nu} \equiv T^{(1)}_{\mu\nu} +T^{(2)}_{\mu\nu} -g\hs \eta_{\mu\nu}\hs \mathcal{O}_1\mathcal{O}_2\,.
\label{eq: T conserved}
\end{equation}
Since the two factors are orthogonal, $\la T^{(1)}T^{(2)}\ra=0$, the two-point coefficient of $T_{\mu\nu}$ is $c_T=c_1+c_2$. In addition, we have an orthogonal combination of the two stress tensors, which we denote by~$S_{\mu\nu}$. It is convenient to introduce the rescaled central charges
\begin{equation}
    \hat c_i \equiv \frac{c_i}{\sqrt{c_1c_2}}\,,\qquad \hat c_1\hat c_2=1\,,
\label{eq: chat}
\end{equation}
in terms of which\footnote{The overall normalization of $S_{\mu\nu}$ is arbitrary, and we fix it so that $c_S=c_T$, as in Section~\ref{ssec:2gravitons}. In~\cite{Aharony:2006hz}, the authors instead use $c_i$ in place of $\hat c_i$, so their $S_{\mu\nu}$ is larger by a factor of $\sqrt{c_1c_2}$.}
\begin{equation}
    S_{\mu\nu} = \hat c_2T^{(1)}_{\mu\nu} -\hat c_1T^{(2)}_{\mu\nu} + g \left( \hat c_1\frac{\Delta_2}{d} -\hat c_2\frac{\Delta_1}{d} \right) \eta_{\mu\nu}\mathcal{O}_1\mathcal{O}_2\,,
\label{eq: T non-conserved}
\end{equation}
where the relative coefficients are fixed by demanding vanishing two-point function with both $T_{\mu\nu}$ and $\mathcal{O}_1\mathcal{O}_2$. The divergence of (\ref{eq: T non-conserved}) is
\begin{equation}\label{eq: non-cons T}
    \partial^\mu S_{\mu\nu} = g\, \frac{(\hat c_1+\hat c_2)}{d} \Big[\Delta_2\hs(\partial_\nu\mathcal{O}_1)\mathcal{O}_2 - \Delta_1\hs\mathcal{O}_1(\partial_\nu\mathcal{O}_2) \Big]\,.
\end{equation}
The operator on the right-hand side is a double-trace vector of dimension $d+1$. It therefore has precisely the quantum numbers required to recombine with $S_{\mu\nu}$ into a long conformal multiplet.

\vskip 4pt
In the bulk, this multiplet recombination is the Higgs mechanism for gravity. The graviton dual to $T_{\mu\nu}$ remains massless, while the graviton dual to $S_{\mu\nu}$ absorbs the vector mode dual to the right-hand side of \eqref{eq: non-cons T} and becomes massive. The mass is generated at one loop through the modified scalar boundary conditions and is related to the anomalous dimension of $S_{\mu\nu}$.

\subsection{Bootstrapping the ACKK Model}
\label{ssec: AdSxAdS bootstrap}
In the following, we will reproduce the structure of the ACKK model from the pseudo-charge conservation identities, and use it to test the formalism of Section~\ref{sec: pseudo}. We assume only the spectrum of operators and the form of the symmetry breaking, and show that the pseudo-charge identities reconstruct the basic structure of the theory. In Appendix~\ref{sec: CPT of AdSxAdS}, we also derive these results through an explicit computation in conformal perturbation theory.

\vskip 4pt
Considering the spectrum $\{T_{\mu \nu}, S_{\mu\nu}, \cO_1, \cO_2\}$ discussed above, we take the scalars to be unit-normalized and orthogonal, and choose the spin-two basis so that $\la TS\ra=0$. We assume that one of the spin-two operators is exactly conserved, while the conservation of the second spin-two operator is broken by double-trace deformations: 
\begin{align}
    & \partial^\mu T_{\mu \nu}= 0\,,\\
    & \partial^\mu S_{\mu\nu}=g_1 \hs(\partial_\nu\cO_1)\cO_2+g_2\hs \cO_1(\partial_\nu\cO_2)\,.
\label{eq: bootstrap breaking}
\end{align}
Nothing is assumed about how the four operators are distributed among factors of a product theory, or indeed that such a factorization exists.

\vskip 4pt
The two breaking parameters in \eqref{eq: bootstrap breaking} are not independent. The right-hand side is equal to the divergence of a traceless spin-two primary operator of dimension $d$, so it has to be primary as well. 
This means the double-trace operator in \eqref{eq: bootstrap breaking} must be annihilated by the action of the special conformal transformation. Using $[K_\mu,\cO_i(0)]=0$ and $[K_\mu,\partial_\nu\cO_i(0)]=2\hs\Delta_i\hs\eta_{\mu\nu}\cO_i(0)$, we find that the action of $K_\mu$ on the right-hand side of (\ref{eq: bootstrap breaking}) vanishes if and only if
\begin{equation}
    \Delta_1\hs g_1+\Delta_2\hs g_2 = 0\,.
\end{equation}
Equation~\eqref{eq: bootstrap breaking} can then be written as
\begin{equation}
    \partial^\mu S_{\mu\nu}=\frac{\hat g}{d}\Big[\Delta_2\hs(\partial_\nu\cO_1)\cO_2-\Delta_1\hs \cO_1(\partial_\nu\cO_2)\Big]\,,
\label{eq: bootstrap breaking primary}
\end{equation}
where $\hat g \equiv g_1-g_2$ and $\Delta_1 + \Delta_2 = d$. We see that a single parameter controls the symmetry breaking.

\paragraph{Current algebra} Given the assumed operator spectrum and the breaking~\eqref{eq: bootstrap breaking primary}, the most general action of the pseudo-charge is
\begin{equation}
\begin{aligned}\label{eq: bootatrap Pt on O1}
    [\widehat Q^\mu, \cO_1] &= a_1\hs \partial^\mu \cO_1 + \hat g\hs \partial^{\hs\mu}\wt\cO_2\,,\\
     [\widehat Q^\mu, \cO_2] &= a_2\hs \partial^\mu \cO_2 - \hat g\hs \partial^{\hs\mu}\wt\cO_1\,,\\
     [\widehat Q^\mu, \mathcal{T}_i] &=  \partial^\mu (\mathcal{A}\,\mathcal{T})_i\,,
\end{aligned}
\end{equation}
where $\mathcal{T}\equiv (T, S)$ and $\mathcal{A}=(a_{ij})$ is a $2\times2$ matrix. The parameters $a_i$ are unfixed current algebra coefficients. The nonlocal transforms appearing in \eqref{eq: bootatrap Pt on O1} are
\begin{equation}
    \wt{\mathcal{O}}_i(x)=\int\d^d\x'\,\frac{\mathcal{O}_i(x')}{|x'-x|^{2\Delta_j}}\,,\quad j\neq i.
\label{eq: shadow scalar}
\end{equation}
Since $\Delta_1+\Delta_2=d$, these are ordinary conformal shadow transforms. Since the stress tensors themselves do not appear in the double-trace deformation~\eqref{eq: bootstrap breaking primary}, there are no nonlocal terms in the action of $\widehat{Q}^\mu$ on $T$ or $S$. Moreover, the matrix ${\cal A}$ is the same as that found for the unbroken case in Section~\ref{ssec:2gravitons}:
\begin{equation}
    \mathcal{A} \equiv \begin{pmatrix}0 & 1\\[2pt]1 & a\end{pmatrix} .
\end{equation}
We will now use pseudo-charge conservation to relate the parameters $a_1$, $a_2$ and~$a$. 

\vskip 4pt
We consider the following constraints:
\begin{align}\label{eq: PTOO}
    0 &= \la [\widehat Q^\mu, \hs T\hs\cO_i\hs \cO_i]\ra\,,\\ \label{eq: PTtOO}
    0&=\la [\widehat Q^\mu, \hs S\hs\cO_i\hs \cO_i]\ra \,.
\end{align}
In these identities, the nonlocal terms either vanish, because $\la T\cO_1\cO_2\ra=0$,\footnote{The unique conformal structure for a spin-two operator and two scalars is only conserved if $\Delta_1=\Delta_2$.} or are of second order in the breaking. Retaining the second-order terms while ignoring the anomalous dimension of $S_{\mu\nu}$, which is of the same order, would not be consistent.\footnote{It would be interesting to include all second-order terms, solve the resulting constraints, and compare them with the corresponding conformal perturbation theory calculation.} Substituting the three-point functions $\la T\cO_i\cO_i\ra=n_i\,\la\!\la T\cO_i\cO_i\ra\!\ra$ and $\la S\cO_i\cO_i\ra=\tilde n_i\,\la\!\la T\cO_i\cO_i\ra\!\ra$ into (\ref{eq: PTOO}) and (\ref{eq: PTtOO}), we get
\begin{align}
    0& = \tilde n_i\hs\partial^{\hs\mu}_{x_1}\la\!\la T\cO_i\cO_i\ra\!\ra +a_i n_i\big(\partial^{\hs\mu}_{x_2}+\partial^{\hs\mu}_{x_3}\big)\la\!\la T\cO_i\cO_i\ra\!\ra\,,\\[4pt]
    0& =\big(n_i+a\tilde n_i\big)\hs\partial^{\hs\mu}_{x_1}\la\!\la T\cO_i\cO_i\ra\!\ra +a_i \tilde n_i\big(\partial^{\hs\mu}_{x_2}+\partial^{\hs\mu}_{x_3}\big)\la\!\la T\cO_i\cO_i\ra\!\ra\,.
\end{align}
Since $\la\!\la T\cO_i\cO_i\ra\!\ra$ is translation invariant, we have $\partial_{x_2}^{\hs \mu}+\partial_{x_3}^{\hs\mu}=-\partial_{x_1}^{\hs \mu}$, so that this becomes
\begin{align}
    0& =\big(\tilde n_i - a_i n_i\big)  \hs\partial^{\hs\mu}_{x_1}\la\!\la T\cO_i\cO_i\ra\!\ra \,,\\[4pt]
    0& =\big(n_i + a\tilde n_i - a_i \tilde n_i
  \big) \partial^{\hs\mu}_{x_1}\la\!\la T\cO_i\cO_i\ra\!\ra \,.
\end{align}
The first constraint is solved by $\tilde n_i = a_i n_i$. Using this in the second constraint, we find
\begin{equation}
    \boxed{a_{1,2}=\frac{1}{2}\left(a\pm\sqrt{a^2+4}\right)}\ ,
\end{equation}
which are precisely the eigenvalues of the matrix ${\cal A}$. We will explain below that this is a manifestation of the equivalence principle.

\paragraph{Pseudo-charge conservation} It is worth highlighting that the breaking has been invisible so far: the identities (\ref{eq: PTOO}) and (\ref{eq: PTtOO}) did not see the nonlocal terms in the action of the pseudo-charge \eqref{eq: bootatrap Pt on O1}. The identities that do see the nonlocal terms are
\begin{align}
\label{eq: PTO1O2}
    0 &= \la [\widehat Q^\mu, \hs T\hs\cO_1 \hs\cO_2]\ra \,,\\ \label{eq: PTtO1O2}
    0 &= \la [\widehat Q^\mu, \hs S\hs\cO_1\hs \cO_2]\ra \,.
\end{align}
We will now show that these constraints are solved for any value of the breaking parameter if the three-point functions are normalized correctly.

\vskip 4pt
To evaluate these constraints, we need various three-point data. First, we note again that $\la T\cO_1\cO_2\ra=0$. Conformal symmetry allows a unique structure for a traceless spin-two operator and two scalars, but this structure is conserved only if $\Delta_1=\Delta_2$, so its coefficient has to vanish for the conserved operator $T_{\mu \nu}$. Since $S_{\mu \nu}$ is also a traceless spin-two primary, the correlator $\la S \cO_1\cO_2\ra$ is proportional to the same structure, but its coefficient is now unconstrained, precisely because $S_{\mu \nu}$ is not conserved. At separated points, we therefore write
\begin{equation}
    \la S_{\mu\nu}\cO_1 \cO_2 \ra = \tau\,\la\!\la S_{\mu \nu}\cO_1\cO_2\ra\!\ra \,,
\label{eq: TtO1O2 parametrisation}
\end{equation}
where $\tau$ is a constant parameter.
Taking the divergence, we get 
\begin{equation}
\begin{aligned}
    \partial^{\hs\mu}\la S_{\mu \nu}\cO_1\cO_2 \ra = \tau\, \frac{(d-1)(\Delta_1-\Delta_2)}{2d\hs \Delta_1\Delta_2}\bigg[&\Delta_2\,\big(\partial^{x_1}_\nu \la \cO_1(x_1)\cO_1(x_2)\ra \big)\la \cO_2(x_1)\cO_2(x_3)\ra  \\ 
    &-\Delta_1\,\la \cO_1(x_1)\cO_1(x_2)\ra \big(\partial^{x_1}_\nu \la \cO_2(x_1)\cO_2(x_3)\ra \big)\bigg]\,.
\end{aligned}
\end{equation}
Comparing this to \eqref{eq: bootstrap breaking primary}, we find
\begin{equation}
    \tau = \frac{2\Delta_1 \Delta_2}{(d-1)\left(\Delta_1- \Delta_2\right)}\hs \hat g\,.
\label{eq: bootstrap tau}
\end{equation}
The normalization of $\la S\cO_1\cO_2\ra$ is therefore fixed in terms of the size of the symmetry-breaking parameter~$\hat g$. 

\vskip 4pt
Explicitly computing the correlators involving the transformed operators $\widetilde {\cal O}_i$ gives
\begin{equation}
\begin{aligned}
    \la T\hs\wt\cO_2\hs\cO_2\ra &= +\hs\mathcal{C}_d\hs \frac{\Delta_1\hs n_2}{\Delta_2-\Delta_1}\,\la\!\la S\cO_1\cO_2\ra\!\ra\,, \\[4pt]  \la T\hs\cO_1\hs\wt\cO_1\ra &=-\hs\mathcal{C}_d\hs \frac{\Delta_2\hs n_1}{\Delta_2-\Delta_1}\,\la\!\la S\cO_1\cO_2\ra\!\ra\,, 
\end{aligned}
\label{eq: TOtO coefficients}
\end{equation}
where $\mathcal{C}_d\equiv 2\pi^{d/2}/\Gamma(\frac{d}{2}+1)$.
Substituting this into \eqref{eq: PTO1O2}, and using $\la T \cO_1 \cO_2 \ra=0$, we get
\begin{equation}
    0= \left[\hs\tau\,\partial^{\hs\mu}_{x_1} +\hat g\hs\mathcal{C}_d\hs\frac{\Delta_1\hs n_2}{\Delta_2-\Delta_1}\,\partial^{\hs\mu}_{x_2} +\hat g\hs \mathcal{C}_d\hs \frac{\Delta_2\hs n_1}{\Delta_2-\Delta_1}\,\partial^{\hs\mu}_{x_3} \right]\la\!\la S\cO_1\cO_2\ra\!\ra\,.
\end{equation}
Note that $\tau \propto \hat g$, so this constraint also does not depend on the value of the breaking parameter. The pseudo-charge conservation identities are therefore satisfied for any size of the symmetry breaking (as long as it remains perturbative). Translation invariance requires that the three coefficients in front of the divergences are equal. Substituting \eqref{eq: bootstrap tau} for $\tau$ and using $\Delta_1 \hs n_2 = n_1 \hs \Delta_2$ (which follows from the translation Ward identity), we find
\begin{equation}
    \boxed{n_i = -\frac{\Gamma\!\big(\tfrac d2+1\big)}{(d-1)\hs\pi^{d/2}}\hs \Delta_i}\ .
\label{eq: bootstrap kinematic}
\end{equation}
This fixes the normalization of $\la T\cO_i\cO_i\ra$ in terms of the scaling dimension $\Delta_i$, and is consistent with~\cite{Osborn_1994}. A similar analysis shows that the same solution also solves the second constraint~\eqref{eq: PTtO1O2}. We have therefore demonstrated that the pseudo-charge conservation identities are nontrivially satisfied for every value of the breaking parameter $\hat g$.

\paragraph{Interpretation} We have seen that the pseudo-charge conservation identities constrain the form of the current algebra:
\begin{equation}
\begin{aligned}
    [\widehat Q^\mu, \cO_1] &= a_1\hs \partial^\mu \cO_1 + \hat g\hs \partial^{\hs\mu}\wt\cO_2\,,\\
    [\widehat Q^\mu, \cO_2] &= a_2\hs \partial^\mu \cO_2 - \hat g\hs \partial^{\hs\mu}\wt\cO_1\,,\\
    [\widehat Q^\mu, \mathcal{T}_i] &=  \partial^\mu (\mathcal{A}\,\mathcal{T})_i\,,
\end{aligned}
\end{equation}
where
\begin{equation}
    \mathcal{A} = \begin{pmatrix}
        0&1\\[2pt]
        1&a
    \end{pmatrix}, \quad a_{1,2}=\frac{1}{2}\left(a\pm\sqrt{a^2+4}\right) .
\end{equation}
In Appendix~\ref{sec: CPT of AdSxAdS}, we show that this matches the structure of the ACKK model after identifying $a_1=\hat c_2$, $a_2=-\hat c_1$, and $\hat g=(\hat c_1+\hat c_2)\hs g$, where the parameters~$\hat c_i$ defined in \eqref{eq: chat} are the rescaled central charges of the two factors in the normalization $c_S=c_T$. Since the eigenvalues $a_1$ and $a_2$ are distinct, we can diagonalize
the matrix ${\cal A}$ and write 
\begin{equation}
    \big[\hs\widehat Q^{\hs\mu},T^{(i)}_{\alpha \beta}\big]=a_i\hs\partial^{\hs\mu}T^{(i)}_{\alpha \beta}\,, \quad {\rm where} \quad T^{(i)}_{\mu \nu} \equiv \frac{a_j T_{\mu \nu} - S_{\mu \nu}}{a_j - a_i} \,, \quad j\ne i.
\end{equation}
Note that $T^{(1)}_{\mu \nu} + T^{(2)}_{\mu \nu} = T_{\mu \nu}$ and $a_1 T^{(1)}_{\mu \nu} + a_2 T^{(2)}_{\mu \nu} = S_{\mu \nu}$.\footnote{This reconstruction captures only the traceless parts of the stress tensors. In the explicit ACKK realization, the full operators in~\eqref{eq: T conserved} and~\eqref{eq: T non-conserved} also contain terms proportional to $\eta_{\mu\nu}\cO_1\cO_2$. These are the trace and contact-term completions required by the deformed Ward identities, and are invisible to the separated-point pseudo-charge identities. Thus, the diagonalization above recovers precisely the traceless parts of the two ACKK stress-tensor combinations; the scalar-bilinear terms must be restored to obtain the full operators quoted in Section~\ref{ssec:ACKK}.} Within each CFT, the scalar and the stress tensor carry the same pseudo-charge eigenvalue $a_i$. This is the boundary form of the {\it equivalence principle}. The different eigenvalues $a_1$ and $a_2$ label the universal couplings in the two decoupled sectors rather than non-universal gravitational couplings inside a single CFT. 

\vskip 4pt
The coupling between the two CFTs breaks the conservation of the second stress tensor~$S_{\mu \nu}$. We have verified explicitly that all pseudo-charge conservation constraints are satisfied for arbitrary values of the breaking parameter. Unlike in the example presented in Section~\ref{sec: Higgsing in CG}, pseudo-charge conservation therefore does not fix the magnitude of the marginal coupling; instead, it fixes how the broken symmetry is realized, reconstructing the two-sector current algebra, the universal stress-tensor normalization, and the characteristic scalar mixing of the ACKK model.

\section{Conformal Gravity in de Sitter}
\label{sec: Higgsing in CG}

We now turn to conformal gravity in four-dimensional de Sitter space~\cite{Weyl:1918ib, Weyl:1919fi, Weyl:1918pdp, Bach:1921zdq}. This is an interesting case where the partial conservation of $X$ is broken by a double-trace deformation of the current itself. In the bulk, this corresponds to a form of {\it self-Higgsing} of the PM gauge symmetry.

\vskip 4pt
The action of conformal gravity is
\begin{equation}
    S = -\frac{\alpha^2}{8} \int \d^4 x \sqrt{-g} \, C^{MNKL} C_{MNKL}\,,
\label{equ:CG-action}
\end{equation}
where $C_{MNKL}$ is the Weyl tensor and $\alpha^2$ is a dimensionless coupling. Expanded around the de Sitter background, the theory propagates a massless graviton and a partially massless spin-2 field~\cite{ Maldacena:2011mk, Deser:2012qg, Deser:2012euu, Joung:2014aba}. These are dual to the conserved stress tensor $T_{\mu\nu}$ and a partially conserved spin-2 current~$X_{\mu\nu}$, respectively. The coupling constant is related to the Planck mass by $\alpha = L M_{\rm Pl}$, where $L$ is the de Sitter radius. In our previous AdS examples, the breaking of current conservation was in a sense optional: it was induced by a choice of boundary conditions that was incompatible with the bulk gauge symmetry. In de Sitter space, on the other hand, we do not have the freedom to pick similar boundary conditions at the de Sitter boundary. Instead, we pick the initial Bunch--Davies state of the fields at early times, which then implies a specific late-time behavior of the fields. Moreover, to derive observables we must integrate over all possible late-time field profiles. In~\cite{Baumann:2025tkm}, it was shown that in conformal gravity this always leads to a breaking of the bulk gauge symmetry. In this section, we will provide a CFT analysis of this phenomenon.

\subsection{Pure Conformal Gravity}
\label{ssec:PureCG}
We will first study pure conformal gravity, where the spectrum of boundary operators consists only of the stress tensor and the partially conserved current. We will show that pseudo-charge conservation fixes all three-point functions of this spectrum up to a single overall normalization, and that the resulting cubic data are exactly those predicted by the action (\ref{equ:CG-action}). 

\vskip 4pt
In the free theory, the PM current satisfies $\partial_\mu\partial_\nu X^{\mu\nu}=0$, but interactions deform this relation. Given the spectrum above, the unique double-trace deformation with the required quantum numbers is $X_{\mu\nu}X^{\mu\nu}$. We are therefore led to the weakly broken conservation equation
\begin{equation}
    \partial_\mu \partial_\nu X^{\mu \nu} = g\hs X_{\mu \nu}X^{\mu \nu}\,,
\label{eq: double trace pure CG}
\end{equation}
where $g$ is a dimensionless coefficient.

\paragraph{Anomalous dimension} The breaking of partial conservation shifts the scaling dimension of the operator $X_{\mu\nu}$, so that $\Delta_X = 2 + \gamma_X$. This shift can be extracted by applying two divergences at each point to the two-point function $\langle XX\rangle$. This calculation was previously carried out in~\cite{Baumann:2025tkm}, and rewriting the result in our current conventions gives 
\begin{equation}
    \gamma_X =\frac{g^2}{4}\,.
\label{eq:CG-anomalous-dimension}
\end{equation}
Matching the coupling $g$ to the bulk normalization of the PM self-interaction gives $g = \sqrt{3}/(\pi M_{\rm Pl})$, in units where the dS radius is set to unity, so that $\gamma_X = 3/(4\pi^2 M_{\rm Pl}^2)$, reproducing the result of~\cite{Baumann:2025tkm}. The PM field thus acquires a {\it positive} anomalous dimension suppressed by the Planck scale, confirming the breaking of partial conservation.\footnote{While the two-point function of $X$ flips sign when going from AdS to dS, the overall minus sign in the exponent of $\Psi_{\rm dS}$ (as opposed to the plus sign multiplying the two-point function in $Z_{\rm AdS}$) compensates for this, so the conclusion that $\gamma_X >0$ remains unaffected.} In de Sitter space, $\gamma_X >0$ implies a \textit{negative} mass shift of the PM field.

\paragraph{Current algebra} Given the assumed operator spectrum and the double-trace deformation~\eqref{eq: double trace pure CG}, it is easy to write down the most general action of the pseudo-charge:
\begin{equation}
\begin{aligned}
    [\pQ, X] & = a_1 T -2g\hs \tX\,,\\[3pt]
    [\pQ, T] & = a_2 \left(\square - \frac{2}{3} (z\cdot \partial )(\partial \cdot D_z)\right) X\,,
\end{aligned}
\label{eq: current algebra CG}
\end{equation}
where we have defined the nonlocal spin-2 operator
\begin{equation}
    \tX^{\mu \nu}(x) \equiv \int \d^3 x' \,\la X_{\rho \sigma}(x') X^{\mu \nu}(x)\ra \,X^{\rho \sigma}(x')\,.
\label{eq: X tilde}
\end{equation}
Note that this operator is {\it not} the shadow transform of the operator $X^{\mu \nu}$, and so three-point functions involving this operator are not necessarily conformal. However, they should still be able to cancel against derivatives of conformal three-point functions in the conservation identities if nontrivial solutions exist.

\vskip 4pt
The coefficient $a_2$ in \eqref{eq: current algebra CG} is not independent of $a_1$. To see this, we consider the two-point functions with general normalizations
\begin{equation}
    \la T(x_1)T(x_2)\ra=c_T\,\frac{H_{12}^2}{(-2P_{12})^5}\,, \quad \la X(x_1)X(x_2)\ra=c_X\,\frac{H_{12}^2}{(-2P_{12})^4}\,.
\end{equation}
Pseudo-charge invariance of the mixed two-point function then gives
\begin{equation}
    0 = \la[\pQ,X(x_1)T(x_2)]\ra = a_1\la T(x_1)T(x_2)\ra +a_2\,\bigg(\square_{x_2} - \frac{2}{3}(z_2\cdot \partial_2)(\partial_2 \cdot D_{z_2})\bigg)\la X(x_1)X(x_2)\ra\,.
\end{equation}
From this, we obtain
\begin{equation}
    8 a_2\hs c_X = -a_1\hs c_T\,,
\label{eq: relation c1 and c2}
\end{equation}
which explains why $a_2$ does not appear as an independent parameter in the three-point solution below. For unit-normalized two-point functions, $c_T=c_X=1$, one simply has $a_2=-a_1/8$.

\paragraph{Conservation identities}
The relevant pseudo-charge identities satisfied by the three-point functions of the minimal spectrum are
\begin{align}
    0&= \la [\pQ, TTT]\ra \,, \label{eq: QTTT}\\
    0&=\la [\pQ, XXX]\ra \,, \label{eq: QXXX}\\
    0&= \la [\pQ, XTT]\ra \,. \label{eq: QXTT}
\end{align}
The remaining identity, $\la [\pQ, TXX]\ra=0$, does not lead to an independent constraint.
 
\vskip 4pt
Using the exactly conserved parity-even ansatz at separated points
\begin{equation}
\begin{aligned}
    \la XTT \ra = n_{XTT}\Bigg(&\frac{V_1^2 H_{23}^2 - 4\,V_1^2 V_2 V_3 H_{23} - 2\,V_2 V_3 H_{12}H_{31} - 20\,V_1^2 V_2^2 V_3^2}{(-2\,P_{12})^2\,(-2P_{23})^3\,(-2P_{31})^2} \\ &- \frac{8\,V_1 V_2 V_3^2 H_{12} + 2\,V_1 V_2 H_{23}H_{31} + (2\leftrightarrow 3)}{(-2\,P_{12})^2\,(-2P_{23})^3\,(-2P_{31})^2}\Bigg)\,,
\end{aligned}
\end{equation}
one finds that \eqref{eq: QTTT} can only be satisfied if either $a_2 = 0$ or $n_{XTT} = 0$. The former would decouple the PM field from gravity and is therefore excluded. We conclude that
\begin{equation}
    \boxed{\rule[-0.9ex]{0pt}{3.2ex}n_{XTT}=0}\ .
\end{equation}
This is consistent with conformal gravity having no coupling between two gravitons and a single PM field (see Appendix~\ref{sec: Details on CG}).

\vskip 4pt
Substituting the current algebra \eqref{eq: current algebra CG} into the remaining constraints \eqref{eq: QXXX} and \eqref{eq: QXTT} then gives 
\begin{align}
\label{eq: conformal gravity identities}
    0&= a_1\la TXX \ra -2g \la \tX XX\ra + (1\leftrightarrow2) + (1\leftrightarrow3) \,,  \\
    0&= a_1\la TTT \ra + \Bigg[ a_2 \bigg(\square_2 - \frac{2}{3} (z_2\cdot \partial_2 )(\partial_2 \cdot D_{z_2})\bigg)\la XXT\ra + (2\leftrightarrow3) \Bigg] \,.
\label{eq: conformal gravity identities2}
\end{align}
The identity \eqref{eq: conformal gravity identities2} contains no term proportional to $g$, because $\la \widetilde XTT\ra$ is obtained by integrating $\la XTT\ra$, which we just showed vanishes. Moreover, after substituting $a_2=-a_1/8$, the constraint becomes independent of the parameters of the current algebra, and therefore only fixes the relative size of $\la TTT\ra$ and $\la XXT\ra$. 

\vskip 4pt
To evaluate the identity~\eqref{eq: conformal gravity identities}, we require the correlator $\la \tX XX\ra$. From \eqref{eq: X tilde}, it takes the form
\begin{equation}
    \la \tX(x_1) X(x_2) X(x_3) \ra = \int \d^3 x\, \la X_{\mu \nu}(x) X(x_1)\ra \la X^{\mu \nu}(x) X(x_2) X(x_3)\ra\,.
\label{eq: X transform}
\end{equation}
We use an ansatz for $\la XXX\ra$ that obeys the depth-0 conservation condition up to semi-analytic terms, as discussed in Section~\ref{sec: Weakly Broken Gauge Symmetry}:
\begin{equation}
\begin{aligned}
    \la XXX\ra & = -\,n^{(1)}_{XXX}\frac{7 V_1^2 H_{23}^2 - 22 V_2 V_3 H_{12}H_{31} - 20 V_1^2 V_2^2 V_3^2 + \text{cyclic}}{3840\, (-2P_{12})^2 (-2P_{23})^2 (-2P_{31})^2} \\[3pt]
    & \quad + \,n^{(2)}_{XXX}\frac{3 V_1^2 H_{23}^2 -8 V_2 V_3 H_{12}H_{31} + 10 V_1^2 V_2 V_3 H_{23} + \text{cyclic}}{640\, (-2P_{12})^2 (-2P_{23})^2 (-2P_{31})^2}\,,
\end{aligned}
\end{equation}
which contains two conformally invariant parity-even structures. Notice that imposing exact conservation would allow only for a single structure. As discussed in Appendix~\ref{app: Semi-Local Terms}, this correlator does not have any semi-local terms that could give contributions at separated points after performing the integral \eqref{eq: X transform}. 

\vskip 4pt
By contrast, our ans\"atze for $\la TXX\ra$ and $\la TTT\ra$ satisfy the homogeneous conservation conditions exactly: a single divergence at the $T$ insertion and two divergences at each $X$ insertion vanish at separated points. Explicitly, we have
\begin{align}
    \langle T X X\rangle = &-n_{TXX}^{(1)}\,\frac{4V_1^2H_{23}^2 - 15V_1^2V_2^2V_3^2 + \big[6V_1V_2H_{23}H_{31} + 2V_2^2H_{31}^2 + (2\leftrightarrow3)\big]}{15\,(-2P_{12})^{5/2}(-2P_{23})^{3/2}(-2P_{31})^{5/2}} \nonumber \\[4pt]
    &+n_{TXX}^{(2)}\,\frac{3V_1^2H_{23}^2 + \big[15V_1V_2V_3^2H_{12} + 12V_1V_2H_{23}H_{31} + 4V_2^2H_{31}^2 + (2\leftrightarrow3)\big]}{15\,(-2P_{12})^{5/2}(-2P_{23})^{3/2}(-2P_{31})^{5/2}} \nonumber\\[4pt]
    &+n_{TXX}^{(3)}\,\frac{V_1^2H_{23}^2 + 2V_1^2V_2V_3H_{23} + \big[V_1V_2H_{23}H_{31} + (2\leftrightarrow3)\big]}{2\,(-2P_{12})^{5/2}(-2P_{23})^{3/2}(-2P_{31})^{5/2}} \nonumber \\[4pt]
    &+n_{TXX}^{(4)}\,\frac{3V_1^2H_{23}^2 + 60V_2V_3H_{12}H_{31} + \big[27V_1V_2H_{23}H_{31} + 14V_2^2H_{31}^2 + (2\leftrightarrow3)\big]}{60\,(-2P_{12})^{5/2}(-2P_{23})^{3/2}(-2P_{31})^{5/2}}\,,  \\[8pt]
    \langle TTT\rangle ={}& - n_{TTT}^{(1)}\frac{10V_1^2H_{23}^2 + 24V_1V_2H_{23}H_{31}-11V_1^2V_2^2V_3^2 + \text{cyclic}}{33\,(-2P_{12})^{5/2}(-2P_{23})^{5/2}(-2P_{31})^{5/2}}\nonumber \\[4pt]
    &+ n_{TTT}^{(2)}\frac{9V_1^2H_{23}^2+11V_1^2V_2V_3H_{23}+26V_1V_2H_{23}H_{31} + \text{cyclic}}{11\,(-2P_{12})^{5/2}(-2P_{23})^{5/2}(-2P_{31})^{5/2}}\,.
\end{align}
Because both correlators contain a stress tensor, their $T\times X$ and $T\times T$ OPE limits are singular enough to require a distributional completion, so $\la TXX\ra$ and $\la TTT\ra$ carry semi-local (contact) terms (see Appendix~\ref{app: Semi-Local Terms}). These are contact terms supported at coincident points; since no nonlocal transform acts on $\la TXX\ra$ or $\la TTT\ra$ in \eqref{eq: conformal gravity identities} they cannot be promoted to separated points and play no role here. The identities \eqref{eq: conformal gravity identities} and \eqref{eq: conformal gravity identities2} then involve the coefficients
\begin{equation}
    \Big\{\,a_1\,, \,g\,,\, n_{TXX}^{(i)}\,,\,n_{XXX}^{(i)}\,, \, n_{TTT}^{(i)}\,\Big\}\,.
\end{equation}
Solving the two equations simultaneously yields the solution
\begin{align}
    &\boxed{n_{XXX}^{(1)} = \frac{2048}{135 \pi^3}\frac{a_1}{g} n_{TTT}^{(1)}}\ , \quad {\rm with} \quad n_{XXX}^{(2)} = \frac{3}{2} n_{XXX}^{(1)} \,, \\[4pt]
    &\boxed{\rule[-1.1ex]{0pt}{4ex} n_{TXX}^{(1)} = 4\hs n_{TTT}^{(1)}}\ , \quad \hspace{1.4cm} {\rm with} \quad  n_{TXX}^{(4)} = - n_{TXX}^{(3)} = - \frac{1}{2} n_{TXX}^{(2)} = - \frac{2}{9} n_{TXX}^{(1)}\,,\\[4pt]
    &\boxed{n_{TTT}^{(2)} = - \frac{9}{4} n_{TTT}^{(1)}}\ .
\end{align}
We see that pseudo-charge conservation fixes the normalizations of $\la XXX \ra$ and $\la TXX \ra$ in terms of the universal gravity correlator $\la TTT\ra$. By matching to the bulk interactions, the latter is fixed by the gravitational coupling $1/M_{\rm Pl}$. It is easy to confirm that the solution exactly matches the bulk computation of the three-point functions in conformal gravity~\cite{Baumann:2025tkm}. This means that the conservation constraints have rediscovered conformal gravity from the boundary point of view as the unique theory of a PM spin-2 and Einstein gravity with a weakly broken PM gauge symmetry.

\subsection{Coupling to Scalars}
\label{sec:CG-coupling-to-scalars}
Having fixed the pure gravitational sector, we next ask whether conformal gravity can consistently couple to scalar matter. The simplest candidate is a scalar primary operator $\cO_2$ of dimension $\Delta=2$, dual to a conformally coupled scalar in dS$_4$. The double-trace operator $\cO_2\cO_2$ then has dimension four and provides a second channel for the breaking of the PM symmetry:
\begin{equation}
    \partial_\mu \partial_\nu X^{\mu \nu} = g\,X_{\mu \nu}X^{\mu \nu} +g'\,\cO_2\cO_2\,,
\label{eq: double trace CG plus scalars}
\end{equation}
where $g$ and $g'$ are both dimensionless.

\vskip 4pt
Before imposing the constraints, it is worth asking which spectrum to impose them on. The bulk suggests two possibilities. As reviewed in Appendix~\ref{app:CouplingScalars}, conformal gravity can be coupled to a conformally coupled scalar, whose dual is a single boundary operator $\cO_2$ of dimension $\Delta=2$. It can also be coupled to a Weyl-invariant scalar with a four-derivative kinetic term $(\square\psi)^2$, which around the dS$_4$ background decomposes into a conformally coupled scalar and a massless scalar, and therefore adds a second boundary operator $\cO_3$ of dimension $\Delta=3$. Guided by this, we will allow for $\cO_3$ in the analysis; the truncation to the minimal spectrum $\{T, X, \cO_2\}$ is recovered by setting the corresponding algebra coefficients to zero. Importantly, the boundary constraints alone do {\it not} force the presence of $\cO_3$: they admit two branches of solutions, one for each spectrum, and both are realized in the bulk.

\paragraph{Anomalous dimension} The two terms in \eqref{eq: double trace CG plus scalars} both contribute to the anomalous dimension of the partially conserved current. At leading nontrivial order in the weak-breaking parameters, one finds
\begin{equation}
    \gamma_X = \frac{g^2}{4}
    +\frac{{g'}^2}{20}\,.
\label{eq:CG-scalars-anomalous-dimension}
\end{equation}
Both channels therefore give a positive contribution to the anomalous
dimension, so the conclusion that the PM field receives a negative mass shift in de Sitter space is robust under the addition of scalar matter. 

\paragraph{Current algebra} Given the assumed operator spectrum, and the broken conservation equation~\eqref{eq: double trace CG plus scalars}, we are led to the following pseudo-charge algebra\footnote{Restricting the theory to have a spectrum $\{T,X,\cO_2\}$ simply corresponds to taking $a_3 = a_4 = 0$.}
\begin{equation}
\begin{aligned}
    [\pQ,X] &= a_1 T-2g\,\tX\,, \\[2pt]
    [\pQ,T] &= a_2\left( \square-\frac{2}{3}(z\cdot\partial)(\partial\cdot D_z) \right)X\,, \\[2pt]
    [\pQ,\cO_2] &=a_3\,\cO_3-2g'\,\tO_2\,,\\[6pt]
    [\pQ,\cO_3] &= a_4\,\square\cO_2\,,
\end{aligned}
\label{eq: current algebra CG plus scalars}
\end{equation}
where $\tX$ was defined in \eqref{eq: X tilde}, and
\begin{equation}
    \tO_2(x) \equiv \int \d^3x'\, \la \cO_2(x')\cO_2(x)\ra\, \cO_2(x')\,.
\label{eq: O2 tilde}
\end{equation}
As for $\widetilde X$, the integral in \eqref{eq: O2 tilde} should be understood as acting within three-point functions, and should be properly regulated using dimensional regularization when needed.

\vskip 4pt
Above we derived the relation \eqref{eq: relation c1 and c2} between the parameters $a_1$ and $a_2$. A similar relation for the coefficients $a_3$ and $a_4$ follows from pseudo-charge invariance of the mixed scalar two-point function:
\begin{equation}
    0 = \la[\pQ,\cO_2(x_1)\cO_3(x_2)]\ra = a_3\la\cO_3(x_1)\cO_3(x_2)\ra +a_4\,\square_{x_2}\la\cO_2(x_1)\cO_2(x_2)\ra\,.
\label{eq: scalar two point charge identity}
\end{equation}
Normalizing the two-point functions as
\begin{equation}
    \la\cO_2(x_1)\cO_2(x_2)\ra=\frac{n_{\cO_2}}{(x_{12}^2)^2}\,, \quad\la\cO_3(x_1)\cO_3(x_2)\ra=\frac{n_{\cO_3}}{(x_{12}^2)^3}\,,
\label{eq: two point normalizations}
\end{equation}
we get 
\begin{equation}
    a_4=-\frac{n_{\cO_3}}{12\hs n_{\cO_2}}\,a_3\,.
\label{eq: c4 scalar relation}
\end{equation}
The operator $\cO_3$ is dual to the massless scalar of the extended bulk spectrum, which carries a wrong-sign kinetic term (see Appendix~\ref{sec: Details on CG}). Anticipating this, we normalize the scalar two-point functions as $n_{\cO_2}=1$ and $n_{\cO_3}=-1$, so that $a_4=a_3/12$.\footnote{We could equally well have chosen unit normalization for $\cO_3$. A negative-norm operator cannot be brought to unit norm by a real rescaling, however, so this amounts to $\cO_3 \to i\hs\cO_3$, and factors of $i$ would then appear in the solution of the pseudo-charge conservation identities below.} The current algebra~\eqref{eq: current algebra CG plus scalars} can then be written in terms of the parameters $\{a_1,a_3,g,g'\}$.

\paragraph{Conservation identities} 
Next, we impose pseudo-charge conservation on three-point functions involving $X$, $T$, and the two scalars. The relevant identities containing one insertion of $X$ or $T$ are
\begin{align} \label{eq: charge identity X O2 O2}
    0 &= \la [\pQ,T\cO_2\cO_2]\ra\,,\\ \label{eq: charge identity T O2 O2}
    0 &= \la [\pQ,X\cO_2\cO_2]\ra\,,\\ \label{eq: charge identity X O3 O3}  
    0 &= \la [\pQ,X\cO_3\cO_3]\ra\,,\\ \label{eq: charge identity X O2 O3}
    0 &= \la [\pQ,X\cO_2\cO_3]\ra\,.
\end{align} 
For the minimal spectrum $\{T,X,\cO_2\}$ only the first two are relevant. The remaining identities of this type, $\la [\pQ, T\cO_2 \cO_3]\ra = 0$ and $\la [\pQ, T\cO_3 \cO_3]\ra = 0$, are redundant once \eqref{eq: c4 scalar relation} and the relations derived below are imposed. Expanding the commutators using \eqref{eq: current algebra CG plus scalars}, these identities become
\begin{align}
    0 &= a_2 \bigg(\square_1 - \frac{2}{3}(z_1\cdot \partial_1)(\partial_1 \cdot D_{z_1})\bigg)\la X\cO_2\cO_2\ra + \Big(a_3\la T\cO_3\cO_2\ra -2g'\la T\tO_2\cO_2\ra + (2\leftrightarrow 3)\Big)\,,
\label{eq: T O2 O2 identity} \\[4pt]
    0 &= a_1\la T\cO_2\cO_2\ra -2g\la\tX\cO_2\cO_2\ra +\Big( a_3 \la X\cO_3\cO_2\ra -2g' \la X\tO_2\cO_2\ra + (2\leftrightarrow 3)\Big)\,, \label{eq: X O2 O2 identity} \\[4pt]
    0 &=a_1\la T\cO_3\cO_3\ra-2g\la\tX\cO_3\cO_3\ra+a_4\Big(\square_2\la X\cO_2\cO_3\ra+\square_3\la X\cO_3\cO_2\ra\Big)\,, \label{eq: X O3 O3 identity}\\[4pt]
    0 &=a_1\la T\cO_2\cO_3\ra-2g\la\tX\cO_2\cO_3\ra+a_3\la X\cO_3\cO_3\ra-2g'\la X\tO_2\cO_3\ra+a_4\,\square_3\la X\cO_2\cO_2\ra \,.\label{eq: X O2 O3 identity}
\end{align}
The insertion of the nonlocal operators leads to convolution integrals of two-point functions with ordinary conformal three-point functions, exactly as in the pure conformal-gravity analysis.

\vskip 4pt
Each conformal three-point structure of one spinning operator $Y$ and two scalars $\{\cO_i, \cO_j\}$ is of the form $\la Y \cO_i \cO_j \ra = N_{Y \cO_i \cO_j} \la\!\la Y \cO_i \cO_j \ra\!\ra$, with $\la\!\la Y \cO_i \cO_j \ra\!\ra$ defined in \eqref{eq: YOO strcuture}. Conservation of the stress tensor forces the mixed correlator $\la T\cO_2\cO_3\ra$ to vanish. The diagonal correlators $\la T\cO_i\cO_i\ra$ are accompanied by a semi-local trace term. Since we contract the identities with auxiliary null vectors, $z_i^2=0$, these terms drop out and play no role here. (In the next subsection, the analogous completion of $\la TJJ\ra$ is not pure trace and does contribute.) A similar statement holds for the mixed correlator $\la X\cO_2\cO_3\ra$, and is discussed in more detail in Appendix~\ref{app: Semi-Local Terms}.

\vskip 4pt
Evaluating the convolution integrals and matching the independent position-space structures gives two branches of solutions. 
\begin{itemize}
\item We first consider the extended scalar sector, so that the operator spectrum is $\{T,X,\cO_2,\cO_3\}$. Solving the pseudo-charge conservation constraints then leads to 
\begin{align}
    &\boxed{n_{X \cO_2 \cO_2} = \frac{12}{\pi^3}\hs \frac{a_1}{g}\hs n_{T \cO_2 \cO_2}\,, \quad n_{X \cO_2 \cO_3} = \frac{3 a_1}{a_3}\hs n_{T \cO_2 \cO_2}\, , \quad n_{X \cO_3 \cO_3} = 0}\ ,\\
    &\boxed{n_{T\cO_3\cO_3}=-\frac{3}{2} \hs n_{T\cO_2\cO_2}\, , \quad n_{T \cO_2 \cO_3} = 0}\quad {\rm and} \quad  \boxed{g' = \frac{g}{3}} \ .
\label{eq: solution CG scalars}
\end{align}
We see that all couplings of the PM field to the bulk scalar fields, reflected in the three-point functions $\la X \cO_i \cO_j \ra$, have been fixed in terms of the size of the coupling to the graviton, as measured by $n_{T\cO_2 \cO_2}$. Moreover, the relative strength of the two symmetry-breaking channels is fixed to $g=3g'$. The corresponding bulk couplings are listed in Appendix~\ref{app:CouplingScalars}. The absence of $f\psi\psi$ and $h\psi\phi$ vertices matches $n_{X\cO_3\cO_3}=n_{T\cO_2\cO_3}=0$, the $f\psi\phi$ vertex matches $n_{X\cO_2\cO_3}\neq0$, and the opposite-sign kinetic terms match $n_{\cO_3}=-n_{\cO_2}$.

\item Interestingly, we have also found a second solution involving only the minimal spectrum $\{T,X,\cO_2\}$. Solving the pseudo-charge conservation constraints with $a_3=a_4=0$, we found 
\begin{equation}
    \boxed{n_{X \cO_2 \cO_2} = \frac{6}{\pi^3} \hs \frac{a_1}{g} \hs n_{T\cO_2 \cO_2}} \quad {\rm and} \quad \boxed{g' = \frac{g}{6}}\ .
\label{eq: solution CG minimal scalars}
\end{equation}
This again matches the results of the bulk analysis presented in Appendix~\ref{sec: Details on CG}. In particular, both $n_{X\cO_2\cO_2}$ and $g'$ are half as large as in the extended spectrum above, which is exactly the ratio of the $f\phi\phi$ vertices in the two bulk theories of Appendix~\ref{app:CouplingScalars}; see \eqref{Lhfphiphi} and~\eqref{Lhfphiphi higher der}.
\end{itemize}

\subsection{Coupling to Vectors}
\label{sec:CG-coupling-to-vectors}
Finally, we investigate whether conformal gravity can be coupled consistently to a massless vector field. With standard boundary conditions, the bulk vector is dual to a conserved spin-one current $J_\mu$ of dimension $\Delta_J=2$ in $d=3$. The double-trace operator $J_\mu J^\mu$ therefore has dimension four and can provide an additional source for the weak breaking of the partially massless symmetry:
\begin{equation}
    \partial_\mu\partial_\nu X^{\mu\nu} = g\,X_{\mu\nu}X^{\mu\nu} +g''\,J_\mu J^\mu\,.
\label{eq: double trace CG plus vector}
\end{equation}
An explicit bulk realization of this coupling is presented in Appendix~\ref{sec: Details on CG}. Here, we derive the corresponding constraints directly from the boundary pseudo-charge identities.

\paragraph{Anomalous dimension} The additional breaking channel shifts the scaling dimension of the partially
massless current. At leading order in the weak-breaking
parameters, the result is
\begin{equation}
    \gamma_X = \frac{g^2}{4} +\frac{3{g''}^2}{20}\,,
\label{eq: CG vector anomalous dimension}
\end{equation}
so that, as before, the anomalous dimension remains positive.

\paragraph{Current algebra} Given the operator spectrum $\{T,X,J\}$ and the non-conservation equation \eqref{eq: double trace CG plus vector}, the most general ansatz for the pseudo-charge algebra is\footnote{One might also allow a local term proportional to $(z\cdot\partial)J$ in $[\pQ,X]$, together with a term proportional to $(\partial\cdot D_z)X$ in $[\pQ,J]$. Within the parity-even sector considered here, these terms would require a nonzero conformal three-point function $\la JXX\ra$ satisfying current conservation, double conservation on the two $X$ insertions, and Bose symmetry under exchange of the two $X$ operators. No such structure exists. We therefore omit these mixing terms. }
\begin{equation}
\begin{aligned}
    [\pQ,X] &= a_1\,T-2g\,\tX\,, \\[2pt]
    [\pQ,T] &= a_2\left(\square-\frac{2}{3}(z\cdot\partial)(\partial\cdot D_z) \right)X\,, \\[2pt]
    [\pQ,J^\mu] &= -2g''\,\widetilde J^\mu\,,
\end{aligned}
\label{eq: current algebra CG plus vector}
\end{equation}
where $\tX$ was defined in \eqref{eq: X tilde}, and
\begin{equation}
    \widetilde J^\mu(x) \equiv \int \d^3x'\, \la J_\nu(x')J^\mu(x)\ra\,J^\nu(x')\,.
\label{eq: J tilde}
\end{equation}
As we will see, this integral is too singular at coincident points of $\la T J J\ra$, which requires the appropriate semi-local completion. Including this term is essential for solving the constraint. This contrasts with all the examples above, in which the semi-local terms in the correlators played no role in the solution.

\paragraph{Conservation identities} The pseudo-charge identities involving only $T$ and $X$ are unchanged from the pure conformal-gravity analysis and therefore do not constrain the coupling to $J$. Instead, we must consider identities containing at least one insertion of $J$. The complete set of constraints may be organized as
\begin{equation}
\begin{aligned}
    0&=\la[\pQ,JJJ]\ra\,, & 0&=\la[\pQ,XJJ]\ra\,, & 0&=\la[\pQ,JXX]\ra\,, \\[2pt]
    0&=\la[\pQ,TJJ]\ra\,, & 0&=\la[\pQ,JTT]\ra\,, & 0&=\la[\pQ,JXT]\ra\,.
\end{aligned}
\end{equation}
Most of these are almost trivially satisfied. There is no conserved parity-even three-point function $\la JJJ\ra$, so the first identity is automatic. Similarly, the absence of an allowed $\la JXX\ra$ structure implies, through $\la[\pQ,JXX]\ra=0$, that $\la JXT\ra=0$. The same conclusion follows from $\la[\pQ,JTT]\ra=0$, since there is no corresponding conserved $\la JTT\ra$ structure. The final identity $\la[\pQ,JXT]\ra=0$ is then also automatic. The only nontrivial vector
constraints therefore are
\begin{align}
    0&=\la[\pQ,XJJ]\ra\,,
\label{eq: vector XJJ commutator identity}
    \\
    0&=\la[\pQ,TJJ]\ra\,.
\label{eq: vector TJJ commutator identity}
\end{align}
Expanding the commutators using
\eqref{eq: current algebra CG plus vector}, we obtain
\begin{align}
    0={}& a_1\la TJJ\ra -2g\la\tX JJ\ra -2g''\la X\widetilde J J\ra -2g''\la XJ\widetilde J\ra\,,
\label{eq: XJJ constraint}
    \\[4pt]
    0={}&a_2\left(\square_1-\frac{2}{3}(z_1\cdot\partial_1)(\partial_1\cdot D_{z_1})\right)\la XJJ\ra -2g''\la T\widetilde J J\ra-2g''\la TJ\widetilde J\ra\,.
\label{eq: TJJ constraint}
\end{align}
To evaluate these, we require the three-point functions $\la XJJ\ra$ and $\la TJJ\ra$. At separated points, the most general structures compatible with the required conservation conditions are
\begin{align}
    \la XJJ\ra ={}& n_{XJJ}^{(1)} \frac{-V_1^2V_2V_3}{(-2P_{12})^2(-2P_{23})(-2P_{31})^2} \nonumber\\
    &+ n_{XJJ}^{(2)} \frac{ -2V_1^2H_{23} -2V_1V_2H_{31} -2V_1V_3H_{12} -H_{12}H_{31}}{2\,(-2P_{12})^2(-2P_{23})(-2P_{31})^2}\,,
\label{eq: XJJ ansatz} 
    \\[6pt]
    \la TJJ\ra'={}&n_{TJJ}^{(1)}\frac{V_1^2H_{23}+2H_{12}H_{31}+5V_1^2V_2V_3}{5\,(-2P_{12})^{5/2}(-2P_{23})^{1/2}(-2P_{31})^{5/2}} \nonumber\\
    &+ n_{TJJ}^{(2)}\frac{3V_1^2H_{23}+5V_1V_2H_{31}+5V_1V_3H_{12}-4H_{12}H_{31}}{5\,(-2P_{12})^{5/2}(-2P_{23})^{1/2}(-2P_{31})^{5/2}}\,.
\label{eq: TJJ separated ansatz}
\end{align}
The separated-point expression \eqref{eq: TJJ separated ansatz} is not the full distributional correlator. The $T\times J$ OPE is sufficiently singular that $\la TJJ\ra$ requires a semi-local completion,
\begin{equation}
    \la TJJ\ra = \la TJJ\ra' +\la TJJ\ra_{\rm c}\,,
\label{eq: TJJ full decomposition}
\end{equation}
where the contact contribution is
\begin{align}
    \la T^{\mu\nu}(x_1)J^\rho(x_2)J^\sigma(x_3)\ra_{\rm c} ={}& \delta^{(d)}(x_{12})\, \mathcal K^{\mu\nu\rho\alpha}\, \la J_\alpha(x_2)J^\sigma(x_3)\ra + (2 \leftrightarrow 3)\,,
\label{eq: TJJ contact term}
\end{align}
with
\begin{equation}
    \mathcal K_{\mu\nu\rho\sigma}
    \equiv
    B\left[
    \frac{2}{3}\,g_{\mu\nu}g_{\rho\sigma}
    -g_{\mu\rho}g_{\nu\sigma}
    -g_{\mu\sigma}g_{\nu\rho}
    \right].
\label{eq: K}
\end{equation}
This is the tensor structure required by the stress-tensor Ward identity~\cite{Osborn_1994}. We keep the coefficient $B$ as a free parameter. Although \eqref{eq: TJJ contact term} vanishes at separated points, it contributes to the integral transforms $\widetilde J$ in \eqref{eq: XJJ constraint} and \eqref{eq: TJJ constraint} and therefore cannot be discarded. In fact, the semi-local piece \eqref{eq: TJJ contact term} is essential: without it, the two constraints \eqref{eq: XJJ constraint} and \eqref{eq: TJJ constraint} admit no simultaneous solution, and it is only for the value of $B$ fixed below that both identities can be satisfied together.

\vskip 4pt
Evaluating the nonlocal transforms, and using the relations already obtained in the pure conformal-gravity sector, the two constraints \eqref{eq: XJJ constraint} and \eqref{eq: TJJ constraint} admit the following solution:
\begin{equation}
\begin{aligned}
    &\boxed{n_{XJJ}^{(1)} = \frac{64\pi}{7}\hs \frac{g}{a_1}\hs n_{TJJ}^{(1)}}\ , \quad {\rm with} \quad n_{XJJ}^{(2)}= n_{XJJ}^{(1)} \quad {\rm and} \quad n_{TJJ}^{(2)} = \frac{6}{7} n_{TJJ}^{(1)}\,,\\
    &\boxed{B = -\frac{2\pi}{21}\, n_{TJJ}^{(1)}} \quad {\rm and} \quad \boxed{\rule[-1.3ex]{0pt}{4.2ex}g'' = 3g}\ .
\end{aligned}
\label{eq: solution CG vectors}
\end{equation}
Again, the coupling of the PM field has been fixed in terms of the gravitational coupling. Moreover, the relative strength of the two symmetry-breaking channels is again fixed. It would be interesting to compute $\la XJJ\ra$ and $\la TJJ\ra$ directly from the bulk couplings \eqref{LhAA} and \eqref{LfAA}, and verify that they reproduce the solution obtained above. This would provide a nontrivial test of the relation $g''=3g$, which, unlike its scalar counterpart, cannot be inferred from the ratio of the two vertices.

\section{Conclusions and Outlook}
\label{sec: conclusions}
Many physically interesting theories contain weakly broken gauge symmetries. In anti-de Sitter space, gauge symmetry can be broken by boundary conditions, while in de Sitter space it can be broken by interactions. In the boundary description, this breaking appears as a violation of the conservation of the corresponding current. The main goal of this work was to determine how much of the constraining power of charge conservation survives in this setting. We used the fact that the weakly broken currents continue to obey integrated charge conservation identities, provided the ordinary charge action is replaced by a pseudo-charge action containing nonlocal terms fixed by the non-conservation equation. These pseudo-charge conservation identities therefore extend current-algebra methods beyond exactly gauge-invariant theories and provide a boundary framework for studying the Higgs mechanism in curved spacetime.

\vskip 4pt
We first used Yang--Mills theory in AdS as an illustrative example. The choice of boundary conditions for the charged scalars can break part of the gauge symmetry, rendering the corresponding currents weakly non-conserved. Nevertheless, the pseudo-charge conservation identities reconstruct the gauge algebra’s full action on the scalars, including the generators that mix the two quantization sectors. We then studied the Higgs mechanism for gravity in ${\rm AdS}\times{\rm AdS}$. While ordinary charge conservation requires a CFT with two conserved stress tensors to factorize into two decoupled theories, a marginal double-trace deformation couples these factors and breaks the conservation of one linear combination of the stress tensors. The pseudo-charge conservation identities hold for any value of the breaking parameter and determine the spin-two current algebra, in agreement with both the ACKK model and conformal perturbation theory. In both examples, symmetry breaking arises entirely from the boundary conditions, so the full gauge symmetry can be restored without modifying the bulk interactions. In Yang--Mills theory, this is achieved by imposing gauge-invariant boundary conditions, and the pseudo-charge identities simply rediscover the Lie-algebra structure found in the unbroken case. In ${\rm AdS}\times{\rm AdS}$, it is achieved by switching off the marginal coupling, consistent with the fact that the pseudo-charge identities leave its size unfixed. 

\vskip 4pt
The situation is different for conformal gravity in de Sitter space. In this case, solving the pseudo-charge conservation identities selects precisely the cubic correlators generated by conformal gravity and its couplings to scalar and vector matter. In the pure theory, the identities require $\la XTT\ra=0$ and fix all remaining three-point functions in terms of $\la TTT\ra$. Coupling to matter introduces additional double-trace operators in the non-conservation equation for the partially conserved spin-2 current: $\cO_2\cO_2$ for scalar matter and $J_\mu J^\mu$ for vector matter. The identities also fix the relative strength of these two breaking channels. Unlike in AdS, there is no freedom to choose boundary conditions at the future boundary, since the late-time behavior of the fields is fixed by the initial state. Moreover, the breaking parameter is not a free coupling, but is tied to the gravitational coupling, $g\propto 1/M_{\rm Pl}$. The PM gauge symmetry therefore cannot be restored without switching off the interactions themselves, which only happens in the decoupling limit $M_{\rm Pl}\to\infty$. In this case, the pseudo-charge identities do not rediscover the structure of an unbroken theory, but instead constrain interactions that exist only in the broken phase.

\vskip 10pt
We conclude with a list of open problems and future directions.
\begin{itemize}
\item Implementing the charge-conservation constraints studied in this paper (and in~\cite{Baumann:2025tkm}) remains somewhat involved. For a given operator spectrum, one must first construct a basis of conformally invariant three-point functions, identify the subset of structures satisfying the Ward--Takahashi identities for conserved currents, and then substitute these expressions into the relevant charge conservation identities. Even for relatively simple examples, this procedure can be computationally cumbersome. It would therefore be valuable either to automate these steps or to identify a simpler diagnostic, perhaps based on a particular limit of the constraints, that yields the restrictions on the CFT data more directly. It is natural to ask whether the recent twistor~\cite{Baumann:2024ttn} or Grassmannian~\cite{Arundine:2026fbr} descriptions of (A)dS correlators might provide such a formulation.

\item It would also be interesting to connect our results with the broader program of the conformal bootstrap~\cite{Poland:2018epd}. The pseudo-charge conservation identities derived here impose algebraic relations among operator dimensions and three-point-function coefficients, and therefore constrain precisely the CFT data that enter the crossing equations. Incorporating these relations into numerical or analytic bootstrap studies could substantially reduce the allowed parameter space, especially in theories containing conserved or weakly broken currents. Conversely, bootstrap bounds may help determine whether the solutions to the pseudo-charge conservation constraints can be completed into fully consistent CFTs. Such a combined approach could provide a systematic way to classify conformal theories admitting weakly broken higher-spin or gauge symmetries.

\item An important class of top-down constructions to study using the bootstrap method developed in this paper is the non-unitary (higher-derivative) generalizations of the critical $O(N)$ vector models. These critical CFTs are obtained by perturbing a free vector model by a relevant double-trace operator and following the resulting RG flow to an interacting IR fixed point. In the original unitary (two-derivative) setting, the spectrum of the three-dimensional critical $O(N)$ model and its bulk interpretation in terms of a weakly Higgsed Vasiliev theory in ${\rm (A)dS}_4$ were studied extensively in \cite{Giombi:2011ya,Maldacena:2012sf}.
Although some aspects of the non-unitary counterparts have been analyzed in \cite{Guo:2023qtt,Guo:2024bll}, a systematic exploration of this broader landscape remains an interesting open problem. In particular, the enlarged set of relevant primary operators available in non-unitary theories suggests a correspondingly richer space of possible RG flows. In analogy with the conjecture formulated for the unbroken case in \cite{Baumann:2025tkm}, one may further ask whether, under suitable additional physical assumptions (such as unitarity of the bulk dual, which would exclude conformal gravity), the critical vector models are the unique consistent solutions to the weakly broken higher-spin charge conservation identities.

\item Another natural future direction is to apply the pseudo-charge identities to correlation functions rather than to wavefunction coefficients. It was recently shown in~\cite{Sleight:2025dmt, Abhishek:2025oki} that the late-time boundary operators in de Sitter space come in shadow pairs, whose dimensions add up to the boundary dimension, and that conservation of the boundary current and stress tensor is weakly broken by double-trace operators built from the two members of a shadow pair. These non-conservation equations are precisely of the form studied here, with a breaking parameter fixed by the bulk coupling. 

\item Our ultimate ambition is to connect these theoretical developments to inflationary phenomenology~\cite{Arkani-Hamed:2015bza,Baumann:2017jvh,Lee:2016vti}. In Section~\ref{sec: Higgsing in CG}, we studied a spectrum of fields that is not too far removed from that encountered in models of the early universe. In particular, it would be intriguing to investigate whether the massless scalar appearing in Section~\ref{sec:CG-coupling-to-scalars} could play the role of the inflaton. Our consistency conditions would then constrain its interactions with both the graviton and the partially massless spin-2 field, thereby restricting the possible inflationary correlation functions. These couplings could generate characteristic signatures in scalar and tensor non-Gaussianities, including distinctive angular dependence and squeezed-limit behavior associated with the exchange of the partially massless field~\cite{Baumann:2017jvh}. More broadly, this setup may provide a concrete arena in which symmetry and consistency conditions determine observable properties of primordial fluctuations, allowing the formal constraints developed here to be confronted with cosmological data.
\end{itemize}

\vspace{0.2cm}
\paragraph{Acknowledgments} We thank Austin Joyce, Hayden Lee and Jiajie Mei for collaboration on related work. We would also like to thank Priyesh Chakraborty, Calvin Chen, Harry Goodhew, Nat Levine and Kamran Salehi Vaziri for enlightening discussions.
 
\vskip 4pt
The research of DB and NM is funded by the European Union (ERC, \raisebox{-2pt}{\includegraphics[height=0.9\baselineskip]{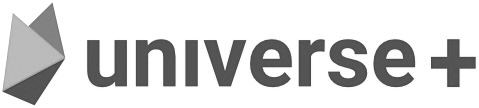}}, 101118787). DB is also grateful to the Avery-Tsui Foundation for funding the Stephen W.~Hawking Professorship of Cosmology at the University of Cambridge. DB is further supported by a Yushan Professorship at National Taiwan University (NTU) funded by the Ministry of Education (MOE). CRTJ is supported by funding from the University of Arizona. We also thank the Munich Institute for Astro-, Particle and BioPhysics (MIAPbP) for its hospitality during the completion of this work.

\vskip 4pt
We acknowledge the use of GPT-5.6 (OpenAI) and Opus 5/Fable 5 (Anthropic) to assist with fact-checking and editing. The authors take full responsibility for the content of the paper.

\newpage
\appendix
\section{Evaluation of Integrals}\label{sec: evaluating tensor integrals}

The analysis in the main text requires the evaluation of integrals of the form
\begin{equation}
    \int \text{d}^d x\, \frac{(x\cdot z_1)^{-a_4}(x\cdot z_2)^{-a_5}(x\cdot z_3)^{-a_6}}{\left[x^2\right]^{a_1}\left[(x+x_{12})^2\right]^{a_2}\left[(x+x_{13})^2\right]^{a_3}}\,,
\label{6indexint}
\end{equation}
where $x_i^\mu$ and $z_i^\mu$ are three-dimensional vectors, with $z_i^2 =0$.
We take $a_{4,5,6}\in \mathds{Z}_{\leq 0}$, while $a_{1,2,3}$ are initially treated as \textit{symbolic} (i.e.~$a_{1,2,3}\in \mathds{C}$). For the physically relevant examples considered in this paper, we evaluate these integrals using dimensional regularization in $d=3-2\epsilon$. This appendix provides a detailed account of the computational method.

\subsection{Tensor Reduction}
\label{ssec: vNV decomposition and IBP}
The numerator in (\ref{6indexint}) can be reduced by decomposing the integration variable $x^\mu$ in a basis adapted to the external vectors \cite{vanNeerven:1983vr,Ellis:2011cr}:
\begin{align}
    x^\mu &= (x\cdot x_{12})\hs \Check{x}_{12}^\mu + (x\cdot x_{13})\hs \Check{x}_{13}^\mu + (x\!\cdot\!n_3) n_3^\mu + x_{[-2\epsilon]}^\mu\,,
\end{align}
where the dual vectors $\Check{x}_i^\mu$ are defined by $x_i\cdot \Check{x}_j = \delta_{ij}$, whereas the unit vector $n_3^\mu$ satisfies $x_{12} \cdot n_3 = x_{13}\cdot n_3 = 0$.
Explicitly, we have
\begin{equation}
\begin{aligned}
    \Check{x}_{12}^\mu &= \frac{ x_{13}^2x_{12}^\mu - (x_{12}\cdot x_{13})x_{13}^\mu}{x_{12}^2 x_{13}^2-(x_{12}\cdot x_{13})^2}\,, \hspace{5mm} \Check{x}_{13}^\mu = \frac{ x_{12}^2x_{13}^\mu - (x_{12}\cdot x_{13})x_{12}^\mu}{x_{12}^2 x_{13}^2-(x_{12}\cdot x_{13})^2}\,, \\ 
    n_3^\mu &= \frac{\epsilon^{\mu \nu \rho} x_{12,\nu} x_{13,\rho}}{\sqrt{x_{12}^2 x_{13}^2 - (x_{12}\cdot x_{13})^2}}\,.
\end{aligned}
\end{equation}
Finally, $x_{[-2\epsilon]}^\mu$ denotes the projection of $x^\mu$ orthogonal to the physical three-dimensional space. Substituting this decomposition into the numerator factors $x \cdot z_i$, using $x_{[-2\epsilon]} \cdot z_i = 0$, and expressing the relevant dot products in terms of inverse propagators,
\begin{equation}
    x\cdot x_{ij} = \frac{1}{2}\left[(x+x_{ij})^2-x^2-x_{ij}^2\right], \quad ij=\{12,13\}\,, 
\end{equation}
reduces the integrals in (\ref{6indexint}) to a sum of simpler integrals of the form
\begin{equation}
    \int \text{d}^d x\, \frac{(x\cdot n_3)^{-a_4}}{\left[x^2\right]^{a_1}\left[(x+x_{12})^2\right]^{a_2}\left[(x+x_{13})^2\right]^{a_3}}\,,
\end{equation}
where $-a_{4}\in \mathds{Z}_{\geq 0}$. If $-a_4 \in 2\mathds{Z}_{\geq 0} +1$, then the integral vanishes because it is odd under reflection of the component of $x^\mu$ along $n_3^\mu$. For $-a_4 \in 2\mathds{Z}_{\geq 0}$, we use the following decomposition of the inverse metric:
\begin{equation}
    g^{\mu\nu} = x_{12}^\mu \Check{x}_{12}^\nu + x_{13}^\mu \Check{x}_{13}^\nu + n_3^{\mu} n_3^\nu + g_{[-2\epsilon]}^{\mu\nu}\,.
\end{equation}
This gives
\begin{equation}
    (n_3\cdot x)^2 = x^2 -(x_{12}\cdot x)(\Check{x}_{12}\cdot x) - (x_{13}\cdot x)(\Check{x}_{13}\cdot x) - \mu^2\,,
\end{equation}
where $\mu^2 \equiv g_{[-2\epsilon]}^{\mu\nu} x_\mu x_\nu$. Expanding this expression again in terms of inverse propagators further reduces the required integrals to the form
\begin{equation}
    G^{(d)}_{a_1,a_2,a_3}\big[\mu^{2k}\big] \equiv \int \text{d}^d x \, \frac{\mu^{2k}}{\left[x^2\right]^{a_1}\left[(x+x_{12})^2\right]^{a_2}\left[(x+x_{13})^2\right]^{a_3}} \,,
\label{triangleint}
\end{equation}
with $k\in \mathds{Z}_{\geq 0}$. For $k=0$, these are standard Euclidean scalar triangle Feynman integrals. If any index is non-positive, $a_i \leq 0$, the integrals reduce further to tensor bubbles and can be evaluated directly using (\ref{tensbubble}). For $k\in \mathds{Z}_{> 0}$, the so-called ``$\mu$-integrals" in (\ref{triangleint}) can be evaluated by \textit{dimension shifting} \cite{Bern:1993kr,Bern:1995db}. The key relation is
\begin{equation}
    G^{(d)}_{a_1,a_2,a_3}\big[\mu^{2k}\big] = \frac{\Omega_{d-3}}{\Omega_{d+2k-3}} G^{(d+2k)}_{a_1,a_2,a_3}\left[1\right], \hspace{8mm} \Omega_d \equiv \frac{2 \pi^{d/2}}{\Gamma\left(\frac{d}{2}\right)}\,.
\label{dimshift}
\end{equation}
In $d=3-2\epsilon$, these integrals are evanescent, as follows from the expansion of the prefactor:
\begin{equation}
    \frac{\Omega_{d-3}}{\Omega_{d+2k-3}} = - \frac{(k-1)!}{\pi^k}\epsilon + O(\epsilon^2)\,.
\end{equation}
They contribute in the limit $\epsilon \rightarrow 0$ only if the triangle integral in $d+2k$ dimensions has a $1/\epsilon$ divergence. Such a divergence can arise only from the region $|x|\rightarrow \infty$. However, the integral is manifestly convergent in this limit when $a_1+a_2+a_3>\frac{1}{2}(3+2k)$; in this range, the $\mu$-integrals may therefore be set to zero. Thus, the original tensor integral (\ref{6indexint}) reduces to a sum of scalar triangle integrals in dimensions $3+2k$.

\vskip 4pt
If the indices are positive integers, $a_i \in \mathds{Z}_{>0}$, they can be reduced further to a small set of master integrals by solving the corresponding integration-by-parts (IBP) identities \cite{Chetyrkin:1981qh}.\footnote{The restriction to integer indices is not intrinsic to the IBP relations. These relations remain well defined for symbolic indices, and the Laporta algorithm \cite{Laporta:2000dsw}, with an appropriately modified ordering on the space of integrals, can be implemented so that it terminates with finitely many master integrals \cite{Smirnov:2010hn}. However, standard off-the-shelf solvers such as FIRE6 \cite{Smirnov:2019qkx} do not currently handle IBP systems with symbolic indices natively without substantial user modification. We therefore use such tools only for integer indices and instead employ the series-expansion method described in Appendix~\ref{ssec: method of regions} for integrals with symbolic indices.} In general dimensions, the master integrals comprise the three bubble integrals $\{G^{(d)}_{1,1,0},G^{(d)}_{0,1,1},G^{(d)}_{1,0,1}\}$, which can be evaluated using (\ref{tensbubble}), and the unit-index triangle integral $G^{(d)}_{1,1,1}$. In $d=3$, the latter is conformal and can be evaluated using the general result
\begin{equation}
    G^{(d)}_{a_1,a_2,a_3} = \pi^{d/2}\frac{\Gamma\left(\frac{d}{2}-a_1\right) \Gamma\left(\frac{d}{2}-a_2\right) \Gamma\left(\frac{d}{2}-a_3\right)}{\Gamma\left(a_1\right) \Gamma\left(a_2\right) \Gamma\left(a_3\right)} \frac{1}{|x_{12}|^{d-2a_3}|x_{13}|^{d-2a_2}|x_{23}|^{d-2a_1}}\,,
\label{conformalint}
\end{equation}
provided that $a_1+a_2+a_3 = d$. For nonconformal integrals with non-integer indices, $a_i \notin \mathds{Z}_{>0}$, the scalar triangle integrals $G^{(d)}_{a_1,a_2,a_3}$ cannot generally be expressed in terms of elementary functions. As discussed in the following subsection, other methods are therefore required to extract the pseudo-charge conservation constraints.
	
\subsection{OPE Series Expansion}\label{ssec: method of regions}
As shown in \cite{Boos:1990rg} using Mellin--Barnes methods, the generic triangle integral (\ref{triangleint}) with $k=0$ can be expressed as a sum of four Appell $F_4$ functions, each of the form
\begin{equation}
    F_4\left[\alpha,\beta;\gamma,\delta;\frac{|x_{12}|^2}{|x_{23}|^2},\frac{|x_{13}|^2}{|x_{23}|^2}\right],
\end{equation}
where $\alpha,\beta,\gamma,\delta$ are linear combinations of the indices $a_i$ and the dimension $d$. The standard hypergeometric-series definition of this function,
\begin{equation}
    F_4\left[\alpha,\beta;\gamma,\delta;x,y\right] \equiv \sum_{m=0}^\infty \sum_{n=0}^{\infty} \frac{(\alpha)_{m+n}(\beta)_{m+n}}{(\gamma)_m(\delta)_n} \frac{x^{m}}{m!} \frac{y^n}{n!}\,,
\label{AppellSeries}
\end{equation}
converges in the domain $\sqrt{|x|}+\sqrt{|y|}<1$. In the present context, this becomes the condition $|x_{12}| + |x_{13}| < |x_{23}|$. However, because the $x_i$ are points in three-dimensional Euclidean space, their relative distances obey the triangle inequality $|x_{12}| + |x_{13}| \geq |x_{23}|$. The two conditions are incompatible. Consequently, the series (\ref{AppellSeries}) cannot be used in its present form and must instead be analytically continued to the region of physical interest \cite{Ananthanarayan:2020xut}.

\vskip 4pt
We instead construct a series expansion around the coincident-point (or OPE) limit. Without loss of generality, we consider
\begin{equation}
    \label{OPElimit}
    |x_{12}| \ll |x_{13}|\,.
\end{equation}
A general prescription for asymptotically expanding Feynman-like integrals before integration is provided by the \textit{method of regions} \cite{Beneke:1997zp}. Applying the method described in \cite{Jantzen:2012mw}, we find that the scalar triangle integral (\ref{triangleint}) has two nontrivial regions: a \textit{soft region} $x \sim x_{12}$ and a \textit{hard region} $x \sim x_{13}$. In each region, we expand the integrand in the small parameter according to the corresponding scaling of $x$ and then integrate term by term. The integrand in (\ref{triangleint}) expands as
\begin{equation}
    \begin{aligned}
    G_{a_1,a_2,a_3}^{(d)}\biggr\vert_{\text{soft}} &= \int \text{d}^d x \left[\frac{1}{\left[x^2\right]^{a_1}\left[(x+x_{12})^2\right]^{a_2}[x_{13}^2]^{a_3}} -\frac{2a_3(x\cdot x_{13})}{\left[x^2\right]^{a_1}\left[(x+x_{12})^2\right]^{a_2}[x_{13}^2]^{a_3+1}}\right. \\[10pt]
    &\hspace{20mm}\left. +\frac{a_3 \left(2(a_3+1)(x\cdot x_{13})^2- x^2 x_{13}^2\right)}{\left[x^2\right]^{a_1}\left[(x+x_{12})^2\right]^{a_2}[x_{13}^2]^{a_3+2}} + \cdots \right] ,\\[6pt]
    G_{a_1,a_2,a_3}^{(d)}\biggr\vert_{\text{hard}} &= \int \text{d}^d x \left[\frac{1}{\left[x^2\right]^{a_1+a_2}\left[(x+x_{13})^2\right]^{a_3}} -\frac{2a_2(x\cdot x_{12})}{\left[x^2\right]^{a_1+a_2+1}\left[(x+x_{13})^2\right]^{a_3}}\right. \\
    &\hspace{20mm}\left. +\frac{a_2\left(2(a_2+1)(x\cdot x_{12})^2-x^2 x_{12}^2\right)}{\left[x^2\right]^{a_1+a_2+2}\left[(x+x_{13})^2\right]^{a_3}} + \cdots \right].
    \end{aligned}
\end{equation}
By applying the dimension-shifting relation (\ref{dimshift}), we may set $k=0$ without loss of generality. Term-by-term integration then requires the tensor bubble integral
\begin{equation}
\begin{aligned}
    \int \text{d}^d x \,\frac{(x\cdot v_2)^{n}}{[x^2]^{b_1}[(x+v_1)^2]^{b_2}} &=  \sum_{r=0}^{\lfloor \frac{n}{2}\rfloor} c_r\, v_2^{2r} v_1^{d+2r-2b_1-2b_2} (v_1\cdot v_2)^{n-2r},
\end{aligned}
\label{tensbubble}
\end{equation}
where $n\in \mathds{Z}_{\geq 0}$ and
\begin{equation}
    c_r(d,b_i,n) \equiv \frac{\pi^{d/2}(-1)^n n!}{4^r r! (n-2r)!} \frac{\Gamma\left(b_1+b_2-\frac{d}{2}-r\right)\Gamma\left(\frac{d}{2}+r-b_2\right)\Gamma\left(\frac{d}{2}+n-r-b_1\right)}{\Gamma(b_1)\Gamma(b_2)\Gamma(d+n-b_1-b_2)}\,.
\end{equation}
In the limit (\ref{OPElimit}), the original integral is the sum of the two region expansions:
\begin{equation}
    G_{a_1,a_2,a_3}^{(d)} = G_{a_1,a_2,a_3}^{(d)}\biggr\vert_{\text{soft}} +G_{a_1,a_2,a_3}^{(d)}\biggr\vert_{\text{hard}}\,,
\end{equation}
where
\begin{align}
    G_{a_1,a_2,a_3}^{(d)}\biggr\vert_{\text{soft}} &= \frac{\pi ^{d/2}\,  \Gamma(\frac{d}{2}-a_1) \Gamma(\frac{d}{2}-a_2)  \Gamma(a_1+a_2-\frac{d}{2})}{\Gamma(a_1) \Gamma (a_2) \Gamma (d-a_1-a_2)}\left(\frac{|x_{12}|}{|x_{13}|}\right)^{d-2 a_1-2a_2}|x_{13}|^{d-2 a_1-2a_2-2a_3} \nonumber \\
   &\hspace{5mm}\times \left[1+\frac{a_3(2a_1-d)}{a_1+a_2-d} \frac{x_{12}\cdot x_{13}}{x_{13}^2}+\frac{a_3 (a_3+1) (2 a_1-d-2) (2 a_1-d)}{2(a_1+a_2-d-1) (a_1+a_2-d)}\frac{(x_{12}\cdot x_{13})^2}{x_{13}^4}\right. \nonumber \\
   &\hspace{12mm}\left.-\frac{a_3 (2 a_1-d) \left(2 a_1^2+2 a_1 (a_2-d-2)-2 a_2 (a_3+2)+(a_3+3) d+2\right)}{2(a_1+a_2-d-1) (a_1+a_2-d) (2 (a_1+a_2-1)-d)}\frac{x_{12}^2}{x_{13}^2}\right. \nonumber \\
   &\hspace{12mm}\left. +\, \mathcal{O}\left(\frac{|x_{12}|^3}{|x_{13}|^3}\right) \right],\\[10pt]
    G_{a_1,a_2,a_3}^{(d)}\biggr\vert_{\text{hard}} &= \frac{\pi ^{d/2} \Gamma(\frac{d}{2}-a_3) \Gamma(\frac{d}{2} -a_1-a_2)  \Gamma(a_1+a_2+a_3-\frac{d}{2})}{\Gamma (a_3) \Gamma (a_1+a_2) \Gamma
   (d-a_1-a_2-a_3)}|x_{13}|^{d-2
   a_1-2a_2-2a_3} \nonumber\\
   &\hspace{5mm}\times \left[1 +\frac{a_2 (2 (a_1+a_2+a_3)-d)}{ a_1+a_2} \frac{x_{12}\cdot x_{13}}{x_{13}^2} \right. \nonumber \\
   &\hspace{12mm}\left.+ \frac{a_2 (a_2+1)  (2
   (a_1+a_2+a_3)-d) (2 (a_1+a_2+a_3+1)-d)}{2(a_1+a_2) (a_1+a_2+1)}\frac{(x_{12}\cdot x_{13})^2}{x_{13}^4}\right. \nonumber\\
   &\hspace{12mm}\left.+ \frac{a_2  (2 (a_1+a_2+a_3)-d) \left(d (2 a_1+a_2+1)-2 (a_1+a_2+1)^2-2 a_1 a_3\right)}{2
    (a_1+a_2) (a_1+a_2+1) (2 (a_1+a_2+1)-d)}\frac{x_{12}^2}{x_{13}^2}\right. \nonumber\\
    &\hspace{12mm} \left. +\, \mathcal{O}\left(\frac{|x_{12}|^3}{|x_{13}|^3}\right) \right].
\end{align}
These explicit expressions reveal several features associated with special values of the indices~$a_i$. Because the soft series contains the prefactor $\left(|x_{12}|/|x_{13}|\right)^{d-2a_1-2a_2}$, the soft and hard series contribute commensurate powers of the expansion parameter only when $d-2a_1-2a_2 \in \mathds{Z}$. If $d-2a_i\in \mathds{Z}_{\leq 0}$, for $i=1,2$, or $d-2a_1-2a_2\in \mathds{Z}_{\geq 0}$, the soft-series contribution diverges and must be regulated by analytically continuing either the dimension or the indices. In either case, the finite part of the series contains factors of $\log\left(|x_{12}|/|x_{13}|\right)$.

\vskip 4pt
We have verified this region expansion in numerous examples for which the integral (\ref{triangleint}) can be evaluated analytically and then re-expanded. A simple example is the conformal integral~(\ref{conformalint}), for which the hard-region contribution vanishes because of the denominator factor $\Gamma(d-a_1-a_2-a_3)$; the soft-region series therefore reproduces the complete result.

\newpage
\section{Semi-Analyticity}
\label{sec:semianalytic}
In Section~\ref{sec: Weakly Broken Gauge Symmetry}, we explained that correlators of weakly non-conserved currents have to satisfy the conservation conditions only up to semi-analytic terms. We illustrated this point with a spin-2, depth-0 field, which played a central role in our treatment of conformal gravity in Section~\ref{sec: Higgsing in CG}. In this appendix, we present an additional example involving a spin-4, depth-0 field. We will denote $X\equiv X_{(4,0)}$ in this appendix.

\subsection{Three-Point Ansatz}
In $d$ dimensions, the three-point function of identical spin-4 fields admits 14 independent Bose-symmetric tensor structures:
\begin{equation}
    \langle X X X\rangle = \frac{\sum_{n=1}^{14} c_n\hs  G^{(4,0)}_n}{\left(-2P_{12}\right)^{\frac{d+3}{2}}\left(-2P_{23}\right)^{\frac{d+3}{2}}\left(-2P_{31}\right)^{\frac{d+3}{2}}}\,,
\label{ansatz444}
\end{equation}
where
\begin{equation}
    G^{(4,0)}_n =  \begin{pmatrix}
    V_1^4 H_{23}^4+\text{cyclic}\\
    V_1^3 V_2 H_{23}^3 H_{31}+\text{cyclic}\\
    V_1^2 V_2^2 H_{23}^2 H_{31}^2+\text{cyclic}\\
    V_1^4 V_2 V_3 H_{23}^3+\text{cyclic}\\
    V_1^3 V_2^2 V_3 H_{23}^2 H_{31}+\text{cyclic}\\
    V_1^4 V_2^2 V_3^2 H_{23}^2+\text{cyclic}\\
    V_1^3 V_2^3 V_3^2 H_{23} H_{31}+\text{cyclic}\\
    V_2^3 V_3^3 V_1^4 H_{23}+\text{cyclic} \\
    V_1^4 V_2^4 V_3^4 \\
    H_{12}^2 H_{23}^2 H_{31}^2 \\
    V_1^2 V_2^2 V_3^2 H_{12} H_{23} H_{31} \\
    V_1 V_2 H_{12} H_{23}^2 H_{31}^2+\text{cyclic} \\
    V_1^2 H_{12} H_{23}^3 H_{31}+\text{cyclic}\\
    V_1^2 V_2 V_3 H_{12} H_{23}^2 H_{31}+\text{cyclic}
    \end{pmatrix}.
    \label{eq:40basisd}
\end{equation}
Acting with the operator $\left(\partial_{x_1} \cdot D_{z_1}\right)^4$ produces a linear combination of the five tensor structures for $\langle \tilde{\mathcal{O}} X X\rangle$, where $\tilde{\mathcal{O}}$ is a spin-0 primary descendant of dimension $\Delta_{\tilde{\mathcal{O}}} = d+3$. These tensor structures satisfy
\begin{equation}
    \left(H_{23}\right)^{4-k}\left(H_{23}+2V_2 V_3\right)^{k} \, \sim \, |x_{23}|^{2k}, \hspace{10mm} k=0,1,2,3,4.
\label{tensorOXX}
\end{equation}
We can now repeat the analysis of Section~\ref{sec: Weakly Broken Gauge Symmetry}. Since $\left(-2P_{23}\right)^{\frac{d+3}{2}} \sim |x_{23}|^{d+3}$, whereas the numerator can vanish at most as $\sim |x_{23}|^{8}$ for any $d$, no semi-analytically conserved correlator that is not fully conserved can exist in $d>5$. Moreover, for $d\in 2\mathds{Z}_{>0}$, the denominator vanishes as a \textit{fractional} power of $x_{23}^2$, whereas the numerator always vanishes as an integer power. Such semi-analytically conserved correlators can therefore exist only for $d\in 2\mathds{Z}_{>0}+1$. Restricting to $d\geq 3$, the only possibilities are $d=3$ and $d=5$.\footnote{Conformal representations in $d=1,2$ are sufficiently different that they should be considered separately.}

\subsection{Broken Conservation}
In $d=5$, the numerator of $\langle (\partial\cdot D)^4 X X X\rangle$ must be proportional to the $k=4$ tensor structure in (\ref{tensorOXX}); the coefficients of the other four tensor structures must vanish. This requirement imposes four conditions on the 14 parameters in the ansatz (\ref{ansatz444}), leaving 10 semi-analytically conserved tensor structures. Requiring instead that $\langle (\partial\cdot D)^4 X X X\rangle$ vanish imposes 5 conditions on the 14 parameters, leaving 9 fully conserved tensor structures. We therefore conclude that a spin-4, depth-0 PM field in (A)dS$_{6}$ admits one anomalous cubic self-interaction. An explicit representative of the corresponding cohomology class is
\begin{equation}
    c_n = \left\{0, 0, 0, 0, 0, 0, 0, 0, 1, 0, -\frac{6991723}{38400}, -\frac{978283}{230400}, \frac{181889}{115200}, -\frac{181909}{7200}\right\}.
\end{equation}
As in the spin-2, depth-0 example considered in Section~\ref{sec: Weakly Broken Gauge Symmetry}, only one double-trace operator can appear in the broken conservation law $ (\partial\cdot D)^4 X \propto X_{\mu\nu\rho\sigma}X^{\mu\nu\rho\sigma}$.

\vskip 4pt
For $d=3$, the numerator of $\langle (\partial\cdot D)^4 X X X\rangle$ may be proportional to any linear combination of the $k=3$ and $k=4$ structures in (\ref{tensorOXX}), yielding three constraints. In addition, five of the original 14 tensor structures in (\ref{ansatz444}) are evanescent, leaving six semi-analytically conserved tensor structures. Full conservation instead imposes five constraints, leaving four fully conserved tensor structures. We therefore conclude that a spin-4, depth-0 PM field in (A)dS$_{4}$ admits two anomalous cubic self-interactions. Explicit representatives of the two cohomology classes are
\begin{equation}
\begin{aligned}
    c_n^{(1)} &= \left\{ 0, 0, 0, 1, 0, -\frac{19}{6}, \frac{271}{9}, \frac{788}{9}, \frac{1978}{9} \right\} \,,\\
    c_n^{(2)} &= \left\{ 0, 0, 0, 1, 0, -\frac{3}{2}, \frac{109}{5}, 64, \frac{798}{5}  \right\} \,,
\end{aligned}
\end{equation}
where we restrict the basis in (\ref{eq:40basisd}) to $n=1,...,9$. Matching these expressions to a conservation law broken by double-trace operators is more involved. A priori, there is a five-parameter family of double-trace operators with the required scaling dimension:
\begin{equation}
\begin{aligned}
    \left(\partial\cdot D\right)^4 X &\,=\, d_1 X_{\mu\nu\rho\sigma}\square X^{\mu\nu\rho\sigma}+ d_2  \left(\partial_\alpha X_{\mu\nu\rho\sigma}\right)\left(\partial^\alpha X^{\mu\nu\rho\sigma}\right)+ d_3 X_{\mu\nu\rho\sigma}\partial^\mu \partial^\alpha {X_\alpha}^{\nu\rho\sigma} \\[4pt]
    &\hspace{5mm} + d_4 \left(\partial^\mu X_{\mu\nu\rho\sigma}\right)\left(\partial_\alpha X^{\alpha \nu\rho\sigma}\right)+ d_5 \left(\partial_\alpha X_{\mu\nu\rho\sigma}\right)\left(\partial^\mu X^{\alpha \nu\rho\sigma}\right).
\end{aligned}
\label{eq:spin4depth0d3}
\end{equation}
However, not all of these operators are consistent with the requirement that the left-hand side be a conformal primary. Explicitly evaluating $\langle (\partial\cdot D)^4 X X X\rangle$ and matching the result to the basis in (\ref{eq:spin4depth0d3}) gives
\begin{equation}
\begin{aligned}
    d_n^{(1)} &= \left\{\frac{5}{36}, -\frac{29}{72}, -\frac{7}{27}, \frac{7}{27}, \frac{7}{18} \right\}\,,\\
    d_n^{(2)} &= \left\{ \frac{1}{6}, -\frac{1}{2}, -\frac{5}{18}, \frac{7}{18}, \frac{1}{2} \right\}\,.
\end{aligned}
\end{equation}
Although the details are not presented here, the pattern exhibited by these examples extends straightforwardly to higher spins. For a spin-6, depth-0 PM field, there are three anomalous cubic self-interactions in (A)dS$_4$, two in (A)dS$_6$, and one in (A)dS$_8$. We leave a systematic analysis of such interactions to future work.

\newpage
\section{Semi-Local Terms}\label{app: Semi-Local Terms}
Conformal symmetry fixes three-point functions only at separated points. The pseudo-charge conservation identities studied in this paper, however, involve nonlocal convolution integrals that may be sensitive to the behavior of correlators at coincident points. Defining these integrals therefore requires a prescription for the coincident-point limit, which may contain ambiguous \textit{semi-local} terms. In this appendix, we explain how {\it differential regularization} resolves these ambiguities. We find that the only required extension is that of $\la TJJ\ra$, first constructed in~\cite{Osborn_1994}. Our results are summarized in Table~\ref{tab: diffreg}.

\subsection{Extension Ambiguities}
\label{ssec: diffreg-criterion}
The objects requiring a definition are the coefficient functions in the operator product expansion,
\begin{equation}
	\cO_i(s)\hs\cO_j(0) = \sum_k C_{ij}{}^{k}(s)\, Y_k(0)\,,
	\qquad s\equiv \x_{ij}=\x_i-\x_j\,,
\label{eq:diffregOPE}
\end{equation}
where the sum runs over all operators $Y_k$ in the theory, and the equality holds within correlation functions at separated points.
Their leading short-distance behavior is
\begin{equation}
	C_{ij}{}^{k}(\lambda s) = \lambda^{-\omega}\, C_{ij}{}^{k}(s)\,, \qquad
	\omega \equiv \Delta_i+\Delta_j-\Delta_k\,.
\label{eq:diffregdegree}
\end{equation}
For $\omega<d$, the coefficient is locally integrable at $s=0$: its integral against any test function converges. Thus, $C_{ij}{}^{k}$ already defines a unique distribution on all of $\mathbb{R}^d$. For $\omega\geq d$, by contrast, the integral diverges at the origin, and an extension must be specified. Any two extensions with the same scaling degree agree away from $s=0$; their difference is therefore supported at the origin and can only be a finite sum of delta functions and their derivatives,
\begin{equation}
	\sum_{|\alpha|\hs =\hs \omega-d} b_\alpha\, \partial^\alpha \delta^{(d)}(s).
\label{eq:diffregfreedom}
\end{equation}
Within a three-point function, such an ambiguity multiplies the two-point function of the remaining pair, $\partial^\alpha\delta^{(d)}(\x_{ij})\hs \la Y_k(0)Y_l(\x_l)\ra$; this is the only structure that appears below.\footnote{Semi-local terms should be distinguished from \textit{ultralocal} terms, which are supported only at total coincidence $\x_1=\x_2=\x_3$. Ultralocal terms are ordinary local counterterms and can be generated by conformal anomalies. They play no role here: in the convolutions of Section~\ref{sec: pseudo}, they force two external points to coincide and therefore vanish for separated insertions.}

\vskip 4pt
We use {\it differential regularization}, which selects a particular extension by expressing the singular coefficient as derivatives of a milder kernel and defining those derivatives through integration by parts~\cite{Freedman:1991tk}. The contact terms then follow from distributional identities such as
\begin{equation}
	\frac{1}{d-2}\,\partial^\mu\partial^\nu \frac{1}{|s|^{d-2}}
	= \operatorname{PV}\left[\frac{d}{|s|^{d}}\left(\frac{s^\mu s^\nu}{s^2}-\frac{g^{\mu\nu}}{d}\right)\right]
	- \frac{\Omega_d}{d}\, g^{\mu\nu}\hs \delta^{(d)}(s)\,,\quad
	\Omega_d \equiv \frac{2\pi^{d/2}}{\Gamma(d/2)}\,,
\label{eq:diffregpv}
\end{equation}
whose traceless part is an ordinary principal value, whereas its trace part is a pure delta function.

\vskip 4pt
Most of the correlators we require have the form $\la Y Z Z\ra$, for which the relevant channel is $Y\times Z\to Z$, and~\eqref{eq:diffregdegree} reduces to $\omega = \Delta_Y$. This criterion singles out the stress tensor: the currents $J$ and $X$ have $\Delta=d-1<d$, so their channels are locally integrable, whereas $\Delta_T = d$ lies exactly at the borderline. This observation does not, however, guarantee that correlators containing $X$ but no $T$ are always unambiguous: mixed channels have $\Delta_k \neq \Delta_i$ and must be examined separately, as we do below.
 
\vskip 4pt
Finally, a semi-local term can affect a constraint only if two conditions are met. First, one of the nonlocal transforms of Section~\ref{sec: pseudo} must act on the correlator. Only then is the $\delta$-function integrated and the term promoted to separated points; otherwise, its support remains at coincidence. Second, the term must survive the projection onto auxiliary null vectors, $\z_i^2=0$, used to contract the identities. A completion proportional to the metric in the indices of a spinning operator does not survive this projection. Everything below follows from these two conditions.
 
\subsection{Semi-Local Completions}
\label{app:diffreg-completions}
Throughout the main text, each three-point function is understood as the sum of its separated-point expression $\la\,\cdots\ra^\prime$ and a semi-local completion $\la\,\cdots\ra_{c}$: $\la Y_1Y_2Y_3\ra = \la Y_1Y_2Y_3\ra' + \la Y_1Y_2Y_3\ra_{c}$. We now specify $\la\,\cdots\ra_{c}$ in each case. The results are summarized in Table~\ref{tab: diffreg}.
\begin{itemize}
\item \textbf{$\la T \cO \cO\ra$} The channel $T\times \cO_i \to \cO_i$ has $\omega = \Delta_T = d$, so~\eqref{eq:diffregfreedom} permits a contact term without derivatives. Conservation and the trace Ward identity fix this term completely~\cite{Osborn_1994}:
\begin{equation}
	\la T^{\mu\nu}(\x_1)\hs \cO_i(\x_2)\hs\cO_i(\x_3)\ra_{c}
	= -\frac{\Delta_i}{d}\, g^{\mu\nu} \Big[\delta^{(d)}(\x_{12})+\delta^{(d)}(\x_{13})\Big] \la \cO_i(\x_2)\cO_i(\x_3)\ra\,.
\label{eq:diffregTOO}
\end{equation}
This term is pure trace in the stress-tensor indices and is therefore removed by contraction with the auxiliary null vectors, $\z_1^2=0$.\footnote{Reference~\cite{Osborn_1994} uses an operator convention in which the trace Ward identity contains $(d-\Delta)$ rather than the $-\Delta$ used here. The two conventions differ by $-g^{\mu\nu}\big(\delta^{(d)}(\x_{12})+\delta^{(d)}(\x_{13})\big)\la \cO\cO\ra$, which is again pure trace and does not affect the discussion below.}

\item \textbf{$\la T JJ\ra$} 
The channel $T\times J\to J$ has $\omega = \Delta_T = d$, and its completion is~\cite{Osborn_1994}
\begin{equation}
	\la T^{\mu\nu}(\x_1)\hs J^\rho(\x_2)\hs J^\sigma(\x_3)\ra_{c}
	= \delta^{(d)}(\x_{12})\, {\cal K}^{\mu\nu\rho}{}_{\alpha}\, \la J^\alpha(\x_2) J^\sigma(\x_3)\ra + (2\leftrightarrow 3)\,,
\label{eq:diffregTJJ}
\end{equation}
with
\begin{equation}
	{\cal K}_{\mu\nu\rho\sigma} = B\left[\frac{2}{d}\, g_{\mu\nu}g_{\rho\sigma} - g_{\mu\rho}g_{\nu\sigma} - g_{\mu\sigma}g_{\nu\rho}\right] ,
	\quad g^{\mu\nu}{\cal K}_{\mu\nu\rho\sigma}=0\,,
\label{eq:diffregK}
\end{equation}
where $B$ is a constant coefficient.
This is the only completion that satisfies both criteria from the previous subsection. It ties the stress-tensor indices to the current indices, so it is not pure trace and survives contraction with the auxiliary null vector $\z_1^2=0$. Moreover, the transform $\wt J$ collapses the $\delta$-function onto separated points. Accordingly, the constant $B$ is retained as a free coefficient in the ansatz and fixed by the constraints. Equations~\eqref{eq:diffregTJJ} and \eqref{eq:diffregK} provide the only instance in which a semi-local term modifies an equation in the main text.

\begin{table}[t!]
\renewcommand{\arraystretch}{1.2}
\centering
\begin{tabular}{| L{0.16\textwidth}| C{0.2\textwidth}| C{0.07\textwidth}| C{0.12\textwidth}| L{0.25\textwidth}|}
\hline
 \cellcolor{gray!30} {Correlator} & \cellcolor{gray!30} {Channel} & \cellcolor{gray!30} $\omega$ & \cellcolor{gray!30} {Completion} & \cellcolor{gray!30} {Fate} \\
\hline
$\la T\cO_i\cO_i\ra$ & $T\times \cO_i\to \cO_i$ & $d$ & \eqref{eq:diffregTOO} & pure trace \\\hline
$\la TJJ\ra$ & $T\times J \to J$ & $d$ & \eqref{eq:diffregTJJ} & contributes \\\hline
$\la TXX\ra$, $\la TTT\ra$ & $T\times X\to X$, $T\times T\to T$ & $d$ & \eqref{eq:diffregTXX} & no transform acts \\\hline
$\la X\cO_2\cO_3\ra$ & $X\times \cO_3\to \cO_2$ & $d$ & \eqref{eq:diffregXOO} & pure trace \\\hline
 & $\cO_2\times\cO_3\to X$ & $d$ & \eqref{eq:diffregXOO} & pure trace \\\hline
$\la J\cO_1\cO_2\ra$ & $J\times\cO_2\to\cO_1$ & $d$ & none & odd index count \\\hline
$\la X\cO_3\cO_3\ra$ & $\cO_3\times\cO_3\to X$ & $d+1$ & none & odd index count \\\hline
all others & --- & $<d$ & none & --- \\\hline
\end{tabular}
\caption{Semi-local completions of  three-point functions entering pseudo-charge conservation identities.} 
\label{tab: diffreg}
\end{table}

\item \textbf{$\la TYY\ra$} The channels $T\times X\to X$ and $T\times T\to T$ also lie at $\omega = \Delta_T = d$, so both $\la TXX\ra$ and $\la TTT\ra$ admit derivative-free completions of the form
\begin{equation}
	\la T^{\mu\nu}(\x_1)\hs Y^{\rho \sigma}(\x_2)\hs Y^{\alpha \beta}(\x_3)\ra_{c}
	= \delta^{(d)}(\x_{12})\, {\cal K}^{\mu\nu \rho \sigma}{}_{\alpha' \beta'}\, \la Y^{\alpha \beta}(x_2)Y^{\alpha' \beta'}(x_3)\ra + (2\leftrightarrow 3)\,,
\label{eq:diffregTXX}
\end{equation}
with tensor structures analogous to~\eqref{eq:diffregK}, as given in~\cite{Osborn_1994}. Unlike~\eqref{eq:diffregTOO}, these completions are not pure trace and therefore survive the null projection. They nevertheless drop out for a different reason: in the conformal gravity identities~\eqref{eq: QTTT}, ~\eqref{eq: QXTT} and ~\eqref{eq: QXXX}, no transform acts on $\la TXX\ra$ or $\la TTT\ra$. The transform $\wt X$ acts only on $\la XXX\ra$, which requires no completion, so the support of these terms never leaves the coincidence locus. Their explicit form is therefore unnecessary and will not be recorded.

\item \textbf{$\la X \cO_2 \cO_3\ra$} The only case that does not involve a stress tensor is $\la X\cO_2\cO_3\ra$. The channel $X\times\cO_2\to\cO_3$ has $\omega=1$ and is locally integrable, whereas the other two channels are borderline: $\omega(X\times\cO_3\to\cO_2)=\omega(\cO_2\times\cO_3\to X)=d$. Because the ambiguity contains no derivatives, the spin-two indices must be carried by the metric:
\begin{equation}
\begin{aligned}
	\la X_{\mu\nu}(\x_1)\hs \cO_2(\x_2)\hs\cO_3(\x_3)\ra_{c} =\ &b_1\hs g_{\mu\nu}\hs \delta^{(d)}(\x_{13})\hs \la \cO_2(\x_2)\cO_2(\x_3)\ra \\
	&+ b_2\hs g^{\rho\sigma} \delta^{(d)}(\x_{23})\hs \la X_{\mu\nu}(\x_1) X_{\rho\sigma}(\x_3)\ra\,.
\end{aligned}
\label{eq:diffregXOO}
\end{equation}
Both structures are pure trace in the indices of $X_{\mu \nu}$: the first is removed by contraction with $\z^2=0$, whereas the second vanishes identically by the tracelessness of $X_{\mu  \nu}$, so that $g^{\rho\sigma}\la X_{\mu\nu}X_{\rho\sigma}\ra=0$. The constants $b_{1,2}$ are therefore unnecessary, and no trace Ward identity for $X$ has been assumed.

\item \textbf{Everything else} The remaining correlators require no completion. For $\la XXX\ra$, $\la XJJ\ra$, $\la J\cO_1\cO_1\ra$ and $\la J\cO_2\cO_2\ra$, every channel lies at $\omega<d$. Two additional correlators have $\omega \geq d$, but nevertheless admit no ambiguity: $\la J\cO_1\cO_2\ra$ through the channel $J\times\cO_2\to\cO_1$ at $\omega=d$, and $\la X\cO_3\cO_3\ra$ through $\cO_3\times\cO_3\to X$ at $\omega=d+1$. In both cases, the set of allowed ambiguities is empty. The coefficients $b_\alpha$ in~\eqref{eq:diffregfreedom} are constant tensors, so a parity-even structure can exist only if the total number of indices they carry (the $\omega-d$ derivatives together with the spins in the channel) is even. This number is odd in both cases: one for $\la J\cO_1\cO_2\ra$ and three for $\la X\cO_3\cO_3\ra$. Consequently, no such tensor can be constructed solely from the metric.
\end{itemize}

\newpage
\section{Perturbative Analysis of ${\rm AdS} \times {\rm AdS}$}\label{sec: CPT of AdSxAdS}
In this appendix, we analyze the model of~\cite{Aharony:2006hz, Kiritsis:2006hy} directly and compare it with the bootstrap results of Section~\ref{ssec: AdSxAdS bootstrap}. The starting point is the action
\begin{equation}
    S_{\rm tot} =  S_1+S_2+g\int\!\d^dx\,\mathcal{O}_1\mathcal{O}_2\,,
\label{equ:Stot}
\end{equation}
where $S_i[T_{\mu \nu}^{(i)}, \cO_i]$ denotes the action of ${\rm CFT}_i$. We use conformal perturbation theory to compute the correlators of the deformed theory to first order in the coupling $g$. Crucially, the pseudo-charge associated with the symmetry need not be postulated; it can be derived explicitly.

\subsection{Conformal Perturbation Theory}
\label{ssec: cpt}
Given the action~\eqref{equ:Stot}, the relevant correlators can be computed perturbatively. To first order in $g$, we have
\begin{equation}
    \la\cdots\ra=\la\cdots\ra_0-g\int\!\d^dy \,\la\cdots\hs\mathcal{O}_1(y )\mathcal{O}_2(y )\ra_0 +O(g^2)\,,
\label{eq: cpt}
\end{equation}
where $\la\cdots\ra_0$ denotes a correlator in the undeformed theory. We use the translation Ward identity \eqref{eq: T Ward identity} throughout:
\begin{equation}
    \partial_\mu\la T^{(i)\hs\mu\nu}(\x)\,Y_1\cdots Y_n\ra_0 =-\sum_{k}\delta^{(d)}(\x-y_k)\,\partial^{\hs\nu}_{y_k}\la  Y_1\cdots Y_n\ra_0\,,
\label{eq: factor ward}
\end{equation}
where $Y_i$ is a generic operator, and the sum on the right-hand side runs over insertions belonging to ${\rm CFT}_i$. We take the scalar two-point functions to be unit-normalized and keep the normalizations $c_i$ of the stress-tensor two-point functions explicit. We also use the rescaled central charges $\hat c_i$ defined in \eqref{eq: chat}.

\vskip 4pt
Equations~\eqref{eq: cpt} and \eqref{eq: factor ward} directly imply the non-conservation equations \eqref{eq: non-conservation eq T1} and \eqref{eq: non-conservation eq T2}. For example, applying a divergence to $\la T^{(1)\hs\mu\nu}\mathcal{O}_1\mathcal{O}_2\ra$ as computed from \eqref{eq: cpt}, the delta-function term collapses the $y$-integral and yields, at separated points,
\begin{equation}
    \partial_\mu\la T^{(1)\hs\mu\nu}(\x)\hs\mathcal{O}_1(x_2)\mathcal{O}_2(x_3)\ra =g\,\partial^{\hs\nu}_{\x}\la\mathcal{O}_1(\x)\mathcal{O}_1(x_2)\ra\, \la\mathcal{O}_2(\x)\mathcal{O}_2(x_3)\ra\,.
\end{equation}
This result implies
\begin{align}
    \partial^\mu T^{(1)}_{\mu\nu} &=g\hs (\partial_\nu\mathcal{O}_1)\mathcal{O}_2\,,
\end{align}
which reproduces the non-conservation equation \eqref{eq: non-conservation eq T1}. The analogous argument for $\la T^{(2)\hs\mu\nu}\mathcal{O}_1\mathcal{O}_2\ra$ gives \eqref{eq: non-conservation eq T2}. Similarly, the trace Ward identity gives $T^{(i)\hs\mu}{}_{\mu}=g\hs\Delta_i\hs\mathcal{O}_1\mathcal{O}_2$. Together with $\Delta_1+\Delta_2=d$, this confirms that $T_{\mu\nu} \equiv T^{(1)}_{\mu\nu} +T^{(2)}_{\mu\nu} -g\hs \eta_{\mu\nu}\hs \mathcal{O}_1\mathcal{O}_2$, defined in \eqref{eq: T conserved}, is both conserved and traceless.

\vskip 4pt
At order $g^2$, the breaking induces an anomalous dimension for $S_{\mu \nu}$, with $\gamma \propto g^2 (c_1+c_2)/(c_1c_2)$. This anomalous dimension was computed in~\cite{Aharony:2006hz} and agrees with the one-loop graviton mass in the bulk. It does not affect the pseudo-charge conservation identities at first order in $g$, so we use the unperturbed dimension $\Delta = d$ throughout.

\subsection{Pseudo-Charge Conservation}
We now determine the action of the pseudo-charge associated with $S_{\mu\nu}$ on the operators of the deformed theory, $\{T_{\mu \nu},S_{\mu \nu},\mathcal{O}_1,\mathcal{O}_2\}$. Instead of postulating an ansatz for the current algebra and fixing its free coefficients through the conservation identities, we derive the action directly from conformal perturbation theory. In order to determine the entire algebra,
the starting point is the identity
\begin{equation}
    0=\int\!\d^d\x\, \partial^\mu\la S_{\mu\nu}(\x)\,\mathcal{T}(x_1)\hs\mathcal{O}_1(x_2)\hs\mathcal{O}_2(x_3)\ra\,,
\label{eq: total derivative}
\end{equation}
where $\mathcal{T} \equiv \{T,S\}$, and we assume that the correlator decays at infinity. Expanding the integrand to first order in $g$ and organizing the result by the support of its $\delta$-functions produces every term in the pseudo-charge identity. The $\delta$-functions at the operator insertions determine the \textit{local} part of the charge action, whereas the separated-point remainder determines the \textit{nonlocal} part.

\vskip 4pt
We abbreviate the undeformed two- and three-point functions of each ${\rm CFT}_i$ by
\begin{equation}\label{eq: K and G definitions}
    \begin{aligned}
    K_i(x,y) &\equiv\la\mathcal{O}_i(x)\mathcal{O}_i(y)\ra_0\,, \\ G^{(i)}(x;y,w) &\equiv\la T^{(i)}(x)\mathcal{O}_i(y)\mathcal{O}_i(w)\ra_0\,.
    \end{aligned}
\end{equation}
We also introduce the general linear combination
\begin{equation}
    \mathcal{T}_{\mu\nu} \equiv A_1\hs T^{(1)}_{\mu\nu} + A_2\hs T^{(2)}_{\mu\nu} - g\hs A_3\hs \eta_{\mu\nu}\mathcal{O}_1\mathcal{O}_2\,.
\end{equation}
The traceless part of this expression reduces to the traceless parts of $T_{\mu \nu}$ and $S_{\mu \nu}$ for $(A_1,A_2)=(1,1)$ and $(\hat c_2,-\hat c_1)$, respectively, where $\hat c_i$ are the rescaled central charges defined in \eqref{eq: chat}. As shown below, the term proportional to $A_3$ does not contribute to the calculation.

\paragraph{Four-point function}
We first compute the four-point functions $\la S T \cO_1 \cO_2\ra$ and $\la S S \cO_1 \cO_2\ra$. These correlators vanish identically in the undeformed product theory. At $\mathcal{O}(g)$, they receive two types of contributions: one from inserting the deformation and another from the explicit double-trace terms in $T$ and $S$; see \eqref{eq: T conserved} and \eqref{eq: T non-conserved}, respectively.

\vskip 4pt
We first consider the double-trace terms, neither of which contributes to the pseudo-charge action. The improvement in $\mathcal{T}_{\mu \nu}(x_1)$ is proportional to $\eta_{\mu\nu}$ and is therefore removed by contraction with the null vector $z^\mu$. The improvement in $S_{\mu \nu}(\x)$ is proportional to $g\hs\eta_{\mu\nu}\hs\mathcal{O}_1(x)\mathcal{O}_2(\x)$. Acting with the divergence produces $g\hs\partial^{\hs\nu}_{\x}F(\x)$, where
\begin{align}
    F(\x) &\equiv\big\la(\mathcal{O}_1\mathcal{O}_2)(\x)\,\mathcal{T}(x_1,\z_1)\,\mathcal{O}_1(x_2)\mathcal{O}_2(x_3)\big\ra_0 \nonumber \\
    &=A_1\,G^{(1)}(x_1;\x,x_2)\,K_2(\x,x_3) +A_2\,K_1(\x,x_2)\,G^{(2)}(x_1;\x,x_3)\,.
\label{eq: F factorised}
\end{align}
After contraction with $\z_1$, the function $F$ has no contact support. As $\x\to x_1$, the only admissible contact term is the semi-local completion \eqref{eq:diffregTOO} of $\la T\mathcal{O}\mathcal{O}\ra$, which is pure trace in the indices of $\mathcal{T}_{\mu \nu}$ and therefore vanishes. As $\x\to x_{2,3}$, matching spins and dimensions excludes any contact term.\footnote{Such a term would correspond to a local double-trace contribution of $\mathcal{O}(g)$ to $[\hs\widehat Q^{\hs\mu},\mathcal{O}_i]$, which would require $\Delta_1+1=\Delta_1+\Delta_2+n$, with $n\in\mathbb{N}$, and hence $\Delta_2=1-n$. Given $\Delta_2>0$, the only possibility is $n=0$, which would require a spin-one structure constructed from two scalars without derivatives.} The contribution $\partial^{\hs\nu}_{\x}F(\x)$ is therefore the total derivative of a function supported at separated points. It integrates to zero and contributes to neither the local nor the nonlocal part of the charge action.

\vskip 4pt
We now retain the $\mathcal{O}(g^0)$ parts of $S$ and $\mathcal{T}$ and evaluate the contribution from the deformation insertion:
\begin{equation}
    -g \int \d^d y\, \big\la (\hat c_2 T^{(1)}-\hat c_1 T^{(2)})(x) \hs (A_1 T^{(1)}+A_2 T^{(2)})(x_1)\hs  \cO_1(x_2)\hs \cO_2(x_3) \hs \cO_1(y)\hs\cO_2(y)\big\ra_0\,.
\end{equation}
The undeformed six-point function factorizes \textit{exactly} into a ${\rm CFT}_1$ correlator and a ${\rm CFT}_2$ correlator, yielding four terms:
\begin{equation}
\begin{aligned}
    I_1&\equiv  \hat c_2A_1\, \la T^{(1)}(\x)T^{(1)}(x_1)\mathcal{O}_1(x_2)\mathcal{O}_1(y)\ra_0\,K_2(x_3,y)\,,\\[4pt]
    I_2&\equiv  \hat c_2A_2\, G^{(1)}(\x;x_2,y)\,G^{(2)}(x_1;x_3,y)\,,\\[4pt]
    I_3&\equiv -\hat c_1A_1\, G^{(1)}(x_1;x_2,y)\,G^{(2)}(\x;x_3,y)\,,\\[4pt]
    I_4&\equiv -\hat c_1A_2\, K_1(x_2,y)\,\la T^{(2)}(\x)T^{(2)}(x_1)\mathcal{O}_2(x_3)\mathcal{O}_2(y)\ra_0\,.
\end{aligned}
\label{eq: channels}
\end{equation}
Each term is multiplied by $(-g)$ and integrated over $y$. 

\paragraph{Taking the divergence}
We next take a divergence with respect to $\x$ in each term of \eqref{eq: channels} and apply the Ward identity \eqref{eq: factor ward} within the factor containing the stress tensor at $\x$. Because $T^{(i)}$ is exactly conserved in the undeformed theory, this procedure produces only contact terms. A separated-point remainder arises when a contact term is supported at the integrated insertion $y$ and is subsequently smeared by the $y$-integral. Combining the four channels and expressing the result in terms of correlators involving $\mathcal{T}$ gives 
\begin{align}\label{eq: divergence result}
    \partial^\mu\la S_{\mu\nu}(\x)\,\mathcal{T}(x_1)\mathcal{O}_1(x_2)\mathcal{O}_2(x_3)\ra =\ &-\,\delta^{(d)}(\x-x_1)\,\partial_{\hs\nu}^{x_1} \big\la \big(\hat c_2A_1T^{(1)}-\hat c_1A_2T^{(2)} \big)(x_1)\hs\mathcal{O}_1(x_2)\mathcal{O}_2(x_3)\big\ra \nonumber  \\[2pt]
    &-\,\delta^{(d)}(\x-x_2)\, \hat c_2\,\partial_{\hs\nu}^{x_2} \la\mathcal{T}(x_1)\hs\mathcal{O}_1(x_2)\mathcal{O}_2(x_3)\ra  \\[2pt]
    &+\,\delta^{(d)}(\x-x_3)\, \hat c_1\,\partial_{\hs\nu}^{x_3} \la\mathcal{T}(x_1)\hs\mathcal{O}_1(x_2)\mathcal{O}_2(x_3)\ra  + \mathcal{R}_\nu(\x)\,,  \nonumber
\end{align}
where the remainder
\begin{equation}
\begin{aligned}
    \mathcal{R}_\nu(\x) \equiv\ &g\,\hat c_2A_1\,\partial_{\hs\nu}^x G^{(1)}(x_1;x_2,x)\,K_2(x_3,\x) \ +\ g\,\hat c_2A_2\,\partial_{\hs\nu}^x K_1(x_2,x)\,G^{(2)}(x_1;x_3,\x)\\[3pt]
    &-g\,\hat c_1A_1\,G^{(1)}(x_1;x_2,\x)\,\partial_{\hs\nu}^x K_2(x_3,x) \ -\ g\,\hat c_1A_2\,K_1(x_2,\x)\,\partial_{\hs\nu}^x G^{(2)}(x_1;x_3,x)\,,
\end{aligned}
\label{eq: remainder}
\end{equation}
collects the terms in which the Ward identity acts on the deformation insertion at $y$. The factor $\delta^{(d)}(\x-y)$ collapses the $y$-integral and leaves a function supported at separated points.

\paragraph{Reading off the charge action}
By \eqref{eq: broken Ward identity}, the coefficient of $\delta^{(d)}(\x-x_1)$ in \eqref{eq: divergence result} gives the local action of the pseudo-charge on $\mathcal{T}$:
\begin{equation}
    \big[\hs\widehat Q^{\hs\mu},\mathcal{T}_{\rho\sigma}\big]_{\rm local} =\hs\partial^{\hs\mu}\big(\hat c_2A_1T^{(1)}_{\rho\sigma}-\hat c_1A_2T^{(2)}_{\rho\sigma}\big)\,.
\label{eq: local on J}
\end{equation}
The right-hand side is precisely the action of the undeformed generator $\widehat Q^{\hs\mu}|_{g = 0}=\hat c_2P^{(1)\mu}-\hat c_1P^{(2)\mu}$: each $P^{(i)\mu}$ translates its own factor and annihilates the other. Because the identity was extracted from a correlator that vanishes identically at $g=0$, it holds at first order in~$g$. Thus, up to pure-trace terms not determined by the null contraction, the local action receives no correction. For $T$ and $S$, \eqref{eq: local on J} gives
\begin{equation}
\begin{aligned}
    \big[\hs\widehat Q^{\hs\mu},T_{\rho\sigma}\big]
    &=\hs\partial^{\hs\mu}\big(\hat c_2T^{(1)}_{\rho\sigma}-\hat c_1T^{(2)}_{\rho\sigma}\big)\,, \\
    \big[\hs\widehat Q^{\hs\mu},S_{\rho\sigma}\big]
     &=\hs\partial^{\hs\mu}\big(\hat c_2^{\hs2}T^{(1)}_{\rho\sigma}+\hat c_1^{\hs2}T^{(2)}_{\rho\sigma}\big)\,.
\end{aligned}
\label{eq: Pt on spin two}
\end{equation}
We return below, in \eqref{eq: Pt full action}, to expressing this action in terms of $T$ and $S$. The second and third lines of \eqref{eq: divergence result} give
\begin{equation}
\begin{aligned}
    \big[\hs\widehat Q^{\hs\mu},\mathcal{O}_1\big]_{\rm local} &=+\hat c_2\hs\partial^{\hs\mu}\mathcal{O}_1\,, \\
    \big[\hs\widehat Q^{\hs\mu},\mathcal{O}_2\big]_{\rm local} &=-\hat c_1\hs\partial^{\hs\mu}\mathcal{O}_2\,,
\end{aligned}
\end{equation}
again with no correction. The relative sign reflects the fact that $\widehat Q^{\hs\mu}$ generates a \textit{relative} translation of the two theories.

\vskip 4pt
Finally, we integrate the remainder \eqref{eq: remainder} over $\x$ and then integrate by parts. Using the relation $\partial^\mu_u K_i(y,u)=-\partial^\mu_y K_i(y,u)$ to move the derivative onto the external points gives
\begin{equation}
\begin{aligned}
    \int\!\d^d\x\ \mathcal{R}_\nu(\x) = g\hs(\hat c_1+\hat c_2)\Bigg[& A_1\,\partial_{\hs\nu}^{x_3}\int\!\d^dy\ G^{(1)}(x_1;x_2,y)\,K_2(x_3,y)\\
    &-A_2\,\partial_{\hs\nu}^{x_2}\int\!\d^dy\ K_1(x_2,y)\,G^{(2)}(x_1;x_3,y)\Bigg]\,.
\label{eq: remainder integrated}
\end{aligned}
\end{equation}
At leading order, the two transforms in the brackets are $\la\mathcal{T}\hs\mathcal{O}_1(x_2)\hs\wt{\mathcal{O}}_1(x_3)\ra$ and $\la\mathcal{T}\hs\wt{\mathcal{O}}_2(x_2)\hs\mathcal{O}_2(x_3)\ra$, respectively. These are undeformed correlators and therefore contain no factor of $g$. Hence, \eqref{eq: remainder integrated} becomes
\begin{equation}
    -g\hs(\hat c_1+\hat c_2)\,\partial^{\hs\nu}_{x_2}\la\mathcal{T}\hs\wt{\mathcal{O}}_2\hs\mathcal{O}_2\ra +g\hs(\hat c_1+\hat c_2)\,\partial^{\hs\nu}_{x_3}\la\mathcal{T}\hs\mathcal{O}_1\hs\wt{\mathcal{O}}_1\ra\,,
\end{equation}
from which we read off the nonlocal terms in the pseudo-charge action on $\mathcal{O}_1$ and $\mathcal{O}_2$. The full action of the pseudo-charge on the spectrum, to first order in $g$, therefore is 
\begin{equation}
\begin{aligned}
    \big[\hs\widehat Q^{\hs\mu},\mathcal{O}_1\big]
    &=+\hat c_2\hs\partial^{\hs\mu}\mathcal{O}_1+g\hs(\hat c_1+\hat c_2)\hs\partial^{\hs\mu}\wt{\mathcal{O}}_2\,, \\[3pt]
    \big[\hs\widehat Q^{\hs\mu},\mathcal{O}_2\big]
    &=-\hat c_1\hs\partial^{\hs\mu}\mathcal{O}_2-g\hs(\hat c_1+\hat c_2)\hs\partial^{\hs\mu}\wt{\mathcal{O}}_1\,, \\[3pt]
    \big[\hs\widehat Q^{\hs\mu},T_{\rho\sigma}\big] &=\hs\partial^{\hs\mu}S_{\rho\sigma}\,, \\[3pt]
    \big[\hs\widehat Q^{\hs\mu},S_{\rho\sigma}\big] &=\hs \partial^{\hs\mu}T_{\rho\sigma} +(\hat c_2-\hat c_1)\hs\partial^{\hs\mu}S_{\rho\sigma}\,, 
\end{aligned}
\label{eq: Pt full action}
\end{equation}
where $\wt\cO_i$ is defined in \eqref{eq: shadow scalar}. Since $\Delta_1+\Delta_2=d$, these are ordinary conformal shadow transforms. The nonlocal part of the pseudo-charge is therefore conformally covariant, in contrast to the partially massless case discussed in Section~\ref{sec: Higgsing in CG}, where $ \wt X$ is not a shadow transform.

\vskip 4pt
Because the identity was contracted with the auxiliary null vector $\z_1$, the calculation determines the action on the spin-two operators only up to pure-trace terms. In particular, it cannot distinguish the full operators $T_{\mu\nu}$ and $S_{\mu\nu}$ from their $\mathcal{O}(g^0)$ parts, which differ by the improvement terms in \eqref{eq: T conserved} and \eqref{eq: T non-conserved}. Consistency fixes the completion displayed in \eqref{eq: Pt full action}: because $T_{\mu\nu}$ and $S_{\mu\nu}$ are traceless, the right-hand sides must also be traceless, which selects the full operators. Equivalently, this completion can be verified by retaining all trace terms in the calculation above. With this completion, \eqref{eq: Pt full action} agrees with the bootstrap solution of Section~\ref{ssec: AdSxAdS bootstrap}. Since $\hat c_1\hat c_2=1$, the spin-two commutators take the form \eqref{eq:ST-action} with $\mathcal{A}=\left(\begin{smallmatrix}0&1\\1&a\end{smallmatrix}\right)$ and $a=\hat c_2-\hat c_1$, whereas the scalar commutators agree upon identifying $a_1=\hat c_2$, $a_2=-\hat c_1$ and $\hat g=(\hat c_1+\hat c_2)\hs g$.

\paragraph{Consistency checks}
It is instructive to verify explicitly that the four contributions cancel. 
All of them are controlled by the two deformed three-point functions $\la T^{(i)}\mathcal{O}_1\mathcal{O}_2\ra$, which at first order in $g$ are given by the convolutions appearing in \eqref{eq: remainder integrated},
\begin{equation}
\begin{aligned}
    \la T^{(1)}(x_1)\mathcal{O}_1(x_2)\mathcal{O}_2(x_3)\ra &=-g\int\!\d^dy\ G^{(1)}(x_1;x_2,y)\,K_2(x_3,y)\,,\\[3pt]
    \la T^{(2)}(x_1)\mathcal{O}_1(x_2)\mathcal{O}_2(x_3)\ra &=-g\int\!\d^dy\ K_1(x_2,y)\,G^{(2)}(x_1;x_3,y)\,.
\end{aligned}
\label{eq: TOO perturbative}
\end{equation}
In particular, $\la\mathcal{T}\hs\mathcal{O}_1\mathcal{O}_2\ra =A_1\la T^{(1)}\mathcal{O}_1\mathcal{O}_2\ra +A_2\la T^{(2)}\mathcal{O}_1\mathcal{O}_2\ra$. Integrating \eqref{eq: divergence result} over $\x$ and adding the remainder \eqref{eq: remainder integrated}, the terms organize according to the two three-point functions, while the derivatives combine into the generator of translations:
\begin{equation}
    -\hat c_2A_1\big(\partial_{x_1}+\partial_{x_2}+\partial_{x_3}\big)^{\nu}\la T^{(1)}\mathcal{O}_1\mathcal{O}_2\ra +\hat c_1A_2\big(\partial_{x_1}+\partial_{x_2}+\partial_{x_3}\big)^{\nu}\la T^{(2)}\mathcal{O}_1\mathcal{O}_2\ra =0\,,
\end{equation}
which vanishes by translation invariance. 
The two structures cancel separately. Moreover, the expression is linear in $A_1$ and $A_2$, so the same computation applies simultaneously to $T$ and $S$.

\newpage
\section{Bulk Analysis of Conformal Gravity}\label{sec: Details on CG} 
In this appendix, we review conformal gravity from a bulk perspective, including its coupling to scalar and vector matter. 

\subsection{Pure Conformal Gravity}
The action for conformal gravity in $D=4$ spacetime dimensions is
\begin{equation} 
    S= -{\alpha^2\over 8} \int \d^4 x\, \sqrt{-g}\, C^{MNKL}C_{MNKL}\,,
\label{conformalglage}
\end{equation}
where $C_{MNKL}$ is the Weyl tensor and $\alpha^2$ is a dimensionless coupling constant. We assume $\alpha^2>0$ for definiteness, but the case $\alpha^2<0$ can be easily obtained by flipping the overall sign of contributions to the Lagrangian. In $D=4$, the theory is invariant under Weyl transformations of the form $\delta g_{MN}=2\sigma g_{MN}$.

\vskip 4pt
The theory has a de Sitter solution for any value of $H$, and expanding around this solution we get a massless spin-2 field and a PM spin-2 field as propagating degrees of freedom. This can be seen by first writing the theory in second-order form by introducing an auxiliary field $f_{MN}$, 
\begin{equation}
    S=\alpha^2 \int \d^4x \sqrt{-g} \left[ {H^2 \over 2} \big(R-6H^2\big)-\big(G_{MN}+3H^2 g_{MN}\big) f^{MN}+f_{MN}f^{MN}-f^2\right].
\label{flageee}
\end{equation}
Eliminating $f_{MN}$ via its equations of motion reproduces \eqref{conformalglage} upon discarding a Gauss--Bonnet total derivative term. We then expand (\ref{flageee}) around the de Sitter solution, $g_{MN}\rightarrow g_{MN}+h_{MN}$. We diagonalize the kinetic terms, $h_{MN}\rightarrow h_{MN}-(2/ H^2)f_{MN}$, and then canonically normalize the fields by defining $h_{MN}\rightarrow 2/(\alpha H) h_{MN}$ and $f_{MN}\rightarrow (H/\alpha) f_{MN}$. The resulting quadratic action is
\begin{equation}
    S_2=\int \d^4x \,\Big( {\cal L}_{\rm FP,0}(h)- {\cal L}_{{\rm FP},2H^2}(f)\Big)\,,
\end{equation}
where the standard Fierz--Pauli Lagrangian for a spin-2 field of mass $m$ is 
\begin{align}
    \frac{{\cal L}_{{\rm FP},m^2}(X)}{\sqrt{-g}} =& -{1\over 2}\nabla_K X_{MN} \nabla^K X^{MN}+\nabla_K X_{MN} \nabla^N X^{MK}-\nabla_M X\nabla_N X^{MN}+\frac{1}{2} \nabla_M X\nabla^M X \nonumber\\ 
    &+3H^2\left( X^{MN}X_{MN}-\frac{1}{2} X^2\right)-\frac{1}{2}m^2(X_{MN}X^{MN}-X^2)\, ,
\label{spin2lage}
\end{align}
showing that the theory propagates a massless graviton, $h_{MN}$, and a PM graviton, $f_{MN}$, with a relative sign between their kinetic terms.

\vskip 4pt
Expanding to cubic order, we find the following interactions between the massless and PM gravitons:
\begin{align} 
    \frac{{\cal L}_{hhh}}{\sqrt{-g}} &= -{1\over M_{\rm Pl}}\Big(  2 h^{MN} \nabla_N h_{KL}\nabla^L h_M^{\ K}+h_{MN} h^{KL}\nabla_K\nabla_L h^{MN}+2H^2 h^M_{\ N}h^N_{\ K}h^K_{\ M}  \Big)\, 
\label{Lhhh},
    \\[4pt] 
    \frac{{\cal L}_{hff}}{\sqrt{-g}} &=  {1\over M_{\rm Pl}}h^{MN}\Big(  f^{KL}\nabla_M\nabla_Nf_{KL}+2 f^{KL}\nabla_K\nabla_Lf_{MN}-2 \nabla_Kf_{ML}\nabla^L f_{N}^{\ K} \\
    & \ \ \ -4f^{KL}\nabla_M\nabla_Kf_{NL}-2H^2f^K_{\ M}f_{NK}  \Big)\,, \nonumber \\[4pt]
    \frac{{\cal L}_{ f  f  f}}{\sqrt{-g}} &=  - {2\over M_{\rm Pl}}\Big(  2 f^{MN} \nabla_N f_{KL}\nabla^L f_M^{\ K}+f_{MN} f^{KL}\nabla_K\nabla_L f^{MN}+2H^2 f^M_{\ N}f^N_{\ K}f^K_{\ M}  \Big)\, ,
\label{Lfff}
\end{align}
where the Planck mass, $M_{\rm Pl}= {\alpha H}$, is obtained by matching the $hhh$ cubic structure to that obtained from the Einstein--Hilbert action. Note that there is {\it no} $fhh$ coupling, so there exists a consistent truncation where we set the PM field to zero, leaving only Einstein gravity. In Section~\ref{ssec:PureCG}, this fact is reproduced from the pseudo-charge conservation identities.

\vskip 4pt
Conformal gravity can be coupled to any matter theory which is Weyl invariant on a curved background. The only bosonic examples of Weyl-invariant matter that contribute a single additional lower-spin particle to the spectrum are Maxwell electromagnetism and the conformally coupled scalar, so let us study these cases in turn.

\subsection{Coupling to Vectors}
\label{app: coupling to vectors}
The action for Maxwell electromagnetism on curved space is
\begin{equation} 
    S= -{1\over 4} \int \d^4 x\, \sqrt{-g}\,  F_{MN}F^{MN} \,,
\label{maxwelllage}
\end{equation}
where $F_{MN}=\nabla_M A_N-\nabla_N A_M$ is the standard Maxwell field strength and $A_M$ is the dynamical photon field. This action is Weyl invariant, where the photon does not transform, i.e.~$\delta g_{MN}=2\sigma g_{MN}$ and $\delta A_M = 0$.

\vskip 4pt
To see how the photon couples to the PM modes, we introduce the auxiliary field $f_{MN}$ just as in pure conformal gravity, and make the same diagonalization and canonical normalization transformations. We obtain the quadratic terms
\begin{equation}
    S_2=\int \d^4x \,\Big( {\cal L}_{\rm FP,0}(h)- {\cal L}_{{\rm FP},2H^2}(f)+{\cal L}_{\rm Maxwell}(A)  \Big)\, ,
\end{equation}
where ${\cal L}_{\rm Maxwell}(A)$ is the Lagrangian in \eqref{maxwelllage} evaluated on dS$_4$. For the cubic terms, we find the same graviton and PM couplings as in \eqref{Lhhh}--\eqref{Lfff}, as well as the following two-derivative on-shell cubic couplings between the photon, graviton and PM field:
\begin{align} 
    \frac{{\cal L}_{hAA}}{\sqrt{-g}} &= -{1\over M_{\rm Pl}}h^{MN}\left(  A^K\nabla_M\nabla_N A_K-2A^K \nabla_M \nabla_K A_N +2 H^2 A_M A_N  \right)\, ,
\label{LhAA}
    \\[4pt] 
    \frac{{\cal L}_{fAA}}{\sqrt{-g}} &=  {1\over M_{\rm Pl}}f^{MN}\left(  A^K\nabla_M\nabla_N A_K-2A^K \nabla_M \nabla_K A_N + H^2 A_M A_N  \right)\,.
\label{LfAA}
\end{align}
One can check that these are invariant on-shell under the massless graviton, PM and Maxwell gauge symmetries. Note the diagonal coupling to the PM field, which is only possible for the massless vector. In Section~\ref{sec:CG-coupling-to-vectors}, we derive the same structure from the pseudo-charge conservation identities.

\subsection{Coupling to Scalars}
\label{app:CouplingScalars}
The action for a conformally coupled scalar on curved space is
\begin{equation}
    S= \int \d^4 x\, \sqrt{-g}\left(  -{1\over 2} (\nabla\phi)^2 -{1\over 12} R\phi^2 +\lambda \phi^4 \right).\label{scallarlllaege}
\end{equation}
This action is Weyl invariant, where $\phi$ transforms nontrivially under the Weyl transformation, $\delta g_{MN}=2\sigma g_{MN}$ and $\delta \phi = - \sigma\phi$. Note the freedom to add a $\phi^4$ potential, which is separately invariant and thus comes with a free coefficient $\lambda$.

\vskip 4pt
Looking for maximally symmetric solutions to \eqref{scallarlllaege} added to \eqref{conformalglage}, we need the scalar to take a constant background value, $\phi=\phi_0$, and we find two branches of solutions,
\begin{align} 
    \phi_0 &=0\,,\\
    \phi_0 &= H/\sqrt{2\lambda}\, ,
\end{align}
where in the second case we have taken the $\phi_0>0$ solution without loss of generality.
We will see that in the trivial branch, $\phi_0=0$, there is a PM graviton, whereas in the nontrivial branch with $\phi_0\not=0$, the PM symmetry is spontaneously broken at tree level and there is a fully massive graviton in the spectrum. Note that the nontrivial branch only exists when $\lambda\not=0$. Moreover, for de Sitter solutions we must have $\lambda>0$, which, looking at \eqref{scallarlllaege}, corresponds to an upside down potential.

\subsubsection*{Nontrivial branch}
Consider first the nontrivial branch $\phi_0= H/\sqrt{2\lambda} >0$, with $\lambda>0$. We expand $\phi$ around this background solution, $\phi\rightarrow \phi_0+\varphi$. The fluctuation $\varphi$ now transforms nonlinearly under the Weyl symmetry, $\delta \varphi=-\sigma \phi_0+\cdots$, and the leading part of this must be a symmetry of the quadratic action.

\vskip 4pt
Making the field redefinition
\begin{equation}
    h_{MN}\rightarrow h_{MN}-c\hs f_{MN}+ \frac{c}{6\sqrt{2\lambda}} \frac{H}{\alpha^2} \hs \varphi \hs g_{MN}\,, \quad c\equiv {24 \alpha^2\lambda\over H^2(12 \alpha^2\lambda-1)}\,,
\end{equation}
we obtain the quadratic action
\begin{eqnarray}
    S_2= \frac{\alpha^2}{2}\int \d^4x \, \bigg[ \frac{1}{c} {\cal L}_{\rm FP,0}(h)- c \left( {\cal L}_{{\rm FP},2H^2}(f)+{H^2\over 12 \alpha^2\lambda} \sqrt{-g}\left(\tilde f_{MN}^{\hskip 1pt 2}-\tilde f^{\hskip 1pt 2} \right) \right)\bigg]\,. 
\label{PMnonsyglee}
\end{eqnarray}
Here, $\tilde f_{MN}\equiv f_{MN}-(\sqrt{2\lambda}/H)(\nabla_M\nabla_N+H^2 g_{MN})\hs \varphi$ is a combination which is invariant under the linearized Weyl gauge transformation, which acts as a PM transformation on $f_{MN}$ and a St\"uckelberg-like shift transformation on $\varphi$:
\begin{equation} 
    \delta f_{MN}= - (\nabla_M\nabla_N+H^2 g_{MN})\hs \sigma\, ,\  \ \delta \varphi=-{H\over \sqrt{2\lambda}} \sigma\,.
\label{stukgsye}
\end{equation}
Note that the scalar $\varphi$ only appears through the Fierz--Pauli mass-like combination $\tilde f_{MN}^{\hskip 1pt 2}-\tilde f^{\hskip 1pt 2}$.

\vskip 4pt
We can fix the gauge symmetry \eqref{stukgsye} by going to a unitary gauge $\varphi=0$. With this choice, we have $\tilde f_{MN}=f_{MN}$, and \eqref{PMnonsyglee} becomes the Lagrangian for a massless spin-2 field $h_{MN}$ and a massive spin-2 field $f_{MN}$, with a mass shifted from its PM value to $m^2=2H^2+\delta m^2$, where
\begin{equation} 
    \delta m^2 =-{1\over 6 \alpha^2\lambda} H^2\,.
\label{massshiftege}
\end{equation}
We thus have a PM Anderson mechanism, in which the PM gauge field eats the Goldstone field $\varphi$ and becomes fully massive.\footnote{Note that, unlike the PM Higgs mechanism in \cite{Hinterbichler:2025ost}, there is no radial mode in the theory, though there is an additional massless graviton.} This theory is related, through the scaling limits of \cite{Paulos:2012xe}, to one of the ghost-free bi-gravity theories of \cite{Hassan:2011zd}. Note that the mass shift \eqref{massshiftege} is negative when $\alpha^2>0$, so this field lies below the Higuchi bound in this case.

\vskip 4pt
By looking at the $hhh$ cubic terms, we can extract the Planck mass, which goes like
\begin{equation}
    M_{ \rm Pl}\sim H\sqrt{12\alpha^2\lambda-1\over \lambda}\,.
\end{equation}
Note that we need $M_{\rm Pl}\gg H$ if we are to have a weakly coupled semiclassical description. There are several regimes depending on the relative values of $\alpha^2$ and $\lambda$: If $\alpha^2 \lambda \gg 1$, then $M_{\rm Pl}\sim \alpha H$ and we must also have $\alpha^2\gg 1$ for a semiclassical description. In this case, the graviton mass shift \eqref{massshiftege} is small, $|\delta m^2|\ll H^2$. In this regime, the PM decoupling limit of \cite{DeRham:2018axr} can be reached, where the massive graviton decouples into a PM graviton and the scalar Goldstone $\varphi$: from expanding \eqref{PMnonsyglee} before going to the unitary gauge, it can be seen that the quadratic terms of $\varphi$ are those of a scalar with a mass $m_\varphi^2=-4H^2$, corresponding to the $k=1$ shift-symmetric scalar in the classification of \cite{Bonifacio:2018zex}. There is a strong-coupling point at $\alpha^2\lambda=1/12$, where the graviton kinetic term passes through zero and the Planck mass goes to zero. On the other side, we have a wrong-sign graviton with
\begin{equation}
    M_{\rm Pl}\sim H\sqrt{1-12\alpha^2\lambda\over \lambda}\,, 
\end{equation}
and if $\alpha^2 \lambda \ll 1$, then $M_{\rm Pl}\sim H/\sqrt{\lambda}$ and we can have a semiclassical description if we also have $\lambda \ll 1$. In this case, the graviton mass correction is large, $|\delta m^2| \sim {M_{\rm Pl}^2/\alpha^2} \gg H^2$. To keep the massive graviton within the semiclassical description, we must have $|\delta m^2| \ll M_{\rm Pl}^2$, which requires also $\alpha^2\gg 1$.

\subsubsection*{Trivial branch}
Consider now the trivial branch with $\phi_0=0$. In this case, the Weyl symmetry acts linearly on the fluctuations around the vacuum (i.e.~$\phi$ itself), and so the symmetry is not visible in the quadratic terms of $\phi$. To see how the scalar couples to the PM modes, we introduce the auxiliary field $f_{MN}$ just as in pure conformal gravity, and make the same diagonalization and canonical normalization transformations. We obtain the quadratic terms for the massless and PM gravitons, and a conformally coupled scalar,
\begin{equation}
    S_2=\int \d^4x \Big( {\cal L}_{\rm FP,0}(h)- {\cal L}_{{\rm FP},2H^2}(f)+{\cal L}_{{\rm scalar},2H^2}(\phi)  \Big)\, ,
\end{equation}
where the Lagrangian for a massive scalar on the background dS$_4$ is
\begin{equation} 
    \frac{{\cal L}_{{\rm scalar},m^2 }(\phi)}{\sqrt{-g}} = -{1\over 2}(\nabla\phi)^2-{1\over 2}m^2\phi^2\,.\label{scalarlagse}
\end{equation}
This trivial branch therefore provides a theory propagating a single conformally coupled scalar, in addition to the graviton and PM graviton.

\vskip 4pt
At cubic order, we obtain the same graviton and PM couplings as in \eqref{Lhhh}--\eqref{Lfff}, as well as the following on-shell cubic couplings between the scalar, graviton and PM field:
\begin{align} 
    \frac{{\cal L}_{h\phi\phi}}{\sqrt{-g}} &= -{1\over M_{\rm Pl}}h^{MN}\phi\nabla_M\nabla_N \phi \, ,\\[4pt] 
    \frac{{\cal L}_{f\phi\phi}}{\sqrt{-g}} &= {1\over M_{\rm Pl}}f^{MN}\phi\nabla_M\nabla_N \phi \,.
\label{Lhfphiphi}
\end{align}
Note the diagonal minimal coupling to the PM field, which is only possible for a conformally coupled scalar.

\subsubsection*{Higher-derivative conformal scalar}
Since we are not constrained by unitarity, we can also consider higher-derivative Weyl-invariant matter theories. The simplest of these is a fourth-order conformal scalar, which can be coupled as follows \cite{Manvelyan:2006bk,Karananas:2015ioa,Brust:2016gjy}:
\begin{equation} 
    S= \int \d^4x\,\sqrt{-g} \, \left[ F(\psi) C^{MNKL}C_{MNKL}+ {1\over 2} \left(\square\psi\right)^2-\left(R^{MN}-{1\over 3}R g^{MN}\right)\nabla_M \psi\nabla_N\psi \right],
\label{conformalglscage}
\end{equation}
where $F(\psi)$ is an arbitrary function.
This is invariant under a Weyl symmetry under which the scalar does not transform,
\begin{equation} 
    \delta g_{MN}=2\sigma g_{MN}\,,\ \ \ \delta\psi=0\,.
\end{equation}
The fact that the scalar does not transform allows for the presence of the arbitrary function $F(\psi)$ in the coupling.

\vskip 4pt
Any maximally symmetric space with constant $\psi=\psi_0$ is a solution. To render the action second order, we introduce an auxiliary tensor field $f_{MN}$ and an auxiliary scalar field $\phi$ as follows:
\begin{align}
    S&= \int \d^4x \, \sqrt{-g} \, \bigg( -8F(\psi)\left[ {H^2 \over 2} (R-6H^2)-(G_{MN}+3H^2 g_{MN}) f^{MN}+f_{MN}f^{MN}-f^2\right] \nonumber \\ 
    &\hspace{3cm} + F(\psi)\left(R_{MNKL}^2-4R_{MN}^2+R^2\right) \nonumber \\
    &\hspace{3cm} -\phi^2-\sqrt{2}\,\nabla_M\phi\nabla^M\psi -\left(R^{MN}-{1\over 3}R g^{MN}\right)\nabla_M \psi\nabla_N\psi \bigg)\, . \label{flageee322}
\end{align}
Eliminating $f_{MN}$ and $\phi$ by their equations of motion recovers \eqref{conformalglscage}, and they both vanish on the background solution.

\vskip 4pt
Because all but the first term in \eqref{conformalglscage} are invariant under shifts in $\psi$, the background value $\psi_0$ shows up only in the Taylor expansion of $F$, so without loss of generality we can expand in $\psi$ and take
\begin{equation} 
    F(\psi)=-{\alpha^2\over 8}+F_1\psi+F_2\psi^2+\cdots,
\label{Fexpansione}
\end{equation}
where we have parametrized the constant term to match the normalization we used for pure Weyl gravity in \eqref{conformalglage}. Expanding the metric around the background, $g_{MN}\rightarrow g_{MN}+h_{MN}$, we diagonalize the $f,h$ kinetic terms by taking $h_{MN}\rightarrow h_{MN}-(2/H^2) f_{MN}$, the scalar kinetic terms by taking $\psi\rightarrow \psi+{\phi/(\sqrt{2}H^2)}$, and we then canonically normalize by taking $h_{MN}\rightarrow {2/(\alpha H)}\hs h_{MN}$, $f_{MN}\rightarrow (H/\alpha)f_{MN}$, $\psi \rightarrow {\psi/(\sqrt{2}H)}$, $\phi\rightarrow H\phi$, after which we get the quadratic terms
\begin{equation} 
    S_2 = \int \d^4x  \Big[ {\cal L}_{\rm FP,0}(h)- {\cal L}_{\rm FP,2H^2}(f)- {\cal L}_{{\rm scalar}, 0}(\psi) + {\cal L}_{{\rm scalar}, 2H^2 }(\phi)   \Big]\,,
\end{equation}
where the scalar Lagrangian is as in \eqref{scalarlagse}. Note that the Gauss--Bonnet part and the $\psi$-dependence of $F$ do not affect the quadratic terms.

\vskip 4pt
We see that in addition to the fields of pure conformal gravity, the spectrum contains two scalars: one massless and one with the conformal mass, with opposite sign kinetic terms (note that the overall sign of the scalar sector can be changed by choosing a different overall sign for all but the first term in the original action \eqref{conformalglscage}).

\vskip 4pt
We get the same cubic terms \eqref{Lhhh}--\eqref{Lfff} involving the tensors and the following cubic couplings involving one tensor and two scalars:
\begin{eqnarray} 
    {1\over \sqrt{-g}} {\cal L}_{ h \psi \psi } &=&  {1\over M_{\rm Pl}} h^{MN}\psi \nabla_M \nabla_N \psi\,, \\
    {1\over \sqrt{-g}} {\cal L}_{ h \phi \phi } &=&  -{1\over M_{\rm Pl}} h^{MN}\phi \nabla_M \nabla_N \phi \,, \\
    {1\over \sqrt{-g}} {\cal L}_{ f  \phi \phi } &=&  {2\over M_{\rm Pl}}f^{MN}\phi \nabla_M \nabla_N \phi \,,
\label{Lhfphiphi higher der}    
    \\
    {1\over \sqrt{-g}} {\cal L}_{ f  \psi \phi } &=&  {2\over M_{\rm Pl}}f^{MN}\psi \nabla_M \nabla_N \phi \,,
\end{eqnarray}
where we fixed the Planck mass $M_{\rm Pl}$ in the same way as before.
Note that the minimal couplings to $h$ have opposite signs, consistent with the opposite sign kinetic terms. Note also that there is {\it no} $f\psi\psi$ coupling: as above, a diagonal coupling to the PM field is only possible for the conformal mass, and $\psi$ is massless. PM invariance does permit the off-diagonal $f\psi\phi$ coupling, and it occurs. The $f\phi\phi$ coupling is twice as large as in the theory with only a conformal scalar. There will also be four-derivative cubic couplings containing one scalar and two tensors: these are all proportional to $F_1$ in the expansion \eqref{Fexpansione}, and will not be important for the consistency computations. There are {\it no} cubic couplings involving three scalars.

\newpage
\phantomsection
\addcontentsline{toc}{section}{References}
\bibliographystyle{utphys}
{\linespread{1.075}
\bibliography{SB-Refs}
}
	
\end{document}